\documentclass[aps,twocolumn,nofootinbib,showpacs,floatfix]{revtex4}
\usepackage{graphics} 
\usepackage{amssymb,amsmath}
\usepackage{amsmath}
\usepackage{psfrag}
\usepackage{graphicx}
\newcommand{\ve}[1]{\mathbf{#1}}
\def\al{\alpha} 
 
\def\de{\delta}

\def\ga{\gamma}

\def\om{\omega}

\def\tr{\mbox{tr}}

\newcommand{\bm}[1]{\mbox{\boldmath $#1$}} 
 
\def\be{\begin{equation}} 
\def\ee{\end{equation}} 
\def\bea{\begin{eqnarray}} 
\def\eea{\end{eqnarray}}  
\def\bean{\begin{eqnarray*}} 
\def\eean{\end{eqnarray*}} 
\def\dd{\partial}

\def\bk{{\bf k}}  
\def\bx{{\bf x}}  
\def\bv{{\bf v}}
\def\br{{\bf r}}

\def\bfo{{\bf f}} 
 
\def\bn{{\bf n}} 
\def\bu{{\bf u}}
\def\bR{{\bf R}}
\def\mD{{\mathcal D}}

\def\bK{{\bf K}}

\def\by{{\bf y}}

\def\bg{{\mathbf g}}

\def\tmD{{\tilde\mD}}
\def\tbu{{\tilde\bu}}

\def\bse{\begin{subequations}}
\def\ese{\end{subequations}}

\def\da{{\dot a}}
\def\dda{{\ddot a}}

\def\bX{{\mathbf X}}
\def\bF{{\mathbf F}}

\def\bm{\mathbf{m}}

\def\mV{\mathcal{V}}

\def\mP{\mathcal{P}}

\def\lsim{\raise 0.4ex\hbox{$<$}\kern -0.8em\lower 0.62ex\hbox{$\sim$}} 
\def\gsim{\raise 0.4ex\hbox{$>$}\kern -0.7em\lower 0.62ex\hbox{$\sim$}} 
\def\mP{\mathcal{P}}
\def\mQ{\mathcal{Q}}

\def\bk{\mathbf{k}}

\def\f0N{f_0^{(N)}}
\def\bec{\begin{center}}
\def\eec{\end{center}}

\begin{document} 
%\draft 
%\twocolumn[\hsize\textwidth\columnwidth\hsize\csname@twocolumnfalse\endcsname
\title{Linear perturbative theory of the discrete cosmological N-body
problem}

\author{B. Marcos, T. Baertschiger}   
\affiliation{Dipartimento di Fisica, Universit\`a ``La Sapienza'',
P.le A. Moro 2,
I-00185 Rome,
Italy,\\
\& ISC-CNR, 
Via dei Taurini 19,
I-00185 Rome,
Italy.} 
\author{M. Joyce}   
\affiliation{Laboratoire de Physique Nucl\'eaire et de Hautes Energies,  
Universit\'e Pierre et Marie Curie-Paris 6, 
UMR 7585, 
%Place Jussieu, 
%Tour 33 -RdC, 75252 Paris Cedex 05, 
Paris, F-75005 France.} 
\author{A. Gabrielli}   
\affiliation{ISC-CNR, 
Via dei Taurini 19,
I-00185 Rome,
Italy,\\\& SMC-INFM,
Dipartimento di Fisica, 
Universit\`a ``La Sapienza'',
P.le A. Moro 2,
I-00185 Rome,
Italy.} 
\author{F. Sylos Labini}   
\affiliation{ ``E. Fermi'' Center, Via Panisperna 89 A, Compendio del 
Viminale, I-00184 Rome, Italy,\\
\& ISC-CNR, 
Via dei Taurini 19,
I-00185 Rome,
Italy.}     
\begin{abstract}   
\begin{center}    
{\large\bf Abstract}   
\end{center}    
We present a perturbative treatment of the evolution under
their mutual self-gravity of particles displaced off an infinite 
perfect lattice, both for a static space and for a homogeneously 
expanding space as in cosmological N-body simulations. The treatment, 
analogous to that of perturbations to a crystal in solid state physics, 
can be seen as a discrete (i.e. particle) generalization of the 
perturbative solution in the Lagrangian formalism of a self-gravitating 
fluid. Working 
to linear order, we show explicitly that this fluid evolution 
is recovered in the limit that the initial perturbations are restricted to 
modes of wavelength much larger than the lattice spacing. The full spectrum of 
eigenvalues of the simple cubic lattice contains both oscillatory
modes and unstable modes which grow slightly faster than  
in the fluid limit.
A detailed comparison of our perturbative treatment, at linear order, 
with full numerical simulations is presented, for two very  different 
classes of initial  perturbation spectra. We find that the range of 
validity is similar to  that of the perturbative fluid approximation
(i.e. up to close to ``shell-crossing''), but 
that the accuracy in tracing the evolution is superior. The formalism
provides a powerful tool to systematically calculate discreteness
effects at early times in cosmological N-body simulations.
\end{abstract}    
\maketitle   
\date{today}  

\twocolumngrid

\section{Introduction}

The standard paradigm for formation of large scale structure in the
universe is based on the growth of small initial density fluctuations
in a homogeneous and isotropic medium (see e.g.
\cite{peebles_80}). In the currently most popular cosmological models,
a dominant fraction (more than 80 \%) of the clustering matter in the
universe is assumed to be in the form of microscopic particles which
interact essentially only by their self-gravity.  At the macroscopic
scales of interest in cosmology the evolution of the distribution of
this matter is then very well described by the Vlasov or
``collisionless Boltzmann'' equations coupled with the Poisson
equation (see e.g. \cite{binney_87}).  A full solution, either
analytical or numerical, of these equations starting from appropriate
initial conditions (IC) is not feasible. There are, on the one hand,
various perturbative approaches to their solution (for reviews see
e.g. \cite{sahni_95,bernardeau_02}), which allow one to understand the
evolution in some limited range (essentially of small to moderate
amplitude fluctuations). On the other hand, there are cosmological
N-body simulations (for reviews see
\cite{bagla_97,bertschinger_98,hockney_99}), which solve numerically
for the evolution of a system of $N$ particles interacting purely 
through gravity,
with a softening at very small scales. The number of particles $N$ in
the very largest current simulations \cite{springel_05} is $\sim
10^{10}$, many more than two decades ago, but still many orders of
magnitude fewer than the number of real dark matter particles ($\sim
10^{80}$ in a comparable volume for a typical candidate).  The
question inevitably arises of the accuracy with which these
``macro-particles''  trace the desired correlation properties
of the theoretical models. This is the problem of discreteness in
cosmological N-body simulations. It is an issue which is of
considerable importance as cosmology requires ever more precise
predictions for its models for comparison with observations.

Up to now the primary approach to the study of discreteness
in N-body simulations  has been through numerical studies of 
convergence (see e.g. \cite{power_03,diemand_04}),  
i.e., one changes the number of particles in a simulation and studies
the stability of the measured quantities. Where results seem fairly 
stable, they are assumed to have converged to the continuum limit. 
While this is a coherent approach, it is far from
conclusive as, beyond the range of perturbation theory, we have no 
theoretical ``benchmarks'' to compare with. Nor is there any systematic
theory of discreteness effects, e.g., we have no theoretical knowledge 
of the $N$ dependence of the convergence. Given that typically $N$ 
is varied over a very modest range (typically one or two orders 
of magnitude) compared to that separating the simulation from the 
model (typically 70 orders of magnitude) there is much room for error.

Different mechanisms by which discreteness effects may make the
evolution of N-body simulations different to that of the fluid limit
have been discussed in the literature. A very basic consideration is
that of the discreteness effects introduced already in the IC, before
any dynamical evolution. Indeed there are necessarily discrepancies
between the correlation properties of the discretised IC and those of
the input theoretical model, as there are intrinsic fluctuations
associated with the particles themselves.  For analysis and discussion
of these effects see, e.g.,
\cite{baertschiger_02,knebe_02,knebe_03,baertschiger_03,joyce_04,joyce_04b}. What
is probably the most obvious effect of discreteness, and certainly the
one most emphasised in the literature, is two body collisionality:
pairs of particles can have strong interactions with one another,
which is an effect absent in the collisionless limit. For analysis and
discussion of these effects see, e.g.,
\cite{melott_93,melott_97,splinter_98,baertschiger_02t,binney_02}.

In this paper we present an approach which allows in principle a
systematic understanding of discreteness effects {\it between these
two regimes}, in the evolution from the IC up to the time when two
body collisions start to occur.  We do so by developing a perturbative
solution to the fully discrete cosmological N-body problem, which is
valid in this regime.  This essentially analytic solution can be
compared to the analogous fluid ($N \rightarrow \infty$) solution, and
one can understand exhaustively the modifications introduced, at a
given time and length scale, by the finiteness of $N$.  While the
usefulness of the approach is restricted to the regime of validity of
this perturbative approach, we can gain considerable insights into the
effects of discreteness and how they introduce error.  Some of the
essential results have already been briefly reported in
\cite{jmgbsl_05}. In this paper we describe in much greater detail the
perturbative method used to describe the evolution, and evaluate its
regime of validity by extensive comparison with numerical
simulations. In a forthcoming paper \cite{joyce_06} we will discuss
the application of this method to the study of discreteness effects in
N-body simulations, providing precise quantifications of these effects
in the regime in which our treatment is valid.

The perturbative scheme we employ is one which is well known and
standard in solid state physics, as it is that used in the treatment
of perturbations of a crystal about the local or global minimum of its
internal energy. Indeed the class of cosmological N-body simulations
we consider are those which start by making very small perturbations
to particles initially placed on a perfect simple lattice
\cite{efstathiou_85,navarro_96,bertschinger_98,power_03,smith_03,diemand_04}. Up
to an overall change in sign, our perturbative scheme is precisely
that one would use for the analysis of the extensively studied Coulomb
lattice or Wigner crystal (see e.g. \cite{clark_57,pines_63}), of $N$
particles on a lattice interacting by a pure unscreened Coulomb
force. At linear order (harmonic analysis) one simply solves a $3N
\times 3N$ eigenvalue problem to determine the eigenmodes and
eigenvalues of the displacements off the crystal.  This can be done at
very low cost in computational resources because of the symmetries of
the lattice. Stable (i.e.  dynamically oscillating) modes in one
problem become unstable (growing and decaying) modes in the other
problem, and vice versa.  One consequence of this which we will
discuss briefly here, and more extensively in \cite{joyce_06}, is
that, for what concerns discreteness in N-body simulations, there are
qualitatively different features on the simple cubic (sc) lattice
compared to the body centered cubic (bcc) or face centered cubic (fcc)
lattice.

A crucial step in our analysis is evidently the comparison of our
solutions for the evolution with those obtained from the treatment of
the continuous self-gravitating system.  This latter problem can also
be solved in a limited range with a perturbative treatment of the
equations of a self-gravitating fluid, which are obtained by
truncation of the collisionless Boltzmann equation. Given that our
perturbation scheme works with the displacements of the particles, one
might anticipate that the appropriate perturbation scheme to compare
with is that given in the Lagrangian formulation of these fluid
equations, in which the evolution is described in terms of the
trajectories of fluid elements \cite{buchert_92}.  We show explicitly,
at linear order, that this is the case: taking the limit in which the
perturbations to the lattice are of wavelengths much larger than the
lattice spacing $\ell$ the evolution described by our scheme maps
precisely on to that at the same order in the Lagrangian description
of the fluid. The Zeldovich approximation, which is simply the
asymptotic form of this solution, can then be understood in very
simple analogy with the long-wavelength coherent ``plasma
oscillations'' in a unscreened charged plasma.

The paper is organized as follows. In the next section we describe the
perturbative treatment for perturbations off a perfect lattice, for
the specific case of gravity. We work firstly, for simplicity, in a
static Euclidean universe, giving the explicit expressions for the
evolution from general IC (i.e. any perturbation from the lattice). In
the following section we explicitly solve the evolution for the case
of a simple cubic lattice, and discuss in detail the structure of the
spectrum of eigenvalues. In Sect.  IV we generalize our results to the
case of an expanding universe, and then show explicitly the recovery
of the fluid limit given by the solution at the same order of the
Lagrangian formulation of the equations of a self-gravitating
fluid. In the next section we present a comparison of the evolution
described by our approximation with that of full numerical N-body
simulations.  We consider both uncorrelated initial perturbations (a
``shuffled lattice'') and a set of highly correlated perturbations
(with a power spectrum of density fluctuations $\sim k^{-2}$), like
that in current cosmological models. We verify that the agreement is
very good, and that the evolution is traced with considerably better
accuracy than by the fluid limit at the same order. Both
approximations break down when particles start to approach one another
(i.e. ``shell-crossing'' in fluid language), which we parametrise
through an appropriate statistical measure. In the final section we
summarise our results and discuss various further developments of the
method presented here which could be pursued, notably the extension of
the perturbation scheme to higher than linear order. This may throw
further light on how insights about discreteness gained using this
formalism, which will be discussed at length in \cite{joyce_06},
extend into the regime of highly non-linear evolution. We also discuss
briefly the possible interest of solving the cosmological N-body
problem on the bcc lattice, as well as the possible extension of our
method to IC generated on ``glassy'' configurations.

\section{Linearization of gravity on a perturbed lattice}
\label{Grav-lat}

In this section we start by defining and studying some general
properties of the gravitational potential and force of an infinite
system of point particles. We then consider the particular case of a
perturbed infinite lattice in a static Euclidean space, the
generalization to an expanding universe being given in
Sect.~\ref{Expanding}. Since the force is zero in the unperturbed
lattice, the dominant contribution to the force in the perturbed case
is linear in the relative displacements of the particles. In the last
subsection, we consider the equations of motion resulting from this
linearized force.

\subsection{Definition of the force and the potential}
 Let us consider carefully first the definition of the gravitational
force in an infinite system of point particles of equal mass $m$.  We
will assume that this system (either stochastic or deterministic) is
characterized by a well defined mean number density $n_0>0$, and mass
density $\rho_0=mn_0$.  The gravitational potential of a particle, per unit mass,  at $\br$, due to the particles in a finite volume $V$, is:
\be
\label{pot-inf}
\phi(\br)=-Gm\sum_{\br'\ne\br} \frac{1}{|\br-\br'|}\mV(V,\br'),
\ee
where the sum is over all the particles contained in the system, and
$\mV(V,\br)$ is the window function for the volume $V$, i.e.,
\be
\mathcal V(V,\br) = \left\{ \begin{array}{ll}
         1 & \text{if }\br\in V,\\ 0, & \text{otherwise.} \end{array} \right. \ee
The force per unit of mass (i.e. the acceleration), due to these same
particles, is given by the gradient of the potential:
\be
\label{force-def}
\bF(\br)=-\nabla \phi(\br).
\ee

Taking the infinite volume limit $V\to\infty$, neither the
 gravitational potential \eqref{pot-inf}, nor the gravitational force
 \eqref{force-def}, are well defined. In the first case the result
 diverges, while in the second it may be finite or infinite, but its
 value depends on how the limit is taken\,\footnote{$\bF(\br)$ is a
 conditionally convergent series.}.

In Euclidean spacetime this behaviour in the infinite volume limit may
be regulated by the introduction of a negative background --- the
so-called Jeans swindle (see e.g. \cite{binney_87,kiessling_99}) ---
so that the potential is defined as
\bea
\label{pot-inf-back}
\nonumber
\phi(\br)&=&-G\lim_{V\to\infty}\bigg[m\sum_{\br'\ne\br} \frac{1}{|\br-\br'|}\mathcal{V}(V,\br')\\
&&-\rho_0\left. \int_{\mathbb R^3} d^3r'\frac{1}{|\br-\br'|}{\mathcal{V}}(V,\br')\right].
\eea
This modifies the usual Poisson equation to
\be
\label{poisson-modi}
\nabla^2\phi(\br)=4\pi G(\rho(\br)-\rho_0).
\ee
The expression \eqref{pot-inf-back} is well defined\,\footnote{For a
more detailed discussion of the gravitational force in infinite
systems see also \cite{gabrielli_05}.}, provided (i) that the limit
$V\to\infty$ is taken in a physically reasonable way\,\footnote{E.g.,
taking the infinite volume limit in compact sets.}, and (ii) that the
fluctuations in the system are sufficiently rapidly decaying at large
scales\,\footnote{If $P(k)$ is the power spectrum of density
fluctuations, it is simple to show, using the modified Poisson
equation Eq.~(\ref{poisson-modi}), that convergence of the
fluctuations in the gravitational potential requires $\lim_{k
\rightarrow 0} k^n P(k)=0$ for $n > 1$. For finite fluctuations in the
force one requires $n>-1$.}. In the cosmological context this negative
background appears naturally as a consequence of the expansion of the
universe (see Sect.~\ref{Expanding}).

The simulations of self-gravitating systems we are interested in are
performed using a {\em finite} cubic simulation box of side $L$ and
volume $V_B=L^3$, subject to periodic boundary conditions. The force
on a particle is thus computed not only from all the other particles
inside the simulation box, but also from all the copies of the
particles contained in the ``replicas''.  The reason for using these
boundary conditions is that they introduce the inevitable finite size
effects without breaking translational invariance: every particle can
be considered to be at the centre of the finite box and therefore sees
the boundary in the same way.  The infinite system we consider is thus
an infinite number of replicas of a finite cubic box, with a negative
background as described above to make the force well
defined\,\footnote{Note also that, because the system is just a
lattice when considered at scales larger than the box size, the
fluctuations are always sufficiently suppressed at large scales so
that the gravitational force is well defined. Thus any possible
divergence in the fluctuations of force will be regulated by the box
size $L$.}. In this case the gravitational potential may be written as
\be
\label{pot-total}
\phi(\br)=\lim_{V\to\infty}\left[\phi_b(\br)+\phi_p(\br)\right],
\ee
where
\be
\label{pot-back}
\phi_b(\br)=G\rho_0 \int_{\mathbb R^3} d^3r'\frac{1}{|\br-\br'|}{\mathcal{V}}(V,\br')
\ee
is the contribution from the background, and
\be
\label{pot-part}
\phi_p(\br)=-Gm\sum_{\bn,\br'}^*\frac{\mV(V,\br'+\bn L)}{|\br-\br'-\bn L|}
\ee
the contribution from the particles.
Here the sum over $\br'$ is restricted to the particles in the box, 
while the other sum, over the three integers $\bn$ (i.e. over the images
of $\br'$), has a ``*'' to indicate that the term $\br'=\br$ is 
excluded when $\bn=\ve 0$.

The gravitational force is:
\be
\label{force-total}
\bF(\br)=\lim_{V\to\infty}\left[\bF_b(\br)+\bF_p(\br)\right],
\ee
where
\be
\label{force-back}
\bF_b(\br)=G\rho_0 \int_{\mathbb R^3} d^3r'\frac{\br-\br'}{|\br-\br'|^3}{\mathcal{V}}(V,\br')
\ee
and
\be
\label{force-part}
\bF_p(\br)=-Gm\sum_{\bn,\br'}^*\frac{\br-\br'-\bn L}{|\br-\br'-\bn L|^3}\mV(V,\br'+\bn L).
\ee
Note that the contribution from the background \eqref{force-back}
is identically zero if one takes a window function with inversion symmetry 
in $\br$ (e.g. a sphere or cube centred on $\br$).

\subsection{Linearization of the gravitational force}
\label{sub-linear}

We consider the infinite lattice generated by the replication of a sc lattice of volume $V_B$ of side $L$ with $N$ elements,
i.e., whose lattice vectors are $\bR=(m_1, m_2, m_3)\ell$ with $m_i\in
[0,N^{1/3}-1]\cap \mathbb N$ and $\ell=L/N^{1/3}$ is the lattice
spacing\,\footnote{The generalization of all the calculations
presented here to any {\em Bravais} lattice is straightforward (see
e.g. \cite{ashcroft_76}).}. This lattice (with a particle at each site) is now perturbed by 
applying displacements $\bu(\bR)$ to each particle $\bR$, so that 
the new positions of the particles can be written as 
 \be
\label{pos}
\br=\bR+\bu(\bR). 
\ee

The ``particle'' term in the gravitational force
[i.e. Eq.~\eqref{force-part}] can then be expanded order by order in
Taylor series about its value in the unperturbed lattice. At linear
order in the relative displacements $\bu(\bR)-\bu(\bR')$ we obtain
\begin{widetext}
\bea
\label{linearized_force}
\nonumber
\bF_p(\br)&=&-Gm \sum_{\bn,\bR'}^*\left\{\frac{\bR-\bR'+\bn
  L}{|\bR-\bR'+\bn L|^3}+\frac{\bu(\bR)-\bu(\bR')}{|\bR-\bR'+\bn L|^3}
-3\frac{[\bu(\bR)-\bu(\bR')]\cdot[\bR-\bR'+\bn L]}{|\bR-\bR'+\bn
  L|^5}(\bR-\bR'+\bn L)\right\}\\
&&\times \,\mV(V,\bR'+\bn L).
\eea
\end{widetext}
The first term in this sum 
\be
\label{first-term}
-Gm \sum_{\bn,\bR'}^*\frac{\bR-\bR'+\bn
  L}{|\bR-\bR'+\bn L|^3}\mV(V,\bR'+\bn L)
\ee
has the poor infinite volume behaviour which is regulated, as
discussed above, by the contribution coming from the background
Eq.~(\ref{force-back}).  The total linearized force is then also well
defined, and given by the infinite volume limit of
Eq.~(\ref{linearized_force}) summed with Eq.~(\ref{force-back}).  In
the case that we choose to calculate using the infinite volume limit
of a volume $V$ with inversion symmetry in $\br$ (i.e. the displaced
position of the particle), the full linearized force is thus given by
Eq.~(\ref{linearized_force}).  If, however, we choose to sum in a
volume with inversion symmetry {\it in the lattice site} $\bR$, it is
simple to show that Eq.~\eqref{first-term} is identically zero.  The
background term then contributes, with the sum
$\left[\eqref{force-back}+\eqref{first-term}\right]$ remaining
invariant.

The convergence criterion for {\em each} term of
\eqref{linearized_force} is
\be
\label{convergence}
|\bR-\bR'|>|\bu(\bR)-\bu(\bR')|.
\ee
Note that the validity of the power expansion does not depend on the
displacement of the particle $\bR$ on which we compute the force, but
on {\em relative} displacements of the particles at the position
$\bR$ and $\bR'$. 
Under the action of the
gravitational interaction, the displacements $\bu(\bR)$ will typically
grow so that the condition Eq.~\eqref{convergence} is violated after
some time. However when some pairs of particles 
no longer satisfy condition \eqref{convergence}, it may nevertheless 
continue to apply for the rest of the particles 
and \eqref{linearized_force} may remain a sufficiently good approximation 
to the force. In order to have a precise characterization of the regime
of validity of the approximation applied to follow the dynamical
evolution of a perturbed lattice, it is necessary to compare
the results with those obtained from 
evolution under full gravity. We will perform such a comparison 
in Sect.~\ref{comp-Nbody} using N-body simulations.

It is convenient to write the linearized force just discussed
in terms of the so-called 
{\em dynamical matrix} $\mathcal D(\bR)$ (see e.g. \cite{ziman_72,ashcroft_76}): 
\be
\label{def-dyn}
\bF(\br)=\sum_{\bR'}\mathcal D(\bR-\bR')\bu(\bR').
\ee
This matrix has the following properties: it is a complete symmetric
operator, i.e., $\mathcal D_{\mu \nu}(\bR)=\mathcal D_{\nu \mu}(-\bR)$
with inversion symmetry, i.e., $\mathcal D_{\mu \nu}(\bR)=\mathcal
D_{\mu \nu}(-\bR)$. Further, since the same displacement applied to
all the particles produces no net force, we have $\sum_\bR \mathcal
D_{\mu \nu}(\bR)= 0$. For any pair interaction potential $v(\ve r)$
it is straighforward to show that it can be written as~\cite{ziman_72,ashcroft_76}
\bse
\label{expr_D}
\begin{align}
\label{expr_D_nonzero}
\mD_{\mu\nu}(\bR\ne\mathbf 0)&=\dd_\mu \dd_\nu w(\bR)\\
\label{expr_D_zero}
\mD_{\mu\nu}(\bR=\mathbf 0)&=-\sum_{\bR'\neq 0} \dd_\mu \dd_\nu w(\bR')
\end{align}
\ese
where
\be
\dd_\mu \dd_\nu w(\br_0)=\left[\frac{\partial^2\,w (\br)}{\partial r_\mu \partial r_\nu}\right]_{\br=\br_0}
\ee
and $w(\br)$ is the periodic function defined as
\be
\label{w-def}
w(\br)=\sum_{\bn}v(\br+\bn L) ,
\ee
i.e., the potential due to a single particle and all its copies.
For gravity we have $v(\ve r)=-Gm/r$ and Eq.~\eqref{w-def} is
implicitly understood to be regulated as discussed at length
above by the addition of a uniform negative background.
We will describe below, and in App.~\ref{ewald-app}, how
we use the well-known Ewald summation technique to explicitly
perform this sum.

\begin{figure}
\includegraphics[width=0.25\textwidth]{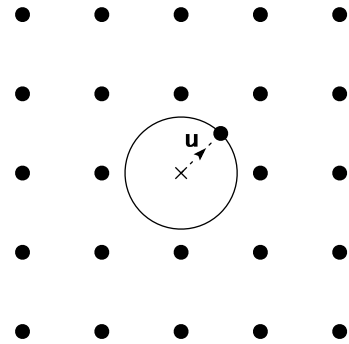}
\caption{Computation of the diagonal terms of the dynamical matrix at $\bR=\mathbf 0$.
\label{d-diag}}
\end{figure}

Equation~\eqref{expr_D_zero} gives the force
on a particle, at first order in the displacements, when it is
displaced and all the others remain unperturbed (see
Fig.~\ref{d-diag}). For gravity it is straightforward 
\cite{gabrielli_05} to show that
\be
\label{4pi3}
\mD_{\mu\nu}(\mathbf 0)=
%\sum_{\bR}\dd_\mu \dd_\nu (\bR)=
\frac{4\pi}{3}G\rho_0\de_{\mu\nu},
\ee 
i.e., the linearized force $\bfo_s(\br)$ on a particle due only to its
own displacement $\bu$ with respect to the rest of the lattice 
is  
\be
\label{4pi3f}
\bfo_s(\br)=\frac{4\pi}{3}G\rho_0\bu(\bR).
\ee 
The simplest way to derive this result is by summing the force
in spheres centred on the {\it unperturbed} position of the 
displaced particle. In this case it is straighforward to show, 
by symmetry, that the linearized direct particle contribution
Eq.~\eqref{linearized_force} is zero and the full force is
given by the background term Eq.~\eqref{force-back}.
The result follows then simply from Gauss' law which
gives that the force comes only from the region inside 
the sphere shown in Fig.~\ref{d-diag}. 

\subsection{Equations of motion in a static Euclidean universe}

In this section we derive the equations of motion of the particles in
the linear approximation, and then solve them.  We treat first a
static Euclidean space, giving the generalization to a cosmological
expanding universe in Sect.~\ref{Expanding}.

Using Newton's second law and Eqs.~\eqref{pos} and \eqref{def-dyn} we
can write the equation of motion of the particles as:
\be
\label{eq-motion}
\ddot\bu(\bR,t)=\sum_{\bR'}\mathcal D(\bR-\bR')\bu(\bR',t),
\ee 
where the double dot denotes a double derivative with respect to time.
The expression \eqref{eq-motion} is a system of vectorial coupled
second order differential equations which can be reduced to an
eigenvalue problem, using standard techniques. From Bloch's theorem
\cite{ashcroft_76} it follows that Eq.~\eqref{eq-motion} can be
diagonalized by the following combination of plane waves:
\be
\label{eigenvector}
\bu(\bR,t)=\frac{1}{N}\sum_{\bk}\tbu(\bk,t)e^{i\bk\cdot\bR}, 
\ee
where the sum over $\bk$ is restricted to the {\em first Brillouin zone}, i.e., for a sc lattice to 
\be
\label{k-vect}
\bk=\frac{2 \pi}{L} \bn,
\ee
with $\bn=(n_1,n_2,n_3)$ such that $n_i\in [-N/2,N/2[\cap\mathbb
Z$. We denote by $\tbu(\bk,t)$ the Fourier transform of $\bu(\bR,t)$:
\be
\label{eigenvector-inv}
\tbu(\bk,t)=\sum_{\bR}\bu(\bR,t)e^{-i\bk\cdot\bR}, 
\ee 
where the sum is restricted to the simulation box (i.e. without
considering the replicas). Inserting Eq.~\eqref{eigenvector} in
Eq.~\eqref{eq-motion}, we obtain for each $\bk$:
\be
\label{eq-motion-k}
\ddot\tbu(\bk,t)=\tilde{\mathcal D}(\bk) \bu(\bk,t),
\ee
where $\tilde{\mathcal D}(\bk)$ is the FT of ${\mathcal D}(\bR)$,
defined analogously to \eqref{eigenvector-inv}. From the properties of
$\mD(\bR)$ given above, it follows that $\tmD(\bk)$ is a real and 
symmetric operator which
satisfies\,\footnote{But note that $\tmD_{\mu\nu}(\bk=0)=\sum_{\bR} \mD_{\mu\nu}
(\bR) = 0$, i.e., $\tmD(\bk)$ is discontinuous at $\bk=\mathbf 0$.}
\be
\lim_{\bk\to0} \tmD_{\mu\nu}(\bk)=\frac{4\pi}{3} G\rho_0\de_{\mu\nu}.
\ee
We can now solve Eq.~\eqref{eq-motion-k} by diagonalizing the
$3\times3$ matrix $\tmD(\bk)$. For each $\bk$, this determines three
orthonormal eigenvectors $\mathbf{\hat e}_n(\bk)$ with three associated
eigenvalues $\om_n^2(\bk)$ ($n=1,2,3$)  satisfying the eigenvalue
equation:
\be
\label{eigen_equation}
\tmD(\bk)\mathbf{\hat e}_n(\bk)=\om_n^2(\bk)\mathbf{\hat e}_n(\bk).
\ee
We can decompose each mode $\tbu(\bk,t)$ in the basis $\{\mathbf{\hat e}_n(\bk)\}$ as
\be
\label{transf-eigen}
\tbu(\bk,t)=\sum_{n=1}^3\mathbf{\hat e}_n(\bk) f_n(\bk,t).
\ee
Using Eqs.~\eqref{eq-motion-k}, \eqref{eigen_equation} and
\eqref{transf-eigen} we get the following equation for the
coefficients $f_n(\bk,t)$:
\be
\label{eq-modes}
\ddot f_n(\bk,t)=\om_n^2(\bk)f_n(\bk,t).
\ee
Depending on the sign of $\om_n^2(\bk)$, we obtain two classes of
solutions $U_n(\bk,t)$ and $V_n(\bk,t)$. We choose them, without any
loss of generality, satisfying 
\bse
\label{mode-boundary}
\begin{align}
&U_n(\bk,t_0)=1,\qquad \dot U_n(\bk,t_0)=0,\\
&V_n(\bk,t_0)=0,\qquad \dot V_n(\bk,t_0)=1.
\end{align}
\ese 
The function $U_n(\bk,t)$ is associated with initial
displacements and $V_n(\bk,t)$ with initial velocities. If
$\om_n^2(\bk)\ge0$ then 
\bse
\label{uv-nonexp-pos}
\begin{align}
&U_n(\bk,t)=\cosh(\om_n(\bk) (t-t_0)),\\
&V_n(\bk,t)=\sinh(\om_n(\bk) (t-t_0))/\om_n(\bk).
\end{align}
\ese
If $\om_n^2(\bk)<0$
\bse
\label{uv-nonexp-neg}
\begin{align}
&U_n(\bk,t)=\cos(\sqrt{|\om_n^2(\bk)|}(t-t_0)),\\
&V_n(\bk,t)=\sin(\sqrt{|\om_n^2(\bk)|} (t-t_0))/\sqrt{|\om_n^2(\bk)|}.
\end{align}
\ese Whereas the modes \eqref{uv-nonexp-pos} with positive eigenvalues
cause an exponential growth of perturbation in the system, the modes
\eqref{uv-nonexp-neg} with negative eigenvalues leads to
oscillations. The evolution of the displacement field from any initial
state $\bu(\bR,t_0)$ is then given by the transformation 
\be
\label{eigen_evol}
\bu(\bR,t)=\frac{1}{N}\sum_{\bk}\left[\mathcal{P}(\bk,t)\tbu(\bk,t_0)+\mathcal{Q}(\bk,t)\dot\tbu(\bk,t_0)\right]e^{i\bk\cdot\bR}
\ee
where the matrix elements of the ``evolution operators'' $\mathcal{P}$ and $\mathcal{Q}$ are
\bse
\label{evol_operators}
\begin{align}
\mP_{\mu\nu}(\bk,t)=&\sum_{n=1}^3 U_n(\bk,t)(\mathbf{\hat e}_n(\bk))_\mu(\mathbf{\hat e}_n(\bk))_\nu,\\
\mQ_{\mu\nu}(\bk,t)=&\sum_{n=1}^3 V_n(\bk,t)(\mathbf{\hat e}_n(\bk))_\mu(\mathbf{\hat e}_n(\bk))_\nu.
\end{align}
\ese
The operator $\mP$ thus evolves the initial displacement field and $\mQ$ the initial velocity field.

\section{Determination and analysis of the spectrum of eigenvalues of $\tilde\mD(\bk)$}
\label{analy_spectrum}
 In this section we describe the determination of the eigenvectors
and spectrum of eigenvalues of the dynamical matrix for gravity. 
We then discuss the physical meaning of the results, notably 
identifying how the fluid limit is obtained and how corrections
to this limit may be calculated. In this discussion we 
will use extensively the strict analogy between the case we
are treating and the Coulomb lattice, or Wigner cystal, studied
in condensed matter physics (see e.g. \cite{pines_63}). This is a system
of positively charged particles embedded in a negative neutralizing
background. The particles interact with a repulsive $1/r$ potential
instead of the attractive $-1/r$ potential of Newtonian gravity. 
Thus all our results are mapped onto those for the corresponding 
Coulomb lattice by making the formal substitution $Gm^2 \rightarrow - e^2$,
where $e$ is the electronic charge\,\footnote{The potential we 
have used here for gravity has been
defined per unit mass, i.e., in our notation $v(\mathbf r)=e^2/mr$ for
the Coulomb lattice.}.

\subsection{Numerical computation of the spectrum of $\tilde\mD(\bk)$}
\label{numerical-spectrum}

 The spectrum of the matrix $\tilde\mD(\bk)$ must be computed
numerically. The matrix $\mD(\bR)$ is constructed using the Ewald sum
method \cite{ewald_21,ziman_72,ashcroft_76,deleeuw_80} to speed up the
convergence of the sum. We continue to work here explicitly, as above,
with a sc lattice of side $L$, with lattice spacing $\ell$ and $N$
elements\,\footnote{The generalization to a parallelepiped box, and to
other Bravais lattices, is straightforward (see
e.g. \cite{ashcroft_76}).}. To determine the dynamical matrix we use
the Ewald method to evaluate $w(\br)$ as given in Eq.~\eqref{w-def},
splitting it into two pieces using an appropriate damping function
$\mathcal C$:
\begin{equation}
\begin{split}
\label{pot_damping}
w(\br)&= \sum_{\bn} v(\br+\bn L){\mathcal C}(|\br+\bn L|,\alpha)\\
&+\sum_{\bn} v(\br+\bn L)[1-{\mathcal C}(|\br+\bn L|,\alpha)], 
\end{split}
\end{equation}
where $\alpha$ is a arbitrary ``damping parameter'' of which the
result is independant. The function ${\mathcal C}(|\br|,\alpha)$ is
chosen to be equal to unity at $\br=\ve 0$ and rapidly decaying to
zero as $|\br|$ goes to infinity. The first sum is then evaluated in
real space and the second one in Fourier space, making use of the
Parseval theorem \cite{nijboer_57}, ${\mathcal C}$ being chosen so
that the second term in Eq.~(\ref{pot_damping}) is analytic at
$\br=\ve 0$ and thus rapidly convergent in Fourier space. A common
choice for a $1/r$ pair potential is
\be 
{\mathcal C}(|\br|,\alpha)={\mathrm{erfc}}(\alpha|\br|).  
\ee 
The expression for the function $w$ is then: 
\be
\label{pot_ewald_div}
w(\br)=w^{(r)}(\br)+w^{(k)}(\br).
\ee
In the gravitational case
\bse
\label{pot_ewald}
\begin{align}
\label{pot_ewald_r}
w^{(r)}(\br)&=-Gm\sum_{\bn}\frac{1}{|\br+\bn L|}{\mathrm{erfc}}(\alpha|\br+\bn L|),\\
\label{pot_ewald_k}
w^{(k)}(\br)&=-Gm\frac{4\pi}{V_B}\sum_{\bk\neq \ve
0}\frac{1}{|\bk|^2}\exp\left(-\frac{|\bk|^2}{4\alpha^2}\right)
\cos\left[\bk\cdot\br\right], 
\end{align}
\ese
where $V_B$ is the volume of the box and the wavevectors $\bk$
are as in  Eq.~\eqref{k-vect}, but with $\bn$ ranging over all
triple integers (i.e. not restricted to the first Brillouin
zone). There is no $\bk=\mathbf 0$ term in the sum \eqref{pot_ewald} 
because of the presence of the negative background: 
when summed over all the particles, this term is equal to 
\be \lim_{\bk\to 0} \tilde\phi_0(\bk)=-\lim_{\bk\to0}\frac{4\pi G\rho_0}{k^2}, \ee 
i.e., the $\ve k=\ve 0$ mode of the potential (calculated from the
Poisson equation in Fourier space) which is cancelled by the contribution coming from the 
negative background.

The Ewald sum for the dynamical matrix can then be calculated 
directly using Eq.~\eqref{expr_D} and \eqref{pot_ewald}. The result, as in
Eq.~\eqref{pot_ewald_div}, is divided in two parts:
\begin{equation}
\label{dynamical_ewald}
\mathcal{D}(\bR)=\mathcal{D}^{(r)}(\bR)+\mathcal{D}^{(k)}(\bR)\,,
\end{equation}
for which the explicit expressions are given in App.~\ref{ewald-app}.

For the results quoted here we have taken $\alpha=2/L$ \cite{hernquist_91}. 
Using this numerical value of $\al$, it is sufficient to sum for 
\be 
|\bn|\le 3 \qquad |\bk|\le \frac{6\pi}{L}.
\ee 
to obtain a well converged determination of the dynamical matrix.
The diagonalization calculation involves essentially $N$ operations
(where $N$ is the number of particles). It is perfectly feasible, 
with modest computer resources,  to perform this diagonalisation 
for particle numbers as large as those used in the largest current 
N-body simulations.

\subsection{Analysis of the spectrum of eigenvalues in a simple cubic lattice}
We now describe the spectrum of eigenvalues of the dynamical matrix $\mD(\bR)$ for a sc lattice. As we have discussed in the introduction,
this is the lattice which is used very widely in N-body simulations of
structure formation in cosmology.

\begin{figure}
\includegraphics[width=0.45\textwidth]{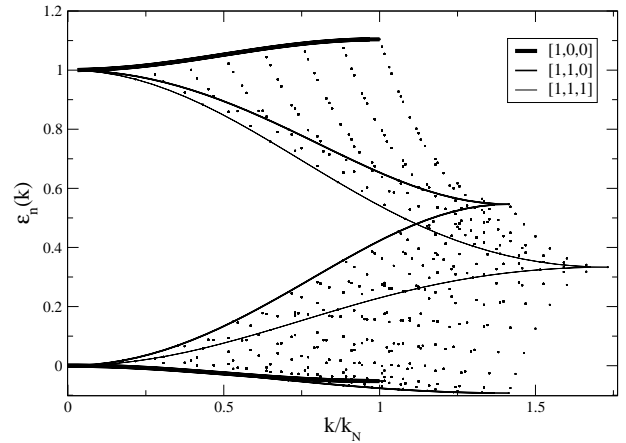}
\caption{Spectrum of eigenvalues for simple cubic lattice with $16^3$
particles. The lines correspond to chosen directions in $k$ space. 
\label{fig1}}
\end{figure}

In Fig.~\ref{fig1} we plot the spectrum of a sc lattice, for
$N=16^3$, obtained with the method outlined in the previous 
subsection. We show the normalized eigenvalues
\be
\label{eigen-norm}
\mathbf{\varepsilon}_n(\bk)=\frac{\om_n^2(\bk)}{4\pi G\rho_0}
\ee
as a function of the modulus of the $\bk$ vectors, normalized to the
Nyquist frequency $k_N=\pi/\ell$. With this normalisation the
spectrum remains substantially the same as we increase 
the number of particles: the only change is that the eigenvalues
become denser in the plot, filling out the approximate functional
behaviours with more points. For our discussion here there is
no interest in considering a greater number of points than
that we have chosen.

For each vector $\bk$ there are three eigenvalues $\om_n^2(\bk)$,
$n=1,2,3$.  Each family of eigenvalues (i.e. with same $n$) defines a
surface, corresponding to the three branches of the frequency-wavevector 
dispersion relation. Sections of these surfaces are plotted
for some chosen directions of the vector $\bk$ in Fig.~\ref{fig1}.

\subsubsection{An expression for $\tilde\mD(\bk)$ and the Kohn sum rule}
\label{general-formalism}
Before proceeding further it is useful to derive some important
results we will employ much in what follows. These are well known in
the context of the application of this formalism in condensed matter
physics (see e.g. \cite{pines_63}). First of all, we derive an
analytical expression for the dynamical matrix in Fourier space. Let
us decompose in Fourier modes the function $w(\br)$ defined in
Eq.~\eqref{w-def}
\be
w(\br)=\frac{1}{V_B}\sum_{\bk}\tilde w(\bk)e^{i\bk\cdot\br},
\ee
where the sum over $\bk$ is performed over {\em all} $k$ space,
i.e., not restricted to the first Brillouin zone and 
\be
\label{FT-pot}
\tilde w(\bk)=\int_{V_B} d^3r\,w(\br) e^{-i\bk\cdot\br}.
\ee
 The derivatives of
the periodic potential are 
\be
\label{der_phi}
w_{\mu\nu}(\br)=-\frac{1}{V_B}\sum_{\bk}k_\mu k_\nu \tilde w(\bk)e^{i\bk\cdot\br}.
\ee
Using the definition of the dynamical matrix 
\be
\label{dynamical}
\tilde\mD_{\mu\nu}(\bk)=\sum_{\bR}\mD_{\mu\nu}(\bR)e^{-i\bk\cdot\bR}
\ee
and Eqs.~\eqref{expr_D} and \eqref{der_phi} we obtain:
\be
\label{dk_first}
\tilde\mD_{\mu\nu}(\bk)=-\frac{1}{V_B}\sum_{\bk',\bR}k'_\mu k'_\nu \tilde w(\bk') \left( e^{i\bR\cdot(\bk'-\bk)}-e^{i\bk'\cdot\bR}\right)
\ee
where we can include the term $\bR=\ve 0$ in the sum because it vanishes. Using the orthogonality relation, we have
\be
\label{ortho}
\sum_{\bR}e^{i(\bk-\bk')\cdot\bR}=N\sum_{\bK}\delta_{\bk',\bk+\bK},
\ee
where the $\bk$ are restricted to the first Brillouin zone and $\bK$ are the reciprocal vectors of $\bR$ satisfying
\be
\bK=2 k_N\bm,
\ee
with $\bm\in\mathbb Z^3$. Substituting Eq.~\eqref{ortho} in \eqref{dk_first} we obtain finally the expression  \cite{pines_63}:
\bea
&&\label{D_of_phi}\tilde \mD_{\mu\nu}(\bk)= - n_0 k_\mu k_\nu\tilde w(\bk)\\\nonumber
&&- n_0\sum_{\bK\ne\ve 0}\left[(k_\mu+K_\mu)(k_\nu+K_\nu)\tilde w(\bk+\bK)-K_\mu
K_\nu\tilde w(\bK)\right],
\eea 
where $n_0$ is the number density of particles. In the gravitational
case, the integral \eqref{FT-pot} cannot be evaluated
analytically. However, neglecting finite size effects, this integral can be
computed over the whole space and the periodic potential $w(\ve r)$
is approximated by the interaction pair potential $v(\mathbf r)=-G m/r$, so that
\be
\label{FT-grav}
\tilde w(\bk)\simeq \tilde v(\bk)=\int_{\mathbb R^3} d^3r \,v(\br) e^{-i\bk\cdot\br}=-\frac{4\pi G m}{k^2}.
\ee
Using this it is straightforward to show (see App.~\ref{kohn-app}) the
following simple result:
\be
\label{kohn}
\sum_{i=1}^3\om_{i}^2(\bk)=-n_0 k^2\tilde w(\bk)=4\pi G\rho_0.
\ee
In the context of the Coulomb lattice this is a well-known result,
the so-called {\em Kohn sum rule}. In this case the quantity which
appears on the r.h.s. of the sum, instead of $4\pi G\rho_0$, is
$-\omega_p^2 = -4 \pi e^2 n_0/m$ where $\omega_p$ is the 
{\it plasma frequency}. We will discuss further below the 
significance of this analogy.

We can use these results and the above sum rule to compute --- in a different
way than in Eqs.~\eqref{4pi3}--\eqref{4pi3f} --- the $\bR=\mathbf 0$
term of the dynamical matrix $\mD(\bR)$ (i.e. the term giving
the force on a particle, at linear order in the relative displacements, 
when it alone is perturbed off the lattice). Using the Kohn sum rule
\eqref{kohn}, the trace of the dynamical matrix is:
\be 
\tr[\mD(\bR)]=4\pi G\rho_0.  
\ee
If the crystal has three equivalent orthogonal directions then the diagonal 
terms of the dynamical matrix will be equal. In the case of lattices with special symmetries (like the sc, bcc and fcc) it is simple to show that when a single particle is displaced along the
direction of an axis, the force acting on it is parallel to the direction of
displacement\,\footnote{This can be explicitly shown e.g. using
Eq.~\eqref{dynamical_ewald_linear_r} (taking the limit $\al\to0$ and
assuming that the sum over the replicas converges).}. This implies that the non-diagonal terms of the dynamical matrix are zero. We can therefore conclude that
\be
\label{D-zero-diag}
\mD_{\mu \nu}(\mathbf 0)=\frac{4}{3}\pi G\rho_0\de_{\mu\nu}.  
\ee

\subsubsection{The branches of the dispersion relation and the fluid limit}
\label{fluid-limit}

We have noted that the spectrum of eigenvalues has a clear
branch structure. To identify the different branches it is useful to 
consider the $\bk\to\ve 0$ limit keeping the interparticle distance 
$\ell$ constant. We expect this to correspond to the fluid limit: 
a plane wave fluctuation $e^{i\bk\cdot\br}$ with $\bk\ll 1/\ell$ has 
a variation scale much larger than the interparticle distance, and
therefore does not ``see'' the particles. 

From Eq.~\eqref{D_of_phi} the limit for $\bk\to\ve 0$ is
straightforward as the contribution of the sum on the r.h.s. vanishes
in this limit\,\footnote{We have assumed that the sum in
Eq.~\eqref{D_of_phi} is well defined ---  which is the case for the
gravitational interaction --- so that it is possible to take the limit
before performing the sum.}  
\be
\label{D_limit_general}
\lim_{\bk\to\ve 0}\tilde \mD_{\mu\nu}(\bk)=-n_0 \hat k_\mu \hat k_\nu\tilde w(\bk).
\ee
Using the eigenvalue equation \eqref{eigen_equation} with
Eqs.~\eqref{D_of_phi} and \eqref{FT-grav}, it follows that the
solutions in the fluid limit are
\begin{enumerate}
\item one {\em longitudinal} eigenvector polarized parallel to $\bk$
  with normalized eigenvalue $\varepsilon_1(\bk\to\ve 0)=1$ and
\item two {\em transverse} eigenvectors polarized in the plane transverse to $\bk$ with normalized eigenvalues $\varepsilon_{2,3}(\bk\to\ve 0)=0$.  
\end{enumerate}
 As the spectrum of eigenvalues $\varepsilon_n(\bk)$ is 
exactly the same, up to an overall negative multiplicative constant,
to that of the Coulomb lattice, we adapt the same terminology as in 
this context. The branch of eigenvalues whose associated eigenvectors converges to the
longitudinal eigenvector
as $\bk\to \mathbf 0$ is called
the {\it optical} or longitudinal branch. The two other branches
whose eigenvectors 
converge to the transverse eigenvectors are called
the {\em acoustic} branches. For finite $k$, the eigenvectors are not
exactly parallel or perpendicular to $\hat\bk$ for all $\bk$ but belong
nevertheless to one of the three branches, which define three-dimensional
hyper-surfaces in the four-dimensional space $(\om,\bk)$ space.

The appearance of an optical branch in a monoatomic crystal is a
characteristic feature of the $1/r$ interaction potential (at large
$r$). In the case of a more rapidly decaying potential at large
scales, i.e., $1/r^{1+\alpha}$ with $\alpha>0$, it becomes a third
acoustic branch. In the case of a potential that decays slower at
large $r$, i.e., $\alpha<0$, the optical branch diverges as $\bk\to\ve
0$. The physical interpretation of the optical branch is that it
represents the coherent excitation of the whole lattice with respect to
the background \cite{clark_57}. In a Coulomb crystal, the
optical mode is produced by the lattice moving against this background
producing a ``plasma oscillation'', at the plasma frequency $\omega_p$
defined above.  This mode is, as we have just seen, purely longitudinal, i.e., the perturbations are parallel to $\bk$, while the tranverse
modes, i.e., the perturbations orthogonal to $\bk$ have zero frequency. 
The reason for this behaviour of long wavelength density fluctuations 
can be easily understood. The density fluctuations are related, in
this fluid limit, to the displacements through the continuity equation:
\be
\de\rho\sim \nabla\cdot\bu,
\ee
which implies in $k$ space that
\be
\label{rho-k}
\de\tilde \rho\sim \bk\cdot\tbu.
\ee
Thus tranverse modes do not source density fluctuations,
and therefore (by the Poisson equation) they do not produce a force. 
In the case of gravity, instead of oscillating as in a plasma, 
the longitudinal mode may be amplified or attenuated (depending
on the initial perturbation), in a way which is independent of
$k$. As we will discuss in detail below, this is just
the well known linear amplification of density fluctuations
in a self-gravitating fluid.

\subsubsection{Corrections to the fluid limit}
\label{smallk}
We have just seen that the fluid limit is obtained by taking the dynamical matrix as 
\be
\label{dyn-fluid}
\tmD(\bk)=\frac{4\pi G\rho_0}{k^2}k_\mu k_\nu.
\ee
We can estimate analytically the corrections to this limit 
for small $k$ (i.e. for large wavelengths) by expanding the 
eigenvalues and eigenvectors of the full dynamical matrix 
about $\bk=\mathbf 0$. We note that this corresponds to calculating
the difference, at large wavelengths, between the evolution
of the perturbed lattice with a finite number of particles
and that of the fluid limit. These are thus what are, in
the context of cosmological simulations, ``discreteness
effects'' introduced by the modelling of the fluid by 
such a system. We will discuss at length this application
of this formalism in \cite{joyce_06}.

\begin{figure}
\includegraphics[width=0.45\textwidth]{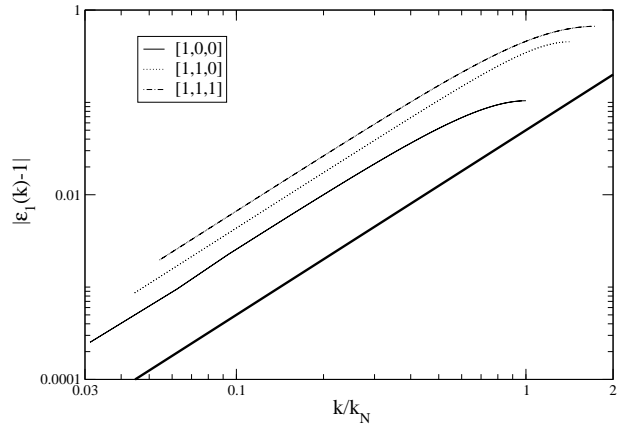}
\caption{Optical branch for different directions of $\bk$. The thick line
is proportional to $k^2$.
\label{bk}}
\end{figure}

When expanding the dynamical matrix in Taylor series about the fluid
limit $\bk \rightarrow 0$, it is simple to show that for $1/r$
interactions this series is in even powers of $k$, because $\mD(\bR)$
is real and $\tmD(\bk)$ analytic for $\bk\to\mathbf 0$ (see
\cite{clark_57,coldwell_60}). It is therefore possible to write the
corrections to the eigenvalues of the optical mode as:
\be
\label{w-smallk-opt}
\om_1^2(\bk)\simeq 4\pi G\rho_0(1- b_1(\hat\bk)k^2),
\ee
where the expression for $b_1(\hat{\bk})$ can be computed by
diagonalizing $\tmD(\bk)$ expanded up to $\mathcal{O}(k^2)$.  The leading correction to the two acoustic modes
may be written
\bse
\label{w-small-k-acoust}
\begin{align}
\om_2^2(\bk)&\simeq 2\pi G\rho_0 b_2(\hat\bk)k^2,\\
\om_3^2(\bk)&\simeq 2\pi G\rho_0 b_3(\hat\bk)k^2.
\end{align}
\ese
The Kohn sum rule implies that
$b_1(\hat\bk)=(b_2(\hat\bk)+b_3(\hat\bk))/2$. In Fig.~\ref{bk} we show
the optical branch, in various different chosen directions. The
approximation with the leading term in the Taylor expansion is very
good up to the Nyquist frequency.

\begin{figure}
\includegraphics[width=0.45\textwidth]{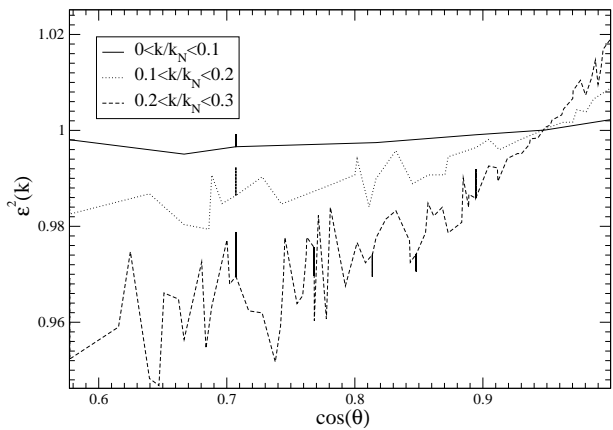}
\caption{Variation of the value of the eigenvalues for various ranges
as a function of the cosine of the angle between $\bk$ and the axes of
the lattice which forms a minimal angle with it. We see that the
effects of anisotropy are strongest for the short-wavelength modes,
and decrease as we go towards the fluid limit.
\label{aniso}}
\end{figure}

In Fig.~\ref{aniso} we show how the anisotropy of the eigenvalues
increases as the modulus of the wave vector increases (i.e. when we look 
at smaller spatial scales). We plot,
for three ranges of values of the modulus of $\bk$, the value of the normalized
eigenvalues as a function of the angle $\theta$ between $\bk$ and the
axis that forms a minimal angle with it. As $\theta$ increases (i.e. 
as $\cos\theta$ decreases with $0< \theta < \pi/2$) there is a clear
trend of decrease in the eigenvalue, in each of the three cases.
The difference as a function of orientation of the vector $\bk$
is, however, much  more marked for larger $k$, i.e., at scales
closer to the Nyquist frequency. This is not unexpected: the 
effects of anistropy (which is completely absent in the
fluid limit, in which the eigenvalues are independent of
the orientation $\bk$) are naturally strongest for the short 
wavelength modes.

\subsubsection{Oscillatory modes}

\begin{figure}
\includegraphics[width=0.45\textwidth]{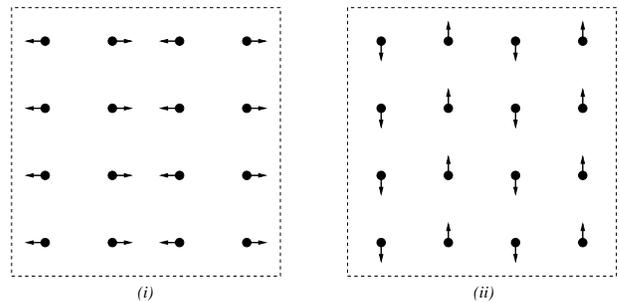}
\caption{Schematic representation of {\em (i)} a mode collapsing faster than fluid limit and {\em (ii)} an oscillating mode.
\label{coll-clust}}
\end{figure}

The spectrum of the sc lattice Fig.~\ref{fig1} includes some modes
[e.g. for $\bk=(k_x,0,0)$] with eigenvalues on the optical branch 
{\em larger than the fluid limit}. For example, this is the case 
for modes with initial
displacement $\bu(\br,0)\propto\hat \bx \exp(ik_x x)$, shown in
Fig.~\ref{coll-clust}-{\em (i)}. Adjacent planes collapse towards one
another, faster than in the fluid limit. The Kohn sum rule Eq.~\eqref{kohn} 
states that the sum of the three eigenvalues $\om_n^2(\bk)$ is equal to $4\pi
G\rho_0$. Therefore, the existence of modes collapsing faster than the
fluid limit implies that there are other modes with negative
eigenvalues $\om_n^2(\bk)$, i.e., which oscillate. This is the case,
e.g., of the mode with initial displacement $\bu(\br,0)\sim\hat \by
\exp(ik_x x)$, shown in the Fig.~\ref{coll-clust}-{\em (ii)}. In this
case, contiguous planes oscillate as indicated in the figure.
 We will study these modes in greater detail using numerical
simulation in Sect.~\ref{sec:oscill}.

The existence of oscillating modes in a perturbed and cold {\it
purely} self-gravitating system (i.e. without any additional
interaction or velocity dispersion giving rise to a restoring
pressure\,\footnote{If there is a non negligible velocity dispersion,
it known that fluctuations at scales smaller than the Jeans length
oscillate \cite{binney_87}.}) is an unexpected curiosity, a behaviour
qualitatively different to that generically expected based on the
analysis of the fluid limit. Translated to the analagous Coulomb
system, the result means that a sc lattice is, in this case, unstable
(as there are growing modes).  While this result has not apparently
been shown in the literature, it is not an unexpected result in this
context.  It has been established \cite{fuchs_35,carr_61} that for the
(classical) Coulomb lattice that the ground state is the bcc
lattice. It has a lower binding energy than the fcc lattice, which in
turn is a lower energy configuration than the sc lattice. Our result
implies that the latter is not only a higher energy state, but that it
is strictly unstable. Indeed we note that the specific modes we have
considered above describe a ``sliding'' of adjacent places in an sc
lattice which deform it towards the lower energy configuration
represented by the fcc lattice.

\section{Generalization to an expanding universe}
\label{Expanding}

In the previous section, we have described the gravitational
evolution of a perturbed lattice in a static Euclidean universe. In
the cosmological context, density fluctuations are a perturbation
around an homogeneous and isotropic Friedmann-Robertson-Walker (FRW)
solution of Einstein's field equations of general relativity.  In
cosmological N-body simulations, since the regions studied are smaller
than the Hubble radius and the velocities are non-relativistic, one
considers the limit in which the equations of motion of the particles
are strictly Newtonian in physical coordinates $\br$ \cite{peebles_80}. These
coordinates are related to the comoving coordinates $\bx$ of the FRW
solution by
\be
\label{comoving}
\br(t)=a(t)\bx(t),
\ee
where $a(t)$ is the scale factor describing the expansion of the
universe. It satisfies the Friedmann equation
\be
\label{friedmann}
\left(\frac{\da}{a}\right)^2=\frac{8\pi G}{3}\rho-\frac{\kappa}{a^2},
\ee
where $\rho$ is the mass density of the universe and $\kappa$ the curvature. 
In the unperturbed FRW model the particles are fixed in comoving coordinates,
all deviation from these positions arising from perturbations to this
model. For this reason it is very natural, and convenient, to work
in comoving coordinates. We therefore start by transforming our
previous Newtonian equations to these coordinates, the only
further difference being that we perturb about a time-dependent solution
describing an expanding FRW universe.

Using Eq.~\eqref{comoving} the acceleration can be written
\be
\label{accel}
\ddot \br=a\ddot \bx+2\da\dot\bx+\dda\bx.  
\ee
The term $\dda\bx$ can be expressed as the background contribution of
the gravitational acceleration. For the specific case of an Einstein de Sitter (EdS)
Universe, i.e., a universe containing only matter without curvature
[$\rho(t)=\rho_0 (a(t)/a(t_0))^3$ and $\kappa=0$], it is given by
\be
\label{accel-back}
\bg_0=\dda\bx=\frac{4\pi}{3a^3}G\rho_0\bx,
\ee
which has exactly the same form (for $a=1$) as the contribution of the
negative background of Eq.~\eqref{4pi3}. We now 
write the position of a particle in comoving coordinates in terms 
of the displacement $\bu$ about the lattice position as
\be
\bx(t)=\bR+\bu(\bR,t).
\ee
The vector $\bR$ is now the position of the lattice sites in {\em
comoving coordinates} (i.e. $\bR$ does not depend on time) 
and $\bu(\bR,t)$ is the displacement of the particle that
was originally at $\bR$ (in fluid theory, this is a {\em Lagrangian
coordinate}, see e.g. \cite{buchert_92}). By using Eq.~\eqref{accel},
we can write Eq.~\eqref{eq-motion} in an expanding universe as
\be
\label{eq-motion-exp}
\ddot\bu(\bR,t)=-2\frac{\da}{a}\dot\bu(\bR,t)+\frac{1}{a^3}\sum_{\bR'}^N\mD(\bR-\bR')\bu(\bR',t),
\ee
where we have implicitly included the background 
term \eqref{accel-back} in the dynamical matrix.
We emphasize that {\it the dynamical matrix is identical to that in the
static case}: it depends only on the kind of lattice and on the
interaction, but not on the dynamics of the background. Therefore all
the analysis of this matrix performed in the preceeding section is 
valid also in this case. From Eq.~\eqref{eq-motion-exp}, the mode equation
\eqref{eq-modes} generalizes simply to
\be
\label{modes-exp}
\ddot f_n(\bk,t)+2\frac{\da}{a}\dot f_n(\bk,t)=\frac{\om_n^2(\bk)}{a^3}f_n(\bk,t).
\ee
This is very similar to the equation of the evolution of a fluid in
Lagrangian coordinates \cite{buchert_92}. The difference is only in 
the factor $\om_n^2(\bk)$ on the r.h.s., which in the fluid limit
is replaced by $4\pi G\rho_0$.

\subsection{Solution in an Einstein--De Sitter universe}

We derive now the solution of the mode equation \eqref{modes-exp} in
the case of an EdS universe. The evolution of
the scale factor is, from Eq.~\eqref{friedmann}: 
\be
a(t)=\left(\frac{t}{t_0}\right)^{2/3},\qquad 6\pi G\rho_0 t_0^2=1, \ee
assuming that $a(0)=0$. Then the mode coefficient equation  \eqref{modes-exp} is
\be
\label{modes-exp-expl}
\ddot f_n(\bk,t)+\frac{4}{3t}\dot f_n(\bk,t)=\frac{2}{3t^2}\varepsilon_n(\bk)f_n(\bk,t),
\ee
where we have used again the adimensional quantity
$\varepsilon_n(\bk)$ defined in Eq.~\eqref{eigen-norm}.  A set of
independent solutions of \eqref{modes-exp-expl} which satisfies the
IC \eqref{mode-boundary} are:
\bse
\label{uv-exp}
\begin{align}
U_n(\bk,t)=&\tilde\al(\bk)
\left[\al_n^{+}(\bk)\left(\frac{t}{t_0}\right)^{\al_n^{-}(\bk)}+
\al_n^{-}(\bk)\left(\frac{t}{t_0}\right)^{-\al_n^{+}(\bk)}\right],\\
V_n(\bk,t)=&\tilde\al(\bk)t_0
\left[\left(\frac{t}{t_0}\right)^{\al_n^{-}(\bk)}-\left(\frac{t}{t_0}\right)^{-\al_n^{+}(\bk)}\right]
\end{align}
\ese
where
\be
\tilde\al(\bk)=\frac{1}{\al_n^{-}(\bk)+\al_n^{+}(\bk)}
\ee
and
\bse
\begin{align}
&\al_n^{-}(\bk)=\frac{1}{6}\left[\sqrt{1+24\varepsilon_n(\bk)}-1\right],\\
&\al_n^{+}(\bk)=\frac{1}{6}\left[\sqrt{1+24\varepsilon_n(\bk)}+1\right].
\end{align}
\ese
If $\varepsilon_n(\bk)>0$ the solution presents a power-law amplification mode
and a power-law decaying mode. If $-1/24<\varepsilon_n(\bk)<0$, there are two
decaying modes. Finally, if $\varepsilon_n(\bk)\leq-1/24$, the solution is
oscillatory and can be written as
\bse
\label{uv-exp-neg}
\begin{align}
U_n(\bk,t)=&\left(\frac{t}{t_0}\right)^{-\frac{1}{6}}\cos\left[\ga_n(\bk)\ln\left(\frac{t}{t_0}\right)\right]\\\nonumber
&+\frac{1}{6\ga_n(\bk)}\left(\frac{t}{t_0}\right)^{-\frac{1}{6}}\sin\left[\ga_n(\bk)\ln\left(\frac{t}{t_0}\right)\right],\\
V_n(\bk,t)=&\frac{t_0}{\ga_n(\bk)}\left(\frac{t}{t_0}\right)^{-\frac{1}{6}}\sin\left[\ga_n(\bk)\ln\left(\frac{t}{t_0}\right)\right]
\end{align}
\ese
where
\be
\ga_n(\bk)=\frac{1}{6}\sqrt{|24\varepsilon_n(\bk)+1|},
\ee
i.e., the static oscillatory behavior of the static universe survives,
but now the oscillation is periodic in the logarithm of time with
decreasing amplitude. The evolution of the displacements is computed
with Eq.~\eqref{eigen_evol}. The effect of the expansion [through the
``viscous'' first term of the r.h.s. of Eq.~\eqref{eq-motion-exp}] is
to slow down the growing and decaying mode of the non-expanding
exponential solution into a power-law solution.

\subsection{Fluid limit and Zeldovich approximation}
Let us calculate the fluid limit of the solution given by
Eqs.~\eqref{eigen_evol}, \eqref{evol_operators} and \eqref{uv-exp}. As
explained in Sect.~\ref{analy_spectrum} this corresponds to taking the
limit $\bk\to0$ at fixed $\ell$ of the dynamical matrix
$\tmD(\bk)$. In this case, as we have seen in
Sect.~\ref{analy_spectrum} one of the eigenvectors is parallel to
$\hat \bk$, with eigenvalue $4\pi G\rho_0$, and the other two are
normal to $\hat \bk$ with eigenvalue equal to zero.  We have then:
\bse
\label{eigen_fluid}
\begin{align}
&\mathbf{\hat e}_1(\bk)=\hat \bk,\, \varepsilon_1(\bk)=1\longrightarrow \al_1^{+}=2/3,\,\al_1^{-}=1,\\
&\mathbf{\hat e}_2(\bk)=\hat \bk_{2\perp},\, \varepsilon_2(\bk)=0\longrightarrow \al_2^{+}=0,\,\al_2^{-}=1/3,\\
&\mathbf{\hat e}_3(\bk)=\hat \bk_{3\perp},\, \varepsilon_3(\bk)=0\longrightarrow \al_3^{+}=0,\,\al_3^{-}=1/3,
\end{align}
\ese
where $\hat \bk_{2\perp}$ and $\hat \bk_{3\perp}$  are orthogonal to $\hat \bk$ chosen so that $\hat \bk_{2\perp}\cdot \hat \bk_{3\perp}=0$. Using \eqref{eigen_fluid} in \eqref{uv-exp}, we get for the mode parallel to $\hat \bk$:
\bse
\label{parallel}
\begin{align}
&U_1(\bk,t)\equiv U_\parallel(t)=\frac{2}{5}\left[\frac{3}{2}\left(\frac{t}{t_0}\right)^{2/3}+\left(\frac{t}{t_0}\right)^{-1}\right],\\
&V_1(\bk,t)\equiv V_\parallel(t)=\frac{3}{5}t_0\left[\left(\frac{t}{t_0}\right)^{2/3}-\left(\frac{t}{t_0}\right)^{-1}\right]
\end{align}
\ese
and for the modes perpendicular to $\hat \bk$:
\bse
\label{perp}
\begin{align}
&U_{2}(\bk,t)=U_{3}(\bk,t)\equiv U_\perp(t)=1,\\
&V_{2}(\bk,t)=V_{3}(\bk,t)\equiv V_\perp(t)=3t_0\left[1-\left(\frac{t}{t_0}\right)^{-1/3}\right].
\end{align}
\ese
The evolution operators \eqref{evol_operators} are then:
\bse
\label{evol_operators_fluid}
\begin{align}
\label{evol_operators_fluid_P}
\mP_{\mu\nu}(\bk,t)&=U_{\parallel}(t)\hat k_{\mu}\hat k_\nu\
+(\hat \bk_{2\perp})_{\mu}(\hat
\bk_{2\perp})_{\nu}+(\hat \bk_{3\perp})_{\mu}(\hat \bk_{3\perp})_{\nu},\\
\label{evol_operators_fluid_Q}
\mQ_{\mu\nu}(\bk,t)&=V_{\parallel}(t)\hat k_{\mu}\hat k_\nu+ \\\nonumber
&+V_{\perp}(t)\left[(\hat \bk_{2\perp})_{\mu}(\hat
\bk_{2\perp})_{\nu}+(\hat \bk_{3\perp})_{\mu}(\hat \bk_{3\perp})_{\nu}\right],
\end{align}
\ese
[where we have used explicitly that $U_\perp(t)=1$]. Using Eq.~\eqref{eigen_evol} we write the evolution of the
displacements in the fluid limit as:
\bea
\label{sol_fluid}
\bu(\bR,t)&=&\bu_\perp(\bR,t_0)+\bu_\parallel(\bR,t_0) U_\parallel(t)\\\nonumber
&&+\bv_\parallel(\bR,t_0) V_\parallel(t)
+ \bv_\perp(\bR,t_0) V_\perp(t),
\eea
where 
\bse
\begin{align}
\bu_\parallel(\bR,t_0)&=\frac{1}{N}\sum_{\bk}(\tbu(\bk,t_0)\cdot\hat\bk)\hat\bk\,e^{i\bk\cdot\bR},\\
\bu_\perp(\bR,t_0)&=\frac{1}{N}\sum_{\bk}(\tbu(\bk,t_0)-(\tbu(\bk,t_0)\cdot\hat\bk)\hat\bk)\,e^{i\bk\cdot\bR},
\end{align}
\ese
and analogously for the velocities $\bv$.  Using the definition of
peculiar gravitational acceleration $\bg$
\be
\label{g-pec}
\bg=\ddot\br-\dda \bx=\ddot\br-\frac{\dda}{a}\br=a\left[\ddot \bu+2\frac{\da}{a}\dot\bu\right],
\ee
we can rewrite Eq.~\eqref{sol_fluid} [with
Eqs.~\eqref{parallel} and \eqref{perp}] as:
\bea
\label{sol_fluid_g}
\nonumber
\bu(\bR,t)&=&\bu_\perp(\bR,t_0)\\\nonumber
&+&\bg(\bR,t_0)t_0^2\left[\frac{9}{10}\left(\frac{t}{t_0}\right)^{2/3}+\frac{3}{5}\left(\frac{t}{t_0}\right)^{-1}\right]\\\nonumber
&+&\bv_\parallel(\bR,t_0) \frac{3}{5}t_0\left[\left(\frac{t}{t_0}\right)^{2/3}-\left(\frac{t}{t_0}\right)^{-1}\right]\\
&+& \bv_\perp(\bR,t_0)3t_0\left[1-\left(\frac{t}{t_0}\right)^{-1/3}\right],
\eea
where $\bv$ is the peculiar velocity defined as
\be
\label{pec-vel}
\bv(\bx,t)=\dot{\ve r} - \frac{\da}{a} \br  = \dot\br-\da\bx.
\ee
This formula is precisely the one found in \cite{buchert_92}. Note
that in this reference, the Lagrangian coordinate $\bX$ corresponds to
the position of the particle at $t=t_0$, i.e., $\bX=\bR+\bu(\bR,t_0)$.
For asymptotically large times the solution \eqref{sol_fluid_g} is
\be
\label{sol_fluid_g_asymp}
\bu(\bR,t)\simeq\frac{3}{5}t_0\left(\frac{t}{t_0}\right)^{2/3}\left[\frac{3}{2}\bg(\bR,t_0)t_0+\bv_\parallel(\bR,t_0)\right].
\ee
This solution, using Eqs.
\eqref{g-pec} and \eqref{pec-vel}, gives the following simple
relation between the displacements and the peculiar velocity with the
peculiar acceleration at any time:
\bse
\label{zeldo-rel}
\begin{align}
\bu(\bR,t)&=\frac{3}{2}\left(\frac{t}{t_0}\right)^{4/3} \bg(\bR,t_0)t_0^2,\\
\bv(\bR,t)&=\bg(\bR,t_0) t.
\end{align}
\ese
By imposing the IC
\bse
\label{zeldo-cond}
\begin{align}
\bu_\perp(\bR,t_0)&=\mathbf 0=\bv_\perp(\bR,t_0),\\
\bv_\parallel(\bR,t_0)&=\bg(\bR,t_0)t_0=\frac{2}{3t_0}\bu_\parallel(\bR,t_0),
\end{align}
\ese
the evolution is
given exactly at any time by Eqs.~\eqref{zeldo-rel}, which is the well known
Zeldovich approximation, in which the decaying mode is zero from the
initial time.  In cosmological N-body simulations of structure
formation IC are canonically imposed
\cite{efstathiou_85,bertschinger_95} using \eqref{zeldo-cond}: given
an initial power spectrum of density fluctuations a Gaussian
realization of the gravitational potential is generated, and used to
derive the initial gravitational field $\bg(\bR, t_0)$ at the
unperturbed particle positions. The particles are then displaced and
given initial velocities as specified by \eqref{zeldo-cond}.

\section{Comparison with N-body simulations}
\label{comp-Nbody}

In this section we compare predictions of the perturbative treatment
we have presented in the previous sections, which we will now refer to
as particle linear theory (PLT), to what one obtains with full gravity
(FG) calculated numerically in N-body simulations.  The aim is to
study the limits of validity of PLT as the relative displacements of
the particles increase. We also compare with the evolution obtained
using standard Lagrangian fluid linear theory (FLT).  At the end of
the section we also study further the oscillating modes which were
identified in the spectrum of eigenvalues discussed in
Sect.~\ref{analy_spectrum}.  Note that time will be expressed
everywhere in this section in units of $1/\sqrt{4\pi G \rho_0}$
(i.e. the characteristic time scale in the fluid limit), unless
otherwise stated.  We also draw the reader's attention to the fact
that the simulations we consider in this section are relatively small
with respect to typical current standard cosmological simulations, but
that this is not a relevant consideration here as PLT works for {\it
any} finite number of particles. For our purposes here it is
sufficient to have a large enough number of particles to separate
large scales (i.e. the box size) from small scales (i.e. the
interparticle distance).

The IC for our study\,\footnote{Except in the the
study of oscillating modes in Sect.~\ref{sec:oscill}.}  are generated
by perturbing in two different ways a sc lattice with $N$ particles.
We choose to restrict ourselves to simulations without space expansion
and with zero initial velocities\,\footnote{This choice is made for
simplicity. Another equally simple case is that of the EdS universe,
with velocities given by the Zeldovich approximation. The difference
between the two cases is simply, as we have seen, in the time
dependence of the modes. Indeed in the asymptotic fluid limit given by
the Zeldovich approximation the two cases can be mapped onto one
another by a simple transformation between time and scale factor:
$\exp(\sqrt{4\pi G \rho_0}t) \leftrightarrow a$. In any case we will
see that the criteria we deduce for validity of the PLT are expressed
in a very simple way which one would expect to be essentially the same
irrespective of the model.}.  In this case it is straightforward to
show, using the results of Sects.~\ref{Grav-lat} and
\ref{analy_spectrum}, that the displacements of the particles
according to PLT are
\begin{multline}
  \ve u(\ve R,t) = \frac{1}{N} \sum_{\ve k} \bigg\{ \exp(i\ve k\cdot \ve
  R) \times \\  
  \sum_{n=1}^3 A_n(\ve k) \cosh\left(\sqrt{4\pi G \rho_0 \varepsilon_n(\ve
    k)} \ t\right) \hat{\ve e}_n(\ve k) \bigg\}  \ ,
  \label{eq:uRt}
\end{multline}
where $\varepsilon_n(\ve k)$ and $\hat{\ve e}_n(\ve k)$ are defined in
Eqs.~\eqref{eigen_equation} and~\eqref{eigen-norm}, while the $A_n(\ve
k)$ are determined by the IC.  In the fluid limit, following the
discussion in Sect.~\ref{analy_spectrum} after
Eq.~\eqref{D_limit_general}, this becomes
\begin{equation}
\begin{split}
\ve u(\ve R,t) &= \frac{1}{N} \sum_{\ve k} \bigg\{ \exp(i\ve k\cdot
\ve R) \times
\\ & \quad \qquad  \bigg[
A_\parallel(\ve k)  \cosh\left(\sqrt{4\pi G \rho_0} \ t\right) \hat{\ve k} \\
& \qquad\qquad +A_{2\perp}(\ve k) 
\hat{\ve k}_{2\perp} + A_{3\perp}(\ve k) \hat{\ve k}_{3\perp} \bigg] \bigg\} \\
&= \ve u(\ve R,0) + \frac{\left[ \cosh\left(\sqrt{4\pi G \rho_0} \ t\right)
  - 1 \right] }{4\pi G \rho_0} \  \ddot{\ve u}(\ve R,0)  \ .\label{eq:flt_chap5}
\end{split}
\end{equation}
This therefore corresponds to what we denote by FLT.  Note that in the 
following, we consider this
last equation with the {\it full} initial acceleration $\ddot{\ve u}(\ve
R,0)$, and not the linearized one [see Eq.~\eqref{linearized_force}
or~\eqref{def-dyn}], since in the Lagrangian fluid approach, the
force on a fluid element is the full gravitational force~\cite{buchert_92}.

We consider two different kinds of initial displacements: spatially 
uncorrelated, and correlated.  In the first case, each particle of 
the lattice is
randomly displaced, with uniform probability, in a small cubic box
centered on its lattice site.  The power spectrum of density fluctuation (PS) $P(\bk,t)\propto|\de\tilde\rho(\bk,t)|^2$ of the resulting
particle distribution is proportional to $k^2$ at small $k$
\cite{glass,GSJP_05}. For the correlated case, the displacements are
obtained from a set of Gaussian variables $\left\{a_\mu(\ve k),
b_\mu(\ve k) \right\}_{\ve k,\mu}\ $: the $\mu$th component of the
displacement of a particle at the lattice point $\ve R$ is then
\begin{equation}
u_\mu(\ve R) = \frac{1}{V} \sum_{\ve k} \exp(i\ve k\cdot \ve R) \underbrace{[a_\mu(\ve
k) + i b_\mu(\ve k) ]}_{\tilde{u}_\mu(\ve k) } \ . 
\label{eq:kmoins2IC}
\end{equation}
The random variables $a_\mu(\ve k)$ and $b_\mu(\ve k)$ have average
$0$ and variance $\sigma^2(\ve k) \propto k^{-4}$ and are
statistically independent of one another\,\footnote{Note that for
$u_\mu(\ve R)$ to be real, the following conditions are required
$a_\mu(-\ve k)=a_\mu(\ve k)$ and $b_\mu(-\ve k)=-b_\mu(\ve k)$.}. This
gives rise, to a good approximation, to a distribution with PS of the
density field proportional to $k^{-2}$ at small $k$\,\footnote{This
can be seen directly from Eq.~\eqref{rho-k}, which is valid in the
continuous limit, i.e., in the limit of small perturbations and when
the effects of the discretisation introduced by the lattice are
neglected. For a detailed analysis of the corrections to this result
see \cite{gabrielli_04,joyce_04,GSJP_05}.}.

\begin{figure*}
\includegraphics[width=0.3\textwidth]{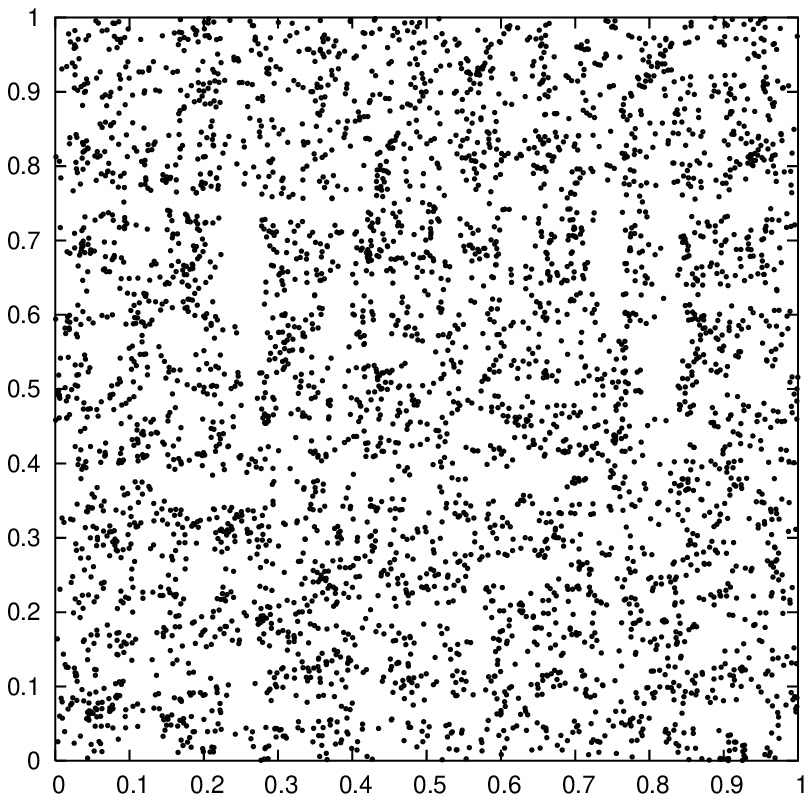}
\includegraphics[width=0.3\textwidth]{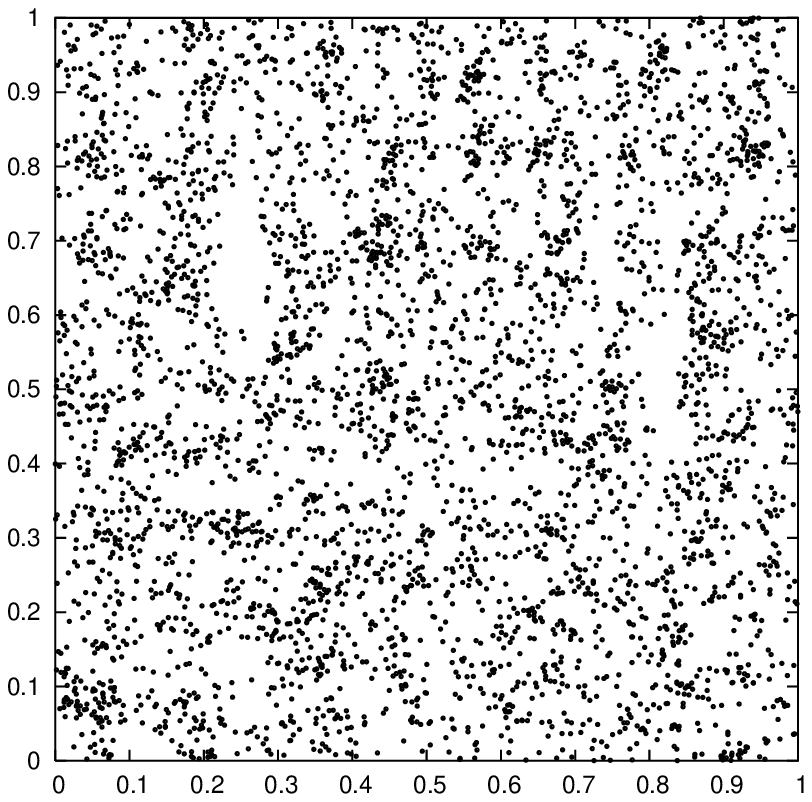}
\includegraphics[width=0.3\textwidth]{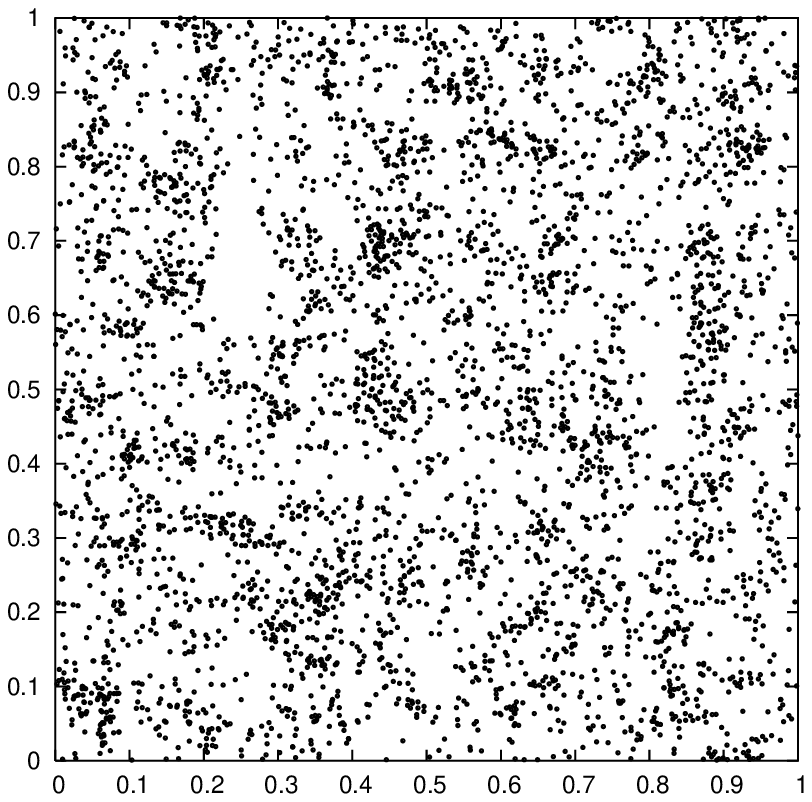}
\\
\includegraphics[width=0.3\textwidth]{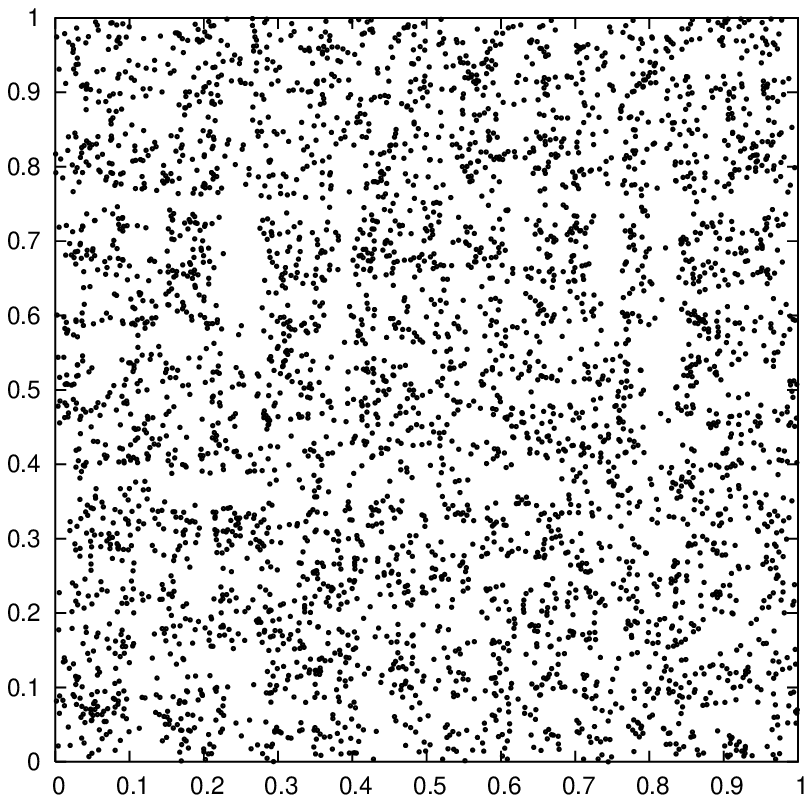}
\includegraphics[width=0.3\textwidth]{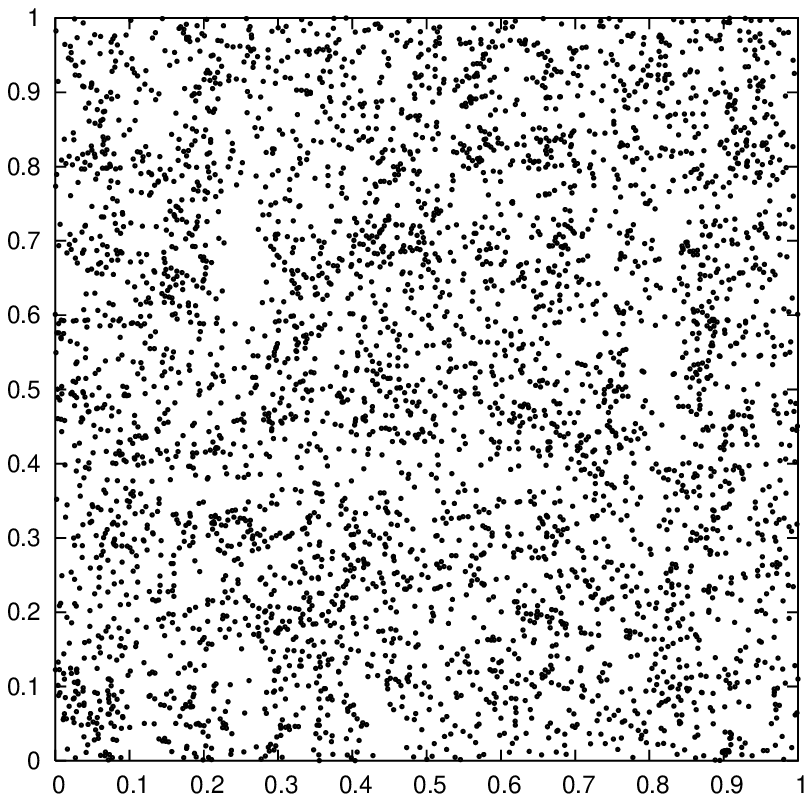}
\includegraphics[width=0.3\textwidth]{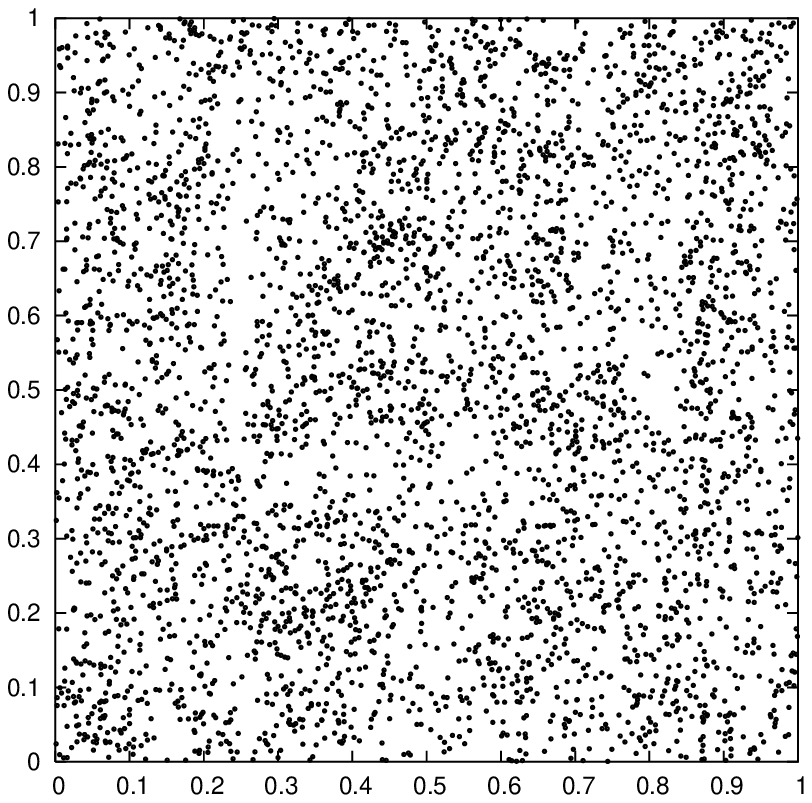}
\\
\includegraphics[width=0.3\textwidth]{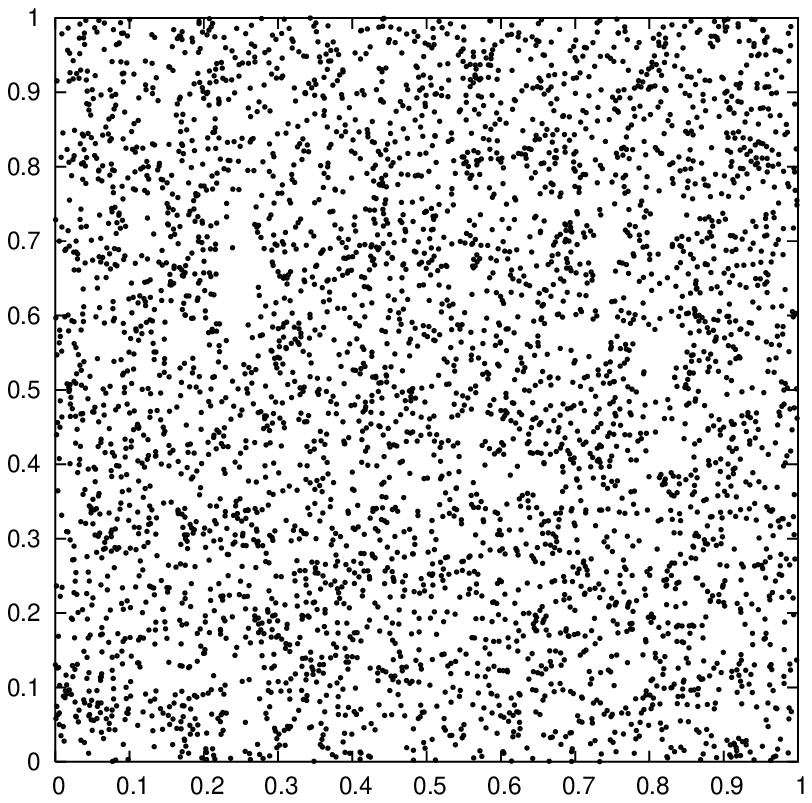}
\includegraphics[width=0.3\textwidth]{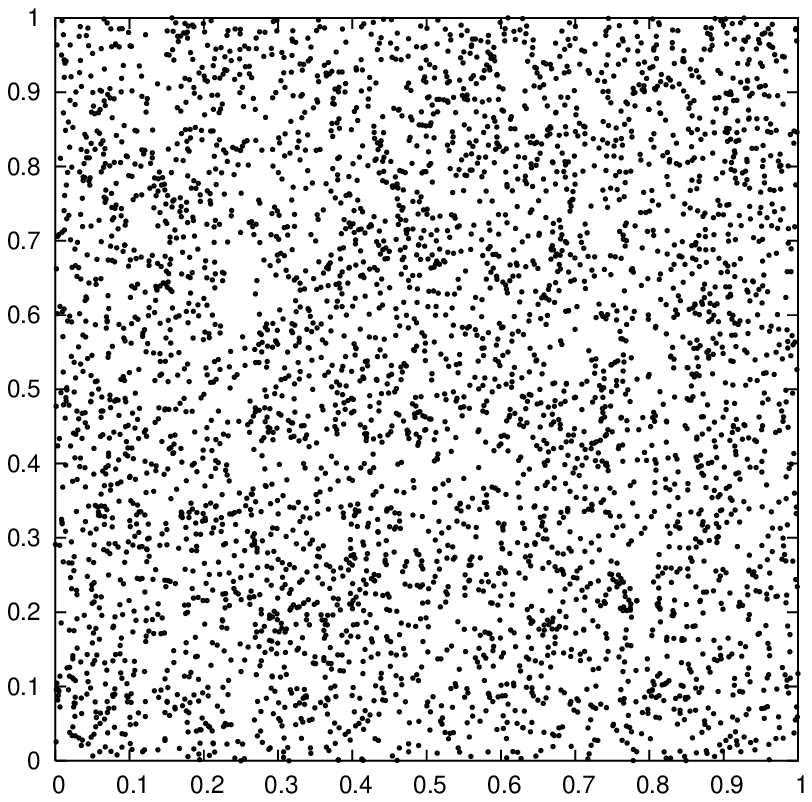}
\includegraphics[width=0.3\textwidth]{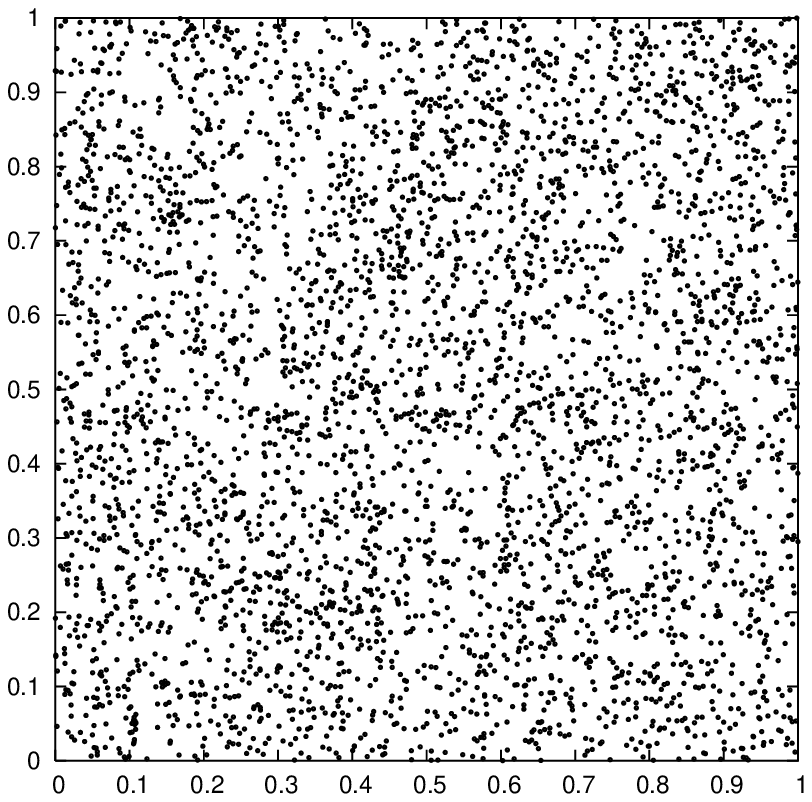}

\caption{Evolution from IC UC16 (projection on the
  plane $z=0$). From left to right,
  times 4.5, 5 and 5.5 and from top to bottom, FG, PLT and FLT. At
  $t=5$, PLT breaks down according to the criterion based on the
  function $\zeta_\text{D}(\ell,t)$ defined in
  Eq.~\eqref{eq:zetacond} (see the time $t_\zeta$ in tab.~\ref{tab:summarysimul}).
 \label{fig:snapSL16}}
\end{figure*}

\begin{figure*}

\includegraphics[width=0.3\textwidth]{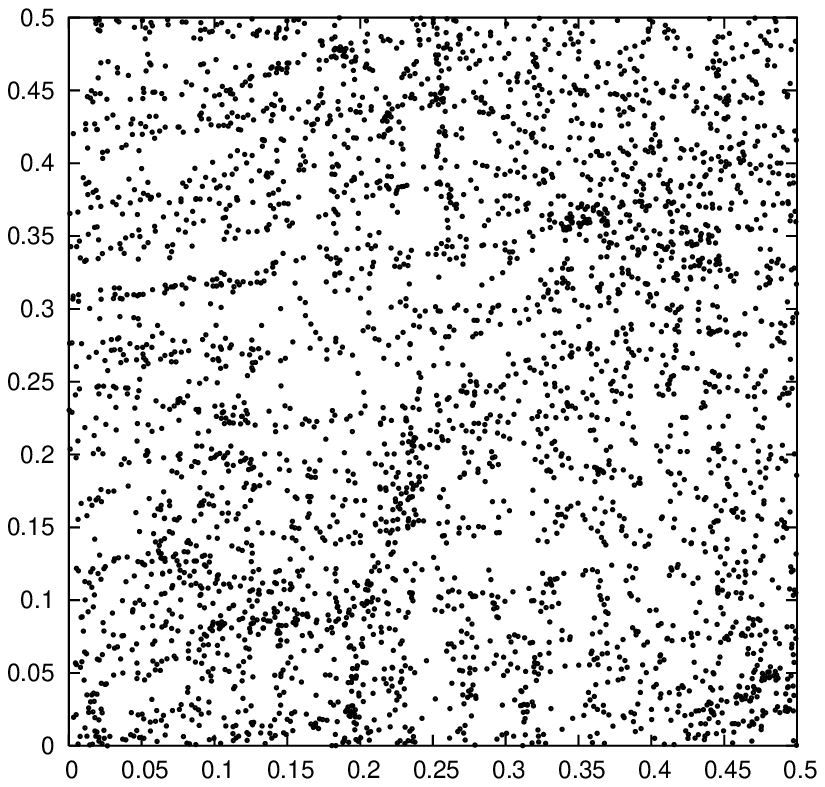}
\includegraphics[width=0.3\textwidth]{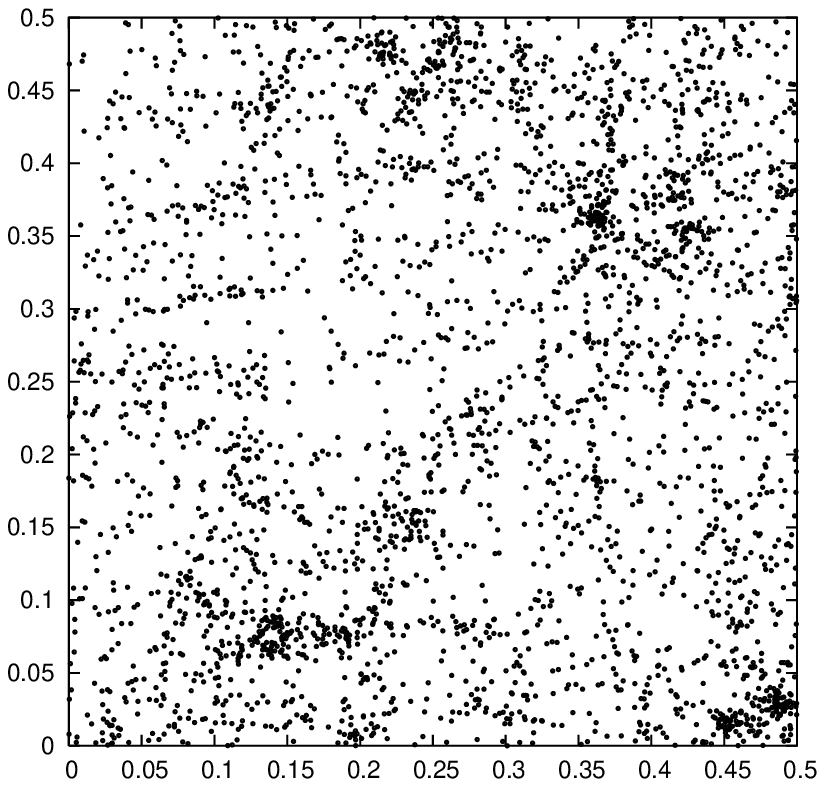}
\includegraphics[width=0.3\textwidth]{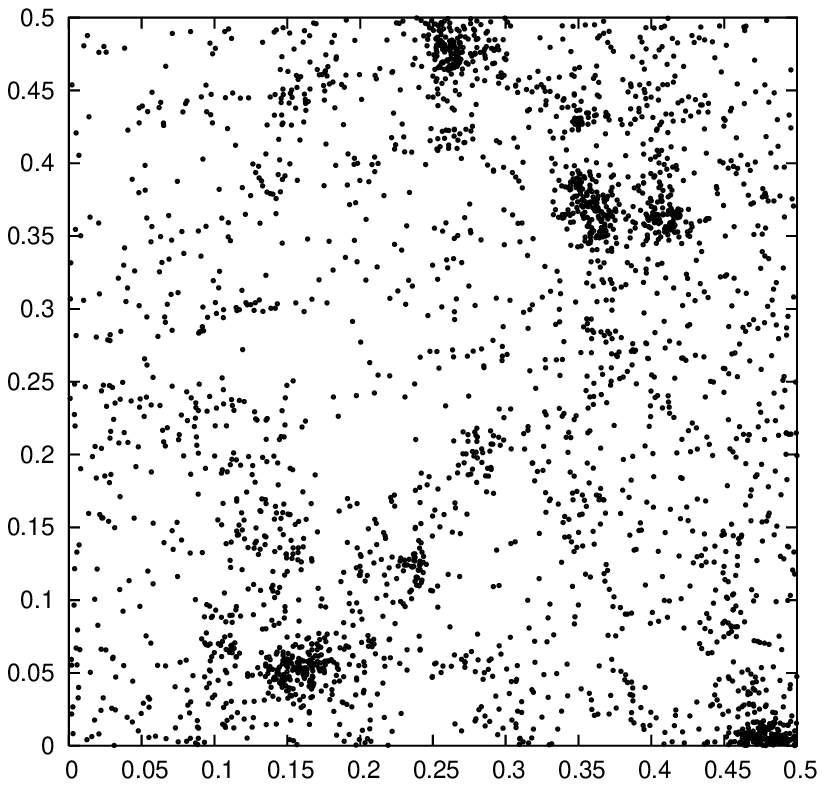}
\\
\includegraphics[width=0.3\textwidth]{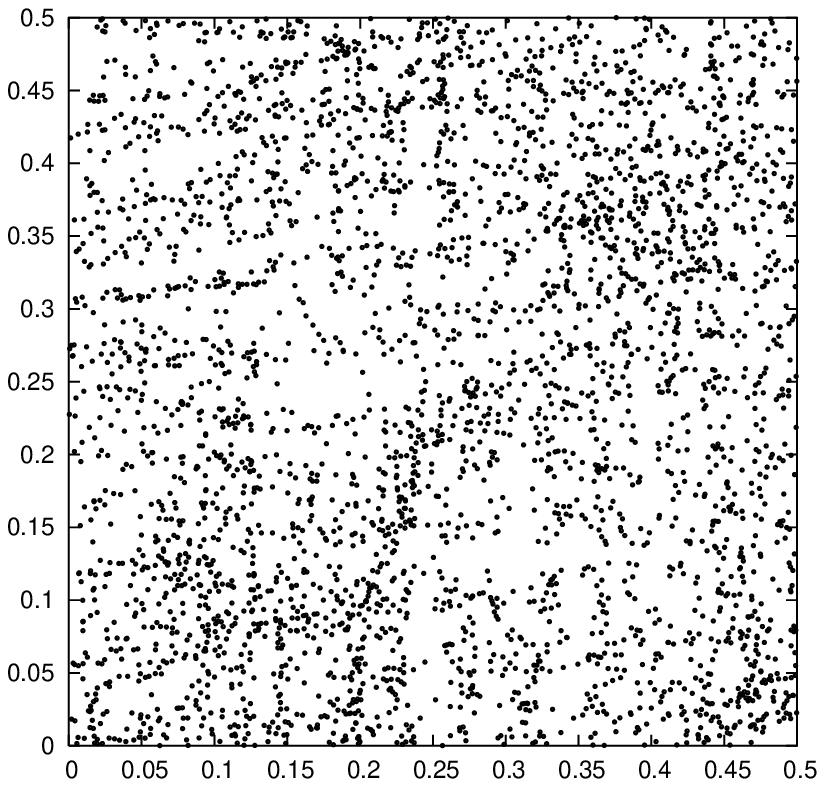}
\includegraphics[width=0.3\textwidth]{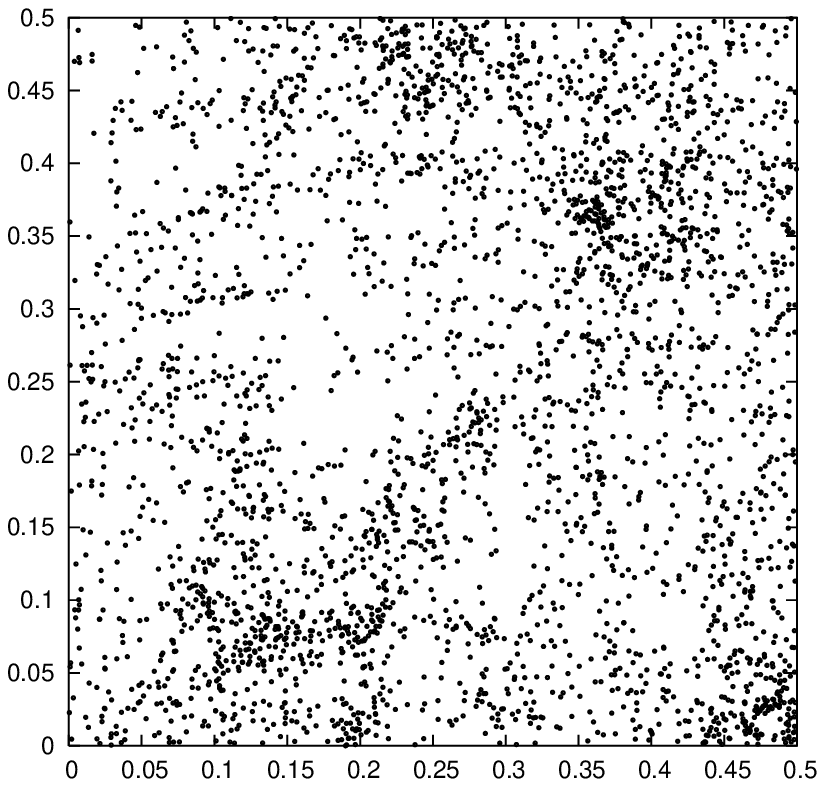}
\includegraphics[width=0.3\textwidth]{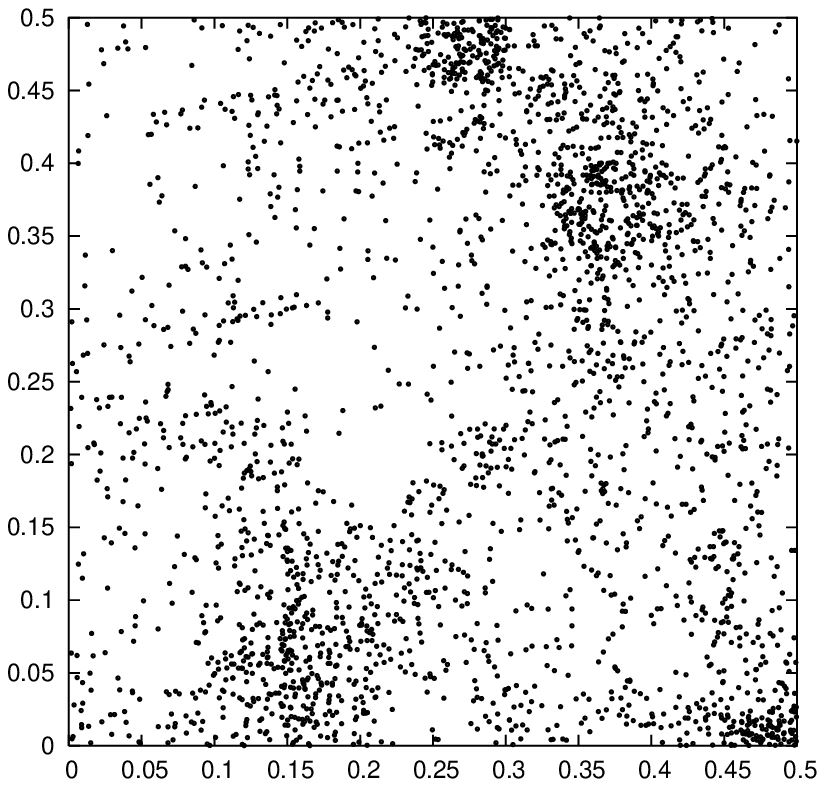}
\\
\includegraphics[width=0.3\textwidth]{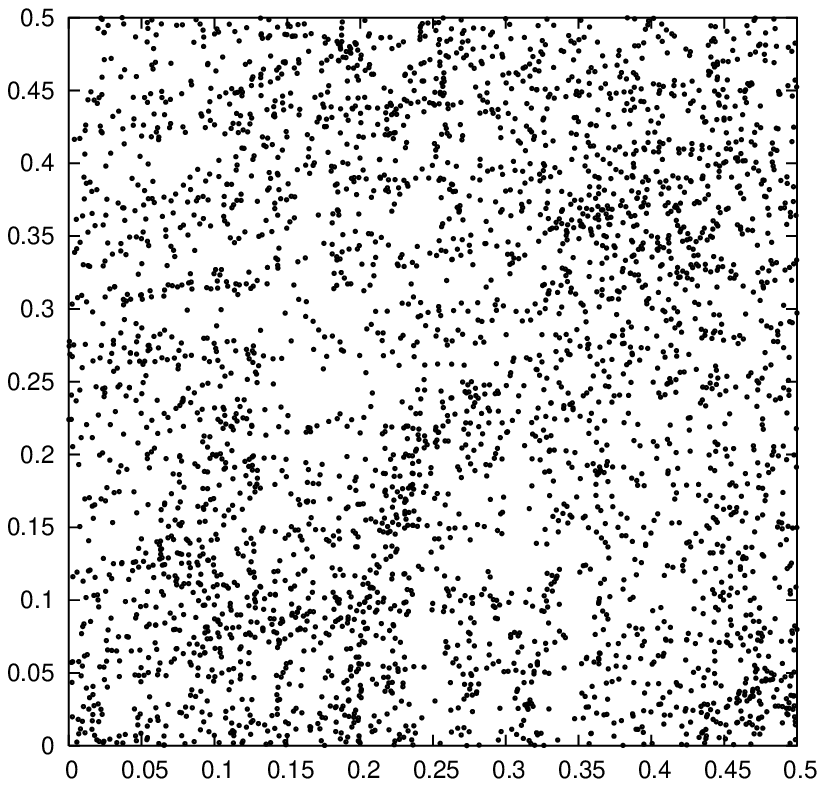}
\includegraphics[width=0.3\textwidth]{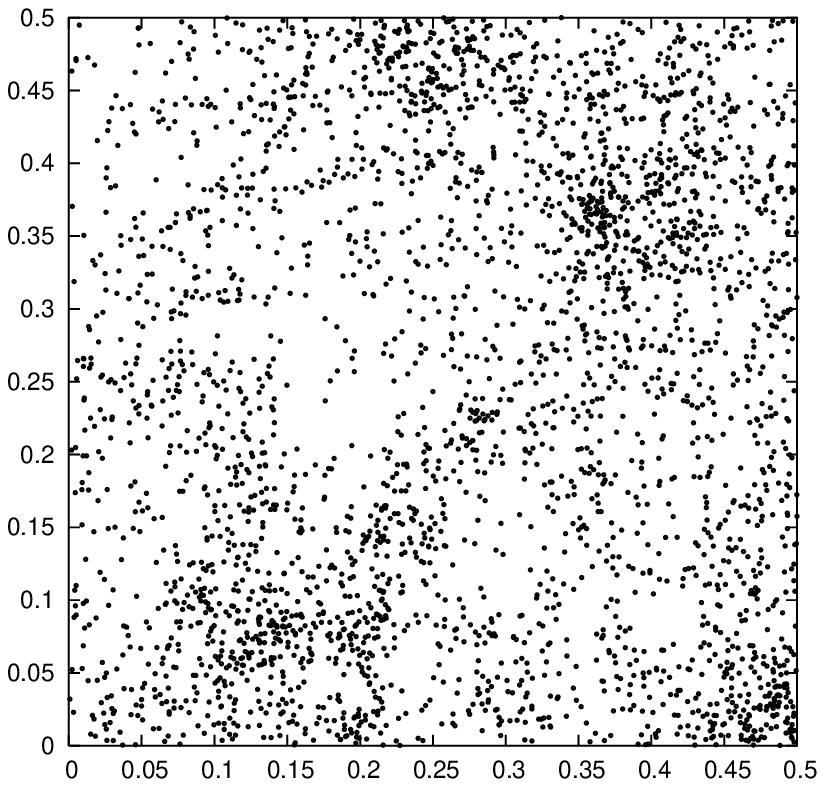}
\includegraphics[width=0.3\textwidth]{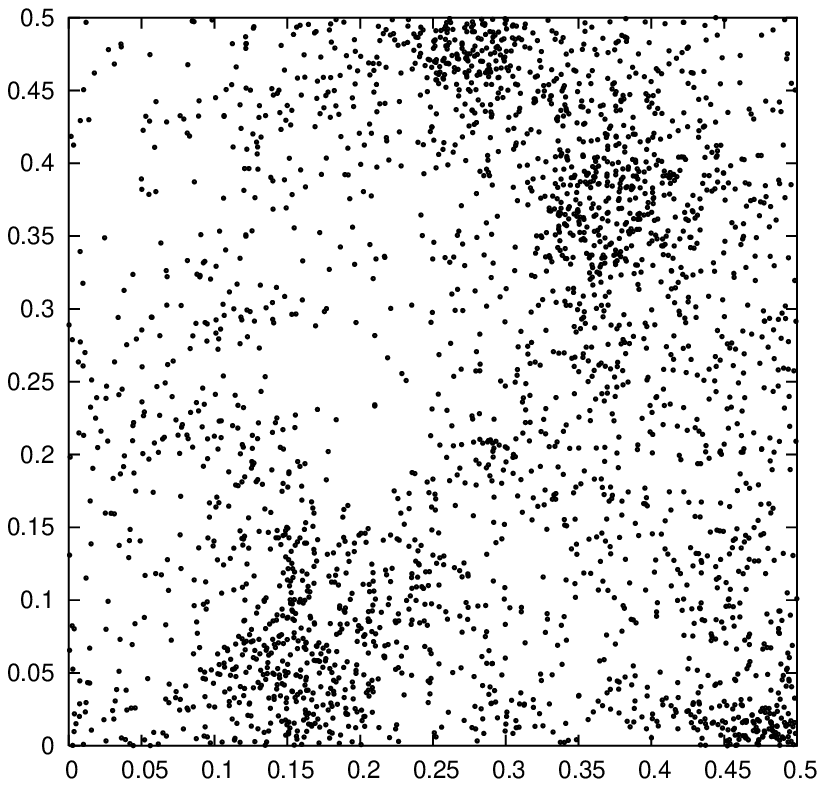}

\caption{Evolution from IC CO32 (projection on the
  plane $z=0$). From left to right,
  times 4.5, 5 and
  5.5 and from top to bottom, FG, PLT and FLT. 
  At $t=5$, PLT breaks
  down according to the criterion based on the function
  $\zeta_\text{D}(\ell,t)$ defined in
  Eq.~\eqref{eq:zetacond} (see the time $t_\zeta$ in tab.~\ref{tab:summarysimul}). Note that this is a cubic subset of $1/8$ of the system to have approximatively $16^3$ particles, as for UC16 in Fig.~\ref{fig:snapSL16}.
\label{fig:snapCO32}}
\end{figure*}

We have performed our FG N-body simulations using the publicly
available treecode \textsc{Gadget}~\cite{gadget}\,\footnote{Version
1.2.  See http://www.mpa-garching.mpg.de/gadget/.}.  In
table~\ref{tab:summarysimul} are summarized the parameters
characterizing the two different IC and simulations.  The number of
particles in the simulations is $16^3$ and $32^3$, respectively. Note
that the softening lengths $\epsilon$ chosen are in both cases much
smaller than the initial inter-particle distance.  This means that
this modification of the gravitational force can be neglected in our
calculation of PLT since it has already broken down when nearest
neighbor particles are so close\,\footnote{The effect of smoothing,
which can be easily implemented in the PLT, will be discussed in a
forthcoming paper~\cite{joyce_06} on the quantification of
discreteness effects.}. For both sets of IC, we have considered three
different evolutions: (i) according to FG, (ii) according to PLT and
(iii) according to FLT. Snapshots are shown in
Figs.~\ref{fig:snapSL16} and~\ref{fig:snapCO32}. In these projected
representations of the configurations at different times, we have
considered in the latter case (CO32) a subvolume of the simulation box
containing initially $16^3$ particles in order to facilitate the
visual comparison of the two cases. From these figures one can already
see how well, to a first approximation, PLT works in predicting the
evolution of the two systems considered.  The improvement given by PLT
over FLT as an approximation to the FG evolution is visually very
clearly manifest in the case of UC16. In the case of CO32 the
difference between the PLT and FLT is less visually evident, but one
can still discern that the former does distinctly better in tracing
the non-linear structures forming in the FG simulation. The reason why
the differences are more pronounced in the case of UC16 is simply that
in the correponding IC there is, compared to CO32, relatively much
more power at small scales than at large scales. As we have seen in
Sect.~\ref{analy_spectrum}, it is at these smaller scales that the two
approximations differ.

The amplitudes of the initial displacements in both simulations
have been chosen sufficiently small so that linear theory 
remains valid during a non-negligible time (in units of $1/ \sqrt{4\pi G\rho_0}$).  Actually
both Eqs.~\eqref{eq:uRt} and~\eqref{eq:flt_chap5}  predict the
same evolution when $t$ is small: $\ve u(\ve R,t) = (t^2/2) \ddot{\ve
u}(\ve R,0)$. In order to distinguish predictions of PLT from those
of FLT, the linear regime must last sufficiently
long, i.e., the initial relative displacements of particles must 
be sufficiently small. We will see below that, for our chosen
initial amplitudes, PLT breaks down approximately at the same 
time in the two systems. This is purely coincidental.
\begin{table}
\begin{ruledtabular}
\begin{tabular}{lcccccc}
Name & $N^{1/3}$ & $P(k)$  & $d_\text{NN}/\ell$  &
$\epsilon/d_\text{NN}$ & $\zeta_\text{D}(\ell,0)$ & $t_\zeta$ \\ 
\hline
UC16 & 16 & $k^2$ &  $0.97 \pm 0.01$  & $0.029$ & 0.002 & $5.0$ \\
CO32 & 32 & $k^{-2}$ & $0.983 \pm 0.009$ & $0.057$ & 0.0006 & $5.0$
\end{tabular}
\end{ruledtabular}
\caption{Summary of the two different distributions used as IC for the
  comparison between linear theory and true gravity (UC stands for
  \emph{uncorrelated} while CO is for \emph{correlated}). The column
  ``$\, d_\text{NN}/\ell\, $'' gives the average distance between
  nearest neighbors and its standard deviation, both normalized by
  $\ell$.  The next column is the softening length $\epsilon$,
  normalized by  $d_\text{NN}$, used in the FG simulation. The
  fifth column gives the initial value of the function $\zeta_D$ at
  the interparticle distance, defined in Eq.~\eqref{zeta-func}. The
  time in the last column is defined later in the same equation: it
  gives the time scale at which PLT is expected to break down.
\label{tab:summarysimul}}
\end{table}

Let us first consider how to characterize quantitatively the regime of
validity of PLT. As discussed in Sect.~\ref{sub-linear}, we anticipate
that the relevant quantity will be the relative displacement of
particles compared to the initial distance separating them. This
motivates the study of the following statistical quantity:
\begin{equation}
\begin{split}
\label{zeta-func}
\zeta_\text{D}(\ve R,t)  &\equiv \frac{\left\langle | \ve u(\ve R',t) - \ve u(\ve R'+\ve R,t) |^2
\right\rangle_{\ve R'}}{|\ve R|^2} \\ &  = 
\frac{2}{|\ve R|^2} \left(  \xi_\text{D}(\ve 0,t) -  \xi_\text{D}(\ve R,t) \right)
\end{split}
\end{equation}
where $\xi_\text{D}(\ve R,t)\equiv \left\langle \ve u(\ve R',t) \cdot
\ve u(\ve R'+\ve R,t) \right\rangle_{\ve R'}$ is the trace of the correlation
tensor of the displacements 
which is related to the PS of displacements
$P_\text{D}(\ve k,t)\equiv |\tilde{\ve u}(\ve k,t)|^2 / N$ by\,\footnote{We note that for the 
case of the spectrum of displacements 
in CO32, $P_D(\ve k,0) \propto |\ve k|^{-4}$, this quantity is actually
not well defined, in the sense that it diverges in the
thermodynamic limit (i.e. $L \rightarrow \infty$ at fixed
number density $n_0=N/L^3$). Indeed for a generic PS of 
displacements $P_\text{D}(\ve k) \propto |\ve k|^n$, we 
have, taking the continuous limit of Eq.~\eqref{eq:xiDpD},
that $\xi_\text{D}(\ve R) =\frac{1}{2\pi^2 n_0 R} \int_0^{k_\text{c}}
k^{n+1} \sin (k R) \, dk$
which is infrared divergent for $n \leq -3$ ($k_c$ being 
an ultraviolet cut-off given by the Nyquist frequency in this context).
However, it can be shown \cite{joyce_04} that the function $\zeta_D(\ve R,t)$ is well defined in the 
same limit for  $n>-5$. This limit in fact just coincides with
the condition that the associated density fluctuations have 
finite variance, since this requires that
$\lim_{k \rightarrow 0} k^3 (k^2 P_\text{D} (\bk))=0$.
These divergences in  $\xi_\text{D}(\ve R)$ are not
a problem provided $\zeta_D(\ve R,t)$ is well defined:
the particles can move an infinite distance from their
lattice positions, but what is important for the validity of 
the approximation used is how much their relative
displacements changes compared to their separation.}
\begin{equation}
\xi_\text{D} (\ve R,t) = \frac{1}{N} \sum_{\ve k} P_\text{D}(\ve k,t)
    \exp\left(i\ve k\cdot \ve R\right) \ .
\label{eq:xiDpD}
\end{equation}
Given the convergence criterion Eq.~\eqref{convergence}, we
expect PLT to break down when 
$\zeta_\text{D}(\ve R,t)\sim 1$ for some $\ve R$. One would
expect that this condition will first be attained for 
$|\ve R| = \ell$, i.e., when a significant number of
nearest neighbors have come close to one another.

\begin{figure}
\psfrag{zetaDellt}{$\zeta_\text{D}(\ell,t)$}
\includegraphics[width=0.5\textwidth]{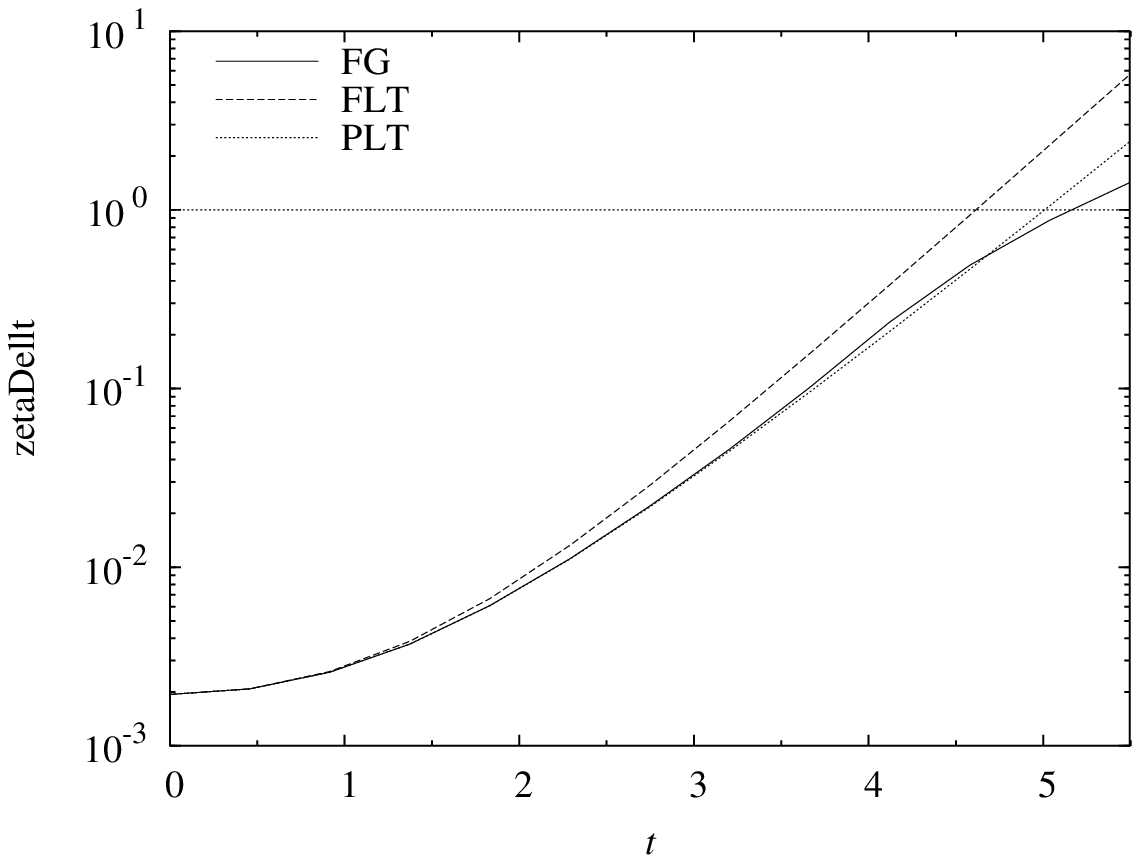}
\caption{Evolution of the function $\zeta_\text{D}(\ell,t)$,
  defined in Eq.~\eqref{eq:zetacond}, according to FG, PLT and
  FLT in the UC16 simulation. The horizontal line is
  $1$ and gives the characteristic time $t_\zeta$ around
  which we expect  PLT to break down (as an approximation to FG).
  \label{fig:evolzetaNNSL16}}
\end{figure}

\begin{figure}
\psfrag{zetaDellt}{$\zeta_\text{D}(\ell,t)$}
\includegraphics[width=0.5\textwidth]{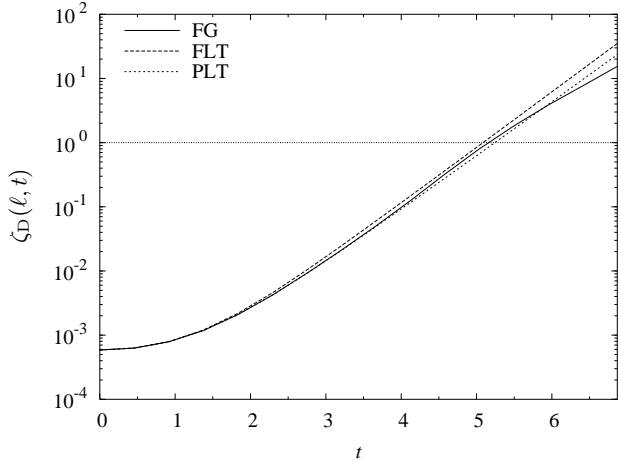}
\caption{Evolution of the function $\zeta_\text{D}(\ell,t)$ according
  to FG, LPT and FLT in the CO32 simulation. The horizontal line is
  $1$ and gives the characteristic time $t_\zeta$ around
  which we expect  PLT to break down (as an approximation to FG). 
  \label{fig:evolzetaNNkm2_2}}
\end{figure}

Figures~\ref{fig:evolzetaNNSL16} and~\ref{fig:evolzetaNNkm2_2} show
the evolution of the function $\zeta_\text{D}(\ve R,t)$ averaged over
the first six nearest neighbors, i.e., for the six vectors $\ve R$ of
norm $\ell\ $\,\footnote{Three of them are actually enough since
$\xi_\text{D}(\ve R,t)$ is symmetric in its first argument.}, for
IC UC16 and CO32 respectively. In both cases, the FG
evolution is compared with the predictions of PLT and FLT. The
horizontal line is at $1$ and allows one to determine the 
characteristic time $t_\zeta$ such that
\begin{equation}
\zeta_\text{D}(\ell,t_\zeta) \equiv \frac{1}{6} \sum_{|\ve R|=\ell}
\zeta_\text{D}(\ve R,t_\zeta)  = 1 \ ,
\label{eq:zetacond}
\end{equation} 
around which we expect PLT to break down, as discussed above.  This
time --- reported in tab.~\ref{tab:summarysimul} --- is of the order
of $5.0$ for both IC\,\footnote{As mentioned above, the fact that this
time is very close in both simulations is purely coincidental.}. We
see from these figures that this expectation turns out to be correct:
the PLT evolution from both sets of IC traces very accurately the FG
evolution of the N-body simulation until a time very close to
$t_\zeta$. For UC16 (Fig.~\ref{fig:evolzetaNNSL16}) the deviation
between the PLT and FG evolutions becomes significant for a time just
slightly earlier, while for CO32 (Fig.~\ref{fig:evolzetaNNkm2_2}) this
time is slightly later. Considering the curves for FLT, we see that it
does less well than PLT in both cases, the improvement given by PLT
over FLT being much more marked in the case of UC16. This confirms
quantitatively the visual impression of Figs. ~\ref{fig:snapSL16} and
\ref{fig:snapCO32}, for the same reason we gave above. Even though
$\zeta_\text{D}(\ell, t)$ is a real space quantity characterising
correlation around the inter-particle distance $\ell$, it is an
integral in $k$ space which picks up a contribution from longer
wave-length modes which are well described (as discussed in
Sects.~\ref{Grav-lat} and \ref{analy_spectrum} and further below) by
FLT. For the much more infra-red dominated spectrum of displacements
of CO32 this contribution is much more significant and so FLT
naturally is a better approximation for this quantity than in the case
of UC16.

\begin{figure}
\psfrag{rl}{$r/\ell$}
\psfrag{omegart}{$\Upsilon(r,t)$}
\includegraphics[width=0.5\textwidth]{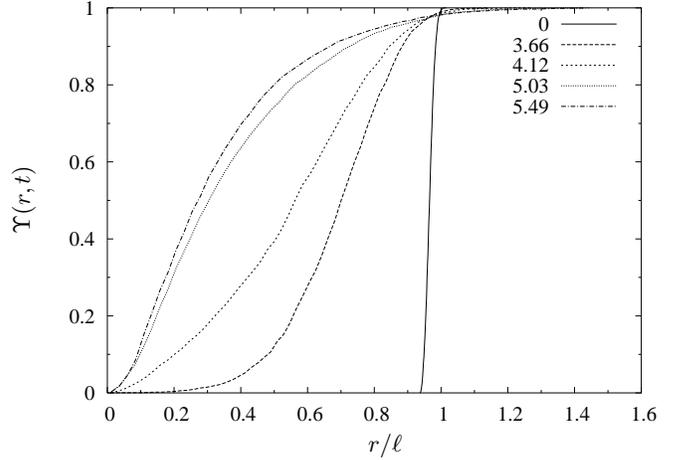}
\caption{Evolution of the cumulative NN distribution 
$\Upsilon(r,t)$ in the UC16 simulation. The times 
are indicated in the legend.
  \label{fig:NNSL16}}
\end{figure}

\begin{figure}
\psfrag{rl}{$r/\ell$} \psfrag{omegart}{$\Upsilon(r,t)$}
\includegraphics[width=0.5\textwidth]{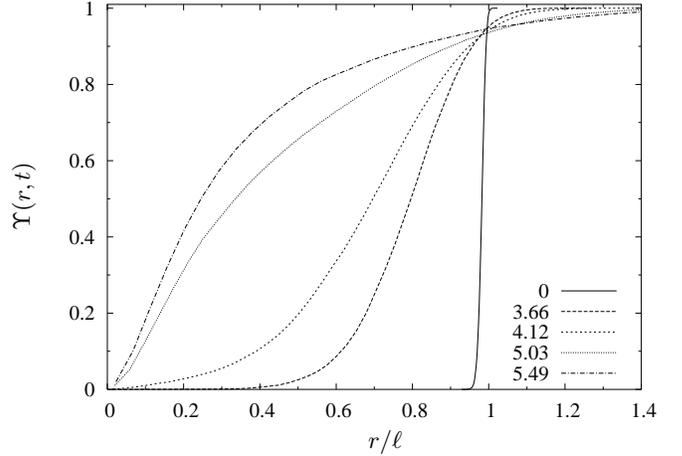}
\caption{Evolution of the NN distribution $\Upsilon(r,t)$ in the CO32
  simulation. The times are indicated in the
  legend.\label{fig:NNCO32}}
\end{figure}

We next give an alternative, perhaps intuititively more direct, 
way of quantifying the regime of validity of PLT. We consider 
the nearest neighbor (NN) distribution $\omega(r,t)$: at a given 
time $t$, this function gives the probability density for a 
particle to have its nearest neighbor at the distance $r$ 
(see e.g. \cite{GSJP_05}).  In Figs.~\ref{fig:NNSL16} and~\ref{fig:NNCO32} 
are shown, for the FG evolution of UC16 and CO32, the cumulative distributions derived from $\omega(r,t)$,
\begin{equation} 
\Upsilon(r,t)= \int_0^r \omega(s,t)\, ds \ .
\end{equation}
These represent the probability that a given particle has its NN
within a distance $r$.  It allows one to determine quite accurately
the time at which ``shell crossing'' occurs\,\footnote{We adopt here
loosely the terminology used in fluid theory to refer to the time when
particles fall on top of one another: if one considers the particles
in the simulation as the centres of fluid elements, this corresponds
to what is called ``shell crossing'', at which point the linearised
Lagrangian fluid theory (i.e. FLT) breaks down as the density
diverges. }. We see that between $t\approx 4$ and $t\approx
5=t_\zeta$, the behavior of the cumulative distributions changes in an
important way: for $t\lesssim 4$, almost no particle has its NN closer
than a distance $r\approx 0.3$, while at time $t\approx 5= t_\zeta$,
$50 \%$ of the particles have their NN at a distance smaller than this
distance. Thus at $t\approx 4$, the first shell crossings occur and at
$t\approx 5$, approximately half of the particles have already had
their own shell crossing or are very close to it.

% ======================================================================

\subsection{Comparison of relative displacements}

\begin{figure}
\psfrag{distNN}{$d(t)/\ell$}
\includegraphics[width=0.5\textwidth]{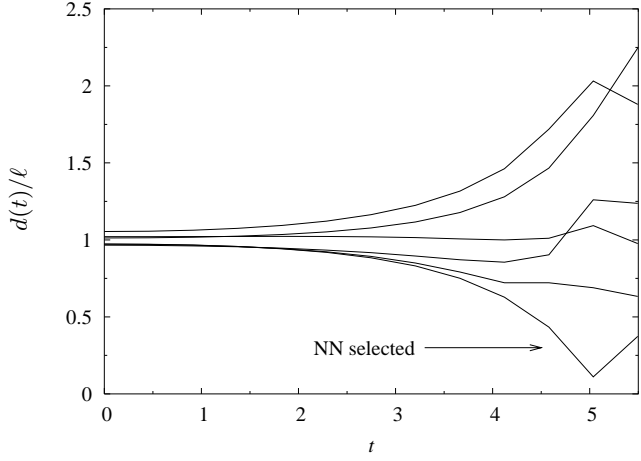}
\caption{Evolution of the distances $d(t)$ (normalized to $\ell$) between a
  randomly chosen particle in UC16 and its six NN in the FG
  simulation. The arrow shows the NN which is used for the study of
  the relative displacement (Fig.~\ref{fig:reldisplNNSL16}).
\label{fig:distNNSL16}
}
\end{figure}

\begin{figure}
\psfrag{distNN}{$d(t)/\ell$}
\includegraphics[width=0.5\textwidth]{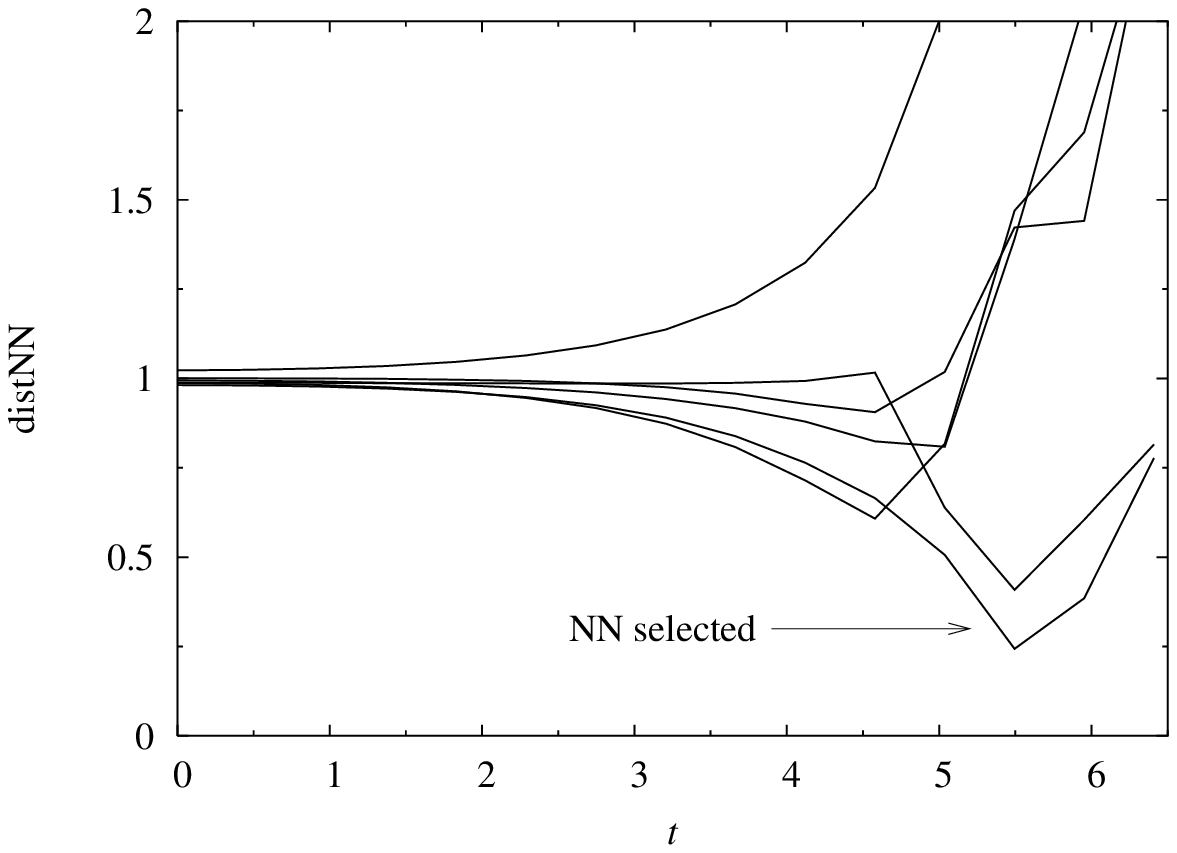}
\caption{Evolution of the distances $d(t)$ (normalized to $\ell$) between a
  randomly chosen particle in CO32 and its six NN in the FG
  simulation. The arrow shows the NN which is used for the study of
  the relative displacement (Fig.~\ref{fig:reldisplNNCO32}).
\label{fig:distNNCO32}
}
\end{figure}

\begin{figure}
\psfrag{absduell}[c]{$|u_\mu(\ve R,t)-u_\mu(\ve R',t)|/\ell$}
\includegraphics[width=0.5\textwidth]{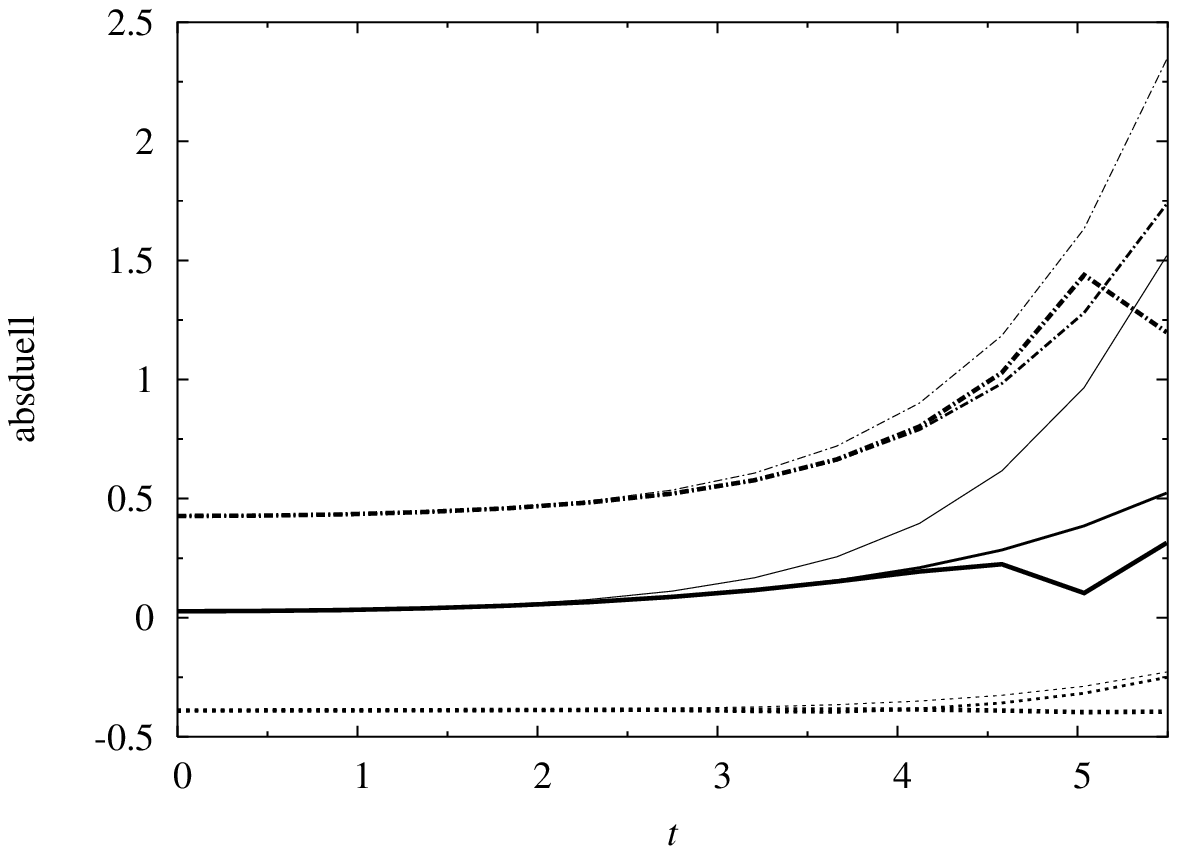}
\caption{Evolution according to FG (thick line), PLT (medium line) and
  FLT (thin line) of the absolute value of each coordinate (denoted by
  a different type line) of the relative displacement $\ve u(\ve R,t)-\ve
  u(\ve R',t)$ of a randomly chosen particle and its NN in UC16. The
  particle is the same as the one chosen for Fig.~\ref{fig:distNNSL16}
  and the NN corresponds to the one indicated by an arrow in this last
  figure. Note that, for clarity, we have shifted two of the
  coordinates by $\pm 0.4$.
  \label{fig:reldisplNNSL16}}
\end{figure}

\begin{figure}
\psfrag{absduell}[c]{$|u_\mu(\ve R,t)-u_\mu(\ve R',t)|/\ell$}
\includegraphics[width=0.5\textwidth]{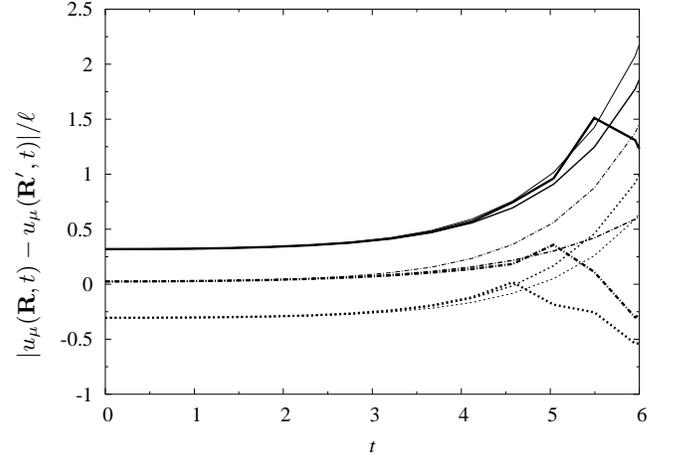}
\caption{ Evolution according to FG (thick line), PLT (medium line)
  and FLT (thin line) of the absolute value of each coordinate
  (denoted by a different type line) of the relative displacement $\ve
  u(\ve R,t)-\ve u(\ve R',t)$ of a randomly chosen particle and its NN
  in CO32. The particle is the same as the one chosen for
  Fig.~\ref{fig:distNNCO32} and the NN corresponds to the one
  indicated by an arrow in this last figure. Note that, for clarity, we
  have shifted two of the coordinates by $\pm 0.3$.
  \label{fig:reldisplNNCO32}}
\end{figure}

It is interesting to study also the accuracy of PLT in tracing the FG evolution
of the trajectories of {\it individual} particles (rather than only averaged
quantities). We consider now the relative displacement of a particle 
with respect to its NN, that is $\ve u(\ve R,t) - \ve u(\ve R',t)$ 
where $\ve R'$ and $\ve R$ are separated by a vector of elementary size $\ell$. To do so we
have chosen randomly a particle in each simulation and selected among its
six initial NN the one which ends up closest to it at the
time at which PLT breaks down in the corresponding FG simulation 
(see Figs.~\ref{fig:distNNSL16} and~\ref{fig:distNNCO32}). 
In Figs.~\ref{fig:reldisplNNSL16}
and~\ref{fig:reldisplNNCO32} are shown the result for UC16 and CO32
respectively. These two figures allow one to see that for both IC, PLT describes very well the evolution of the relative
displacement of the two particles considered up to a time of the order
of $4.5 = 0.9 t_\zeta$. In both systems, this time is slightly smaller
than the time at which the particle selected is the closest to its
NN: from Figs.~\ref{fig:distNNSL16} and~\ref{fig:distNNCO32} we may
estimate that this time is approximately $5.0$ in the system UC16
and $5.5$ in CO32.

We also note that PLT does again also in this case better 
than FLT in describing the relative displacements. Indeed FLT 
already breaks down at $t\approx 3$ in UC16 and $t\approx 3.5$ in CO32.


% ================================================================================================

\subsection{Evolution of modes}

In this subsection we consider the comparison of PLT with FG and FLT
in reciprocal space. We first consider the evolution of the PS of displacements
$P_\text{D}(\ve k,t)$ for a few specific vectors $\ve k$.  Then
we study the evolution of this quantity, but now averaged on vectors $\ve
k$ of similar modulus. Besides verifying the conclusions drawn in
Sect.~\ref{analy_spectrum}, this numerical study allows us 
to assess the validity of PLT for different wavenumbers as a function
of time. 

Following the results and discussion in Sect.~\ref{analy_spectrum},
we expect firstly to see the PLT evolution to differ less and
less from FLT as we go to longer wavelength modes ($|\ve k| \ll 1/\ell$), 
since the eigenvalues and eigenvectors of the modes approach those 
of the fluid limit in this case.  For such modes we expect that PLT and
FLT should both follow the FG evolution accurately up to at least 
$t\sim t_\zeta$. For short wavelength modes ($|\ve k| \sim 1/\ell$)
we expect PLT to be significantly more accurate than FLT. 
As for the time of breakdown of PLT for a given mode, we
would expect that long wavelength mode evolution should be described
accurately by PLT for a time longer than that of short 
wavelength modes: the breakdown of the approximation in the sense
we have characterised it above, i.e., in terms of the approach
of NN particles, would not be expected to affect significantly
the evolution of longer wavelength modes. We will discuss
this point further below.

\begin{figure}
\includegraphics[width=0.5\textwidth]{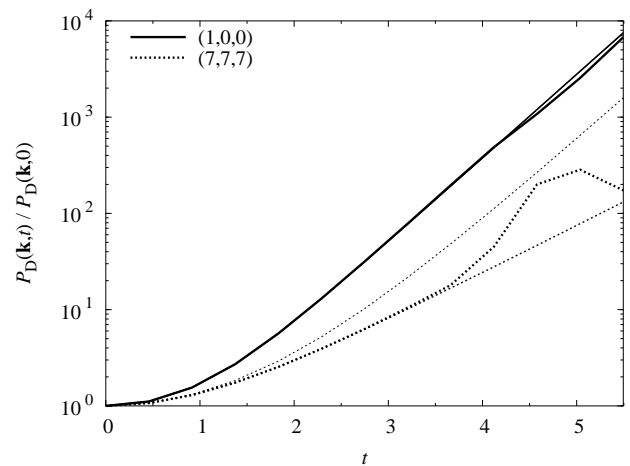}
\caption{Evolution of $P_\text{D}{\ve u}(\ve k,t)$ normalized to its
  initial value for two chosen
  vectors $\ve k$ according to FG (thick lines), PLT (medium lines)
  and FLT (thin lines) with IC UC16.  The vectors $\ve
  k$ are $2\pi (1,0,0)/ L$ and $2\pi (7,7,7)/L$.  
  \label{fig:evoluk2SL16}}
\end{figure}

\begin{figure}
\includegraphics[width=0.5\textwidth]{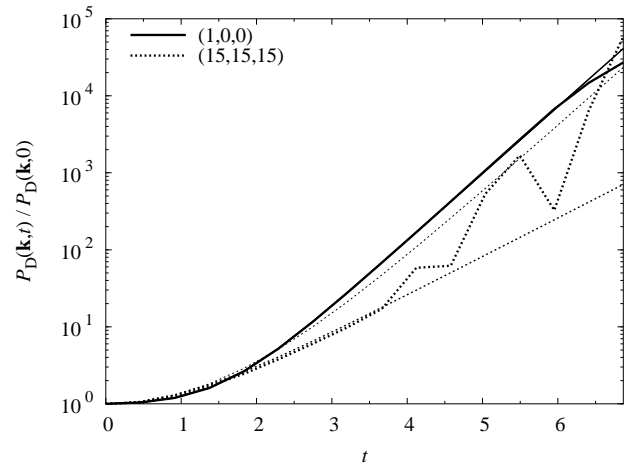}
\caption{Plot similar to the one in Fig.~\ref{fig:evoluk2SL16} but for
  IC CO32 and vectors $\ve k=2\pi (1,0,0)/L$ and $\ve
  k=2\pi (15,15,15)/L$.    \label{fig:evoluk2km2_2}}
\end{figure}

Figures \ref{fig:evoluk2SL16} and \ref{fig:evoluk2km2_2} show the
evolution of $P_\text{D}(\ve k,t)$, normalized to its initial value
(i.e. at $t=0$), for two chosen vectors $\ve k$ with very different 
lengths and different orientations with respect to the lattice,
for IC UC16 and CO32. In both cases, PLT follows
very accurately the FG evolution up to a certain time for both 
long and small wavelength modes. The time up to which
the agreement is good depends, as has been anticipated, 
on the wavelength. It breaks down first for the large 
$k$ (small wavelength) mode, and significantly later for
the small $k$ (large wavelength) mode. For both
IC the effects of non-linearity become
significant for the large $k$ mode chosen (one of the modes 
of largest modulus in each case), at $t\approx
3.5$, i.e., slightly before the time of the first shell crossings
as determined above from Figs.~\ref{fig:NNSL16} and~\ref{fig:NNCO32}. 
In this case we see also clearly that FLT is a poor approximation
to the evolution, corresponding to an evolution with an exponent
which is significantly too large. For the long wavelength mode
(one of the modes of smallest modulus in each case), 
on the other hand, we see that FLT and PLT, as expected
predict  the same evolution  (at the level of precision allowed by the
figures). And in this case we see that they both trace 
the FG evolution very accurately for a very significantly longer time,
up to $t \approx 6$ in the case of C032.

We now consider the PS of displacements $P_\text{D}(\ve k,t)$,
averaged in spherical shells, which we denote simply by
$P_\text{D}(k,t)$:
\begin{equation}
P_\text{D}(k,t) = \frac{1}{n_k} \sum_{k-\mathrm{d}k \leq |\ve k| <
  k+\mathrm{d}k} P_\text{D}(\ve k,t)  
\label{eq:PDaveraged}
\end{equation}
where $n_k$ is the number of vectors $\ve k$ in the spherical
shell $[k-\mathrm dk,k+\mathrm dk[$.  This analysis of
$P_\text{D}(k,t)$ allows one to observe in further detail
how PLT fails in describing FG evolution after some time and 
at different scales. Moreover, it shows how it is more 
accurate than FLT in approximating FG.

\begin{figure}
\includegraphics[width=0.5\textwidth]{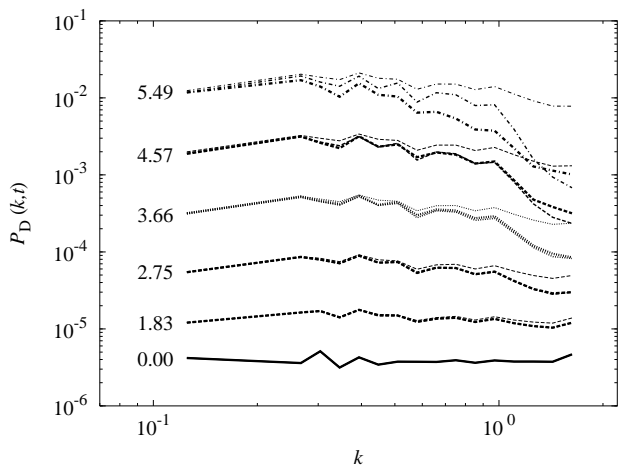}
\caption{Evolution of the averaged PS of displacements $P_\text{D}(k,t)$,
  Eq.~\eqref{eq:PDaveraged}, according to FG (thick lines), PLT (medium
  lines) and FLT (thin lines) from IC UC16. The times
  are indicated at the left of the curves. The values of $k$ are given
  in unit of the Nyquist frequency $k_\text{N}$.
  \label{fig:evolPDSL16}}
\end{figure}

Figure \ref{fig:evolPDSL16} shows the evolution of the averaged
PS of displacements according to FG (thick lines), PLT (medium lines) and
FLT (thin lines) with IC UC16. The first difference
between PLT and FG appears at large $k$ at $t\approx 4.5$, slightly before
$t_\zeta$. This difference propagates to smaller $k$ at later times.
FLT is already discernibly different from 
FG at $t\approx 1.8$ at large $k$ and the
difference propagates to smaller $k$ at later times.

\begin{figure}
\includegraphics[width=0.5\textwidth]{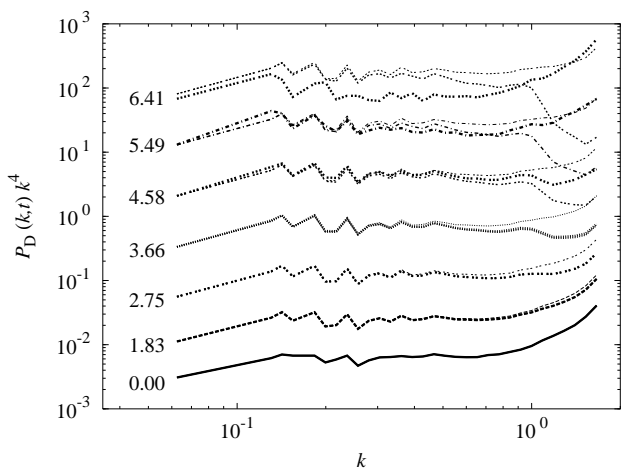}
\caption{Evolution of the averaged PS of displacements $P_\text{D}(k,t)$,
  Eq.~\eqref{eq:PDaveraged}, according to FG (thick lines), PLT
  (medium lines) and FLT (thin lines) from IC
  CO32. The times are indicated at the left of the curves. The values
  of $k$ are given in unit of $k_\text{N}$. Note that it is actually
  $P_\text{D}(k,t)k^4$ which is represented. 
  \label{fig:evolPDkm2_2}}
\end{figure}

Figure \ref{fig:evolPDkm2_2} is similar to Fig.~\ref{fig:evolPDSL16}
but concerns IC CO32. Note that, since
$P_\text{D}(k,0)\propto k^{-4}$, it is
$P_\text{D}(k,t)k^4$ which is shown\,\footnote{A slight excess of power
over the expected PS is clearly visible around the Nyquist frequency.
This is a small aliasing effect due to the fact that we have
included some $\bk$ in our sum outside the first Brillouin zone: 
we have summed in Eq.~\eqref{eq:kmoins2IC} over the modes 
$(L/ 2\pi )\bk \in [-30,30]^3$ rather than $( L/2\pi) \bk \in [-16,15]^3$.
This has no bearing on the conclusions drawn here.}.
Similar conclusions can be drawn to those in the uncorrelated case.
The differences between PLT and FG become visible at
$t\approx 4.5$ and propagate at later times. Actually at $t\approx 6.5$,
there is no longer good agreement at any $k$. FLT starts to 
deviate discernibly from FG at $t\approx 1.8$.

Figs.~\ref{fig:evolPDSL16} and \ref{fig:evolPDkm2_2} thus confirm
clearly what was already observed in Figs.~\ref{fig:evoluk2SL16} 
and \ref{fig:evoluk2km2_2}: the breakdown of PLT starts at the
largest $k$ and propagates progressively in time to smaller $k$.
Further PLT (and FLT) remain a good approximation to the
evolution at the smaller $k$ in our simulations at times
significantly longer than the time $t_\zeta$ which we used
to characterize the global validity of PLT. The reason is
simple\,\footnote{The argument given is, in the context of
cosmology, attributed to Zeldovich. For an extensive 
discussion see \cite{peebles_80} \S27-28, and also
\cite{causality}.}, and it is the same one which explains, e.g., why 
linear fluid theory successfully describes the evolution
of small perturbations to a self-gravitating system 
even when there are strong non-linearities at smaller
scales: re-arranging matter in any way, subject only to
the constraint that matter and momentum are conserved,
up to some real space finite scale, $\ell_c$ say, 
can produce, at most, fluctuations at small $k$ 
(i.e. $k \ll 1/\ell_c$) with a PS of density fluctuations
$\propto k^4$. The perturbative approximation to
full gravity represented by PLT breaks down globally,
as we have seen, when NN start to approach one another. 
When this happens, however, the full gravitational
force can still be approximated for a longer time
by the full gravitational force on each particle
due its NN particle, plus the force from
all other particles still linearised, as in PLT, in the
relative displacements from their starting positions.
This means that FG can continue to be approximated 
by PLT plus an effective interaction {\it of finite
range}. This latter interaction can produce at most
a term in the PS in $k^4$, which will always be
dominated by the simple amplification given by PLT
of the initial PS, which in the cases we have considered
has a small k behaviour $\propto k^n$ with $n <4$
($n=2$ and $n=-2$ for UC16 and CO32 respectively).
An interesting question is whether, at these longer
times, PLT is a good approximation in so far as 
it agrees well with FLT, or whether it can actually
continue to trace the FG evolution better than FLT.
From the numerical results we have given here it
is not possible to distinguish FLT from PLT in the
corresponding regime. Larger simulations would
be required to answer this question.

% ===============================================================================
% ===============================================================================
% ===============================================================================

\subsection{Oscillating modes}
\label{sec:oscill}

We have noted in Sect.~\ref{analy_spectrum} that the spectrum of
eigenvalues of the sc lattice contains some negative eigenvalues,
which give oscillating modes. These modes are not of practical 
importance in typical cosmological simulations since they are relatively few,
and in a short time  of little importance compared to the unstable
modes (which also have considerably larger exponents\,\footnote{Further,
the IC of cosmological simulations, because
of the Zeldovich approximation Eqs.~\eqref{sol_fluid_g_asymp}-\eqref{zeldo-cond},
are purely longitudinal, while the oscillating modes are on the
accoustic branches which are close to purely transversal for most
$\bk$.}). It is nevertheless interesting to study them briefly
since they are a peculiarity of the discrete system. Indeed, as
discussed in Sect.~\ref{analy_spectrum} such oscillating modes 
do not exist in fluid theory without initial velocities.

To study these modes we consider a sc
lattice with $N^{1/3}$ {\it even}\,\footnote{Some eigenvectors of the
oscillating modes have in this case the particularly simple 
form we will treat. The fact that these modes differ so
significantly for the case that $N^{1/3}$ is odd or even
illustrates again that their presence is entirely 
associated with the discreteness of the system.} and the following initial displacement:
\begin{multline}
\ve u(\ve R,t) =\\ \delta  \cos\left(\frac{2\pi}{L}  \frac{n}{2}
R_x\right) \hat{\ve y} \ = 
\begin{cases}
+\delta \hat{\ve y}  &  \text{if }  R_x/\ell \text{ is even,}\\
 -\delta \hat{\ve y} & \text{if } R_x/\ell \text{ is odd,}
\end{cases}
\label{eq:oscillationdispl}
\end{multline}
where $\delta$ is a constant.  This corresponds to displacing each plane
of particles with constant $R_x$ in the direction $+\hat{\ve y}$ and
$-\hat{\ve y}$ alternately (see Fig.~\ref{coll-clust}). The only excited mode is $\ve k=
\ (0,-1/\ell,0)$. According to PLT, the eigenvalue
associated to this displacement is $-0.156\cdot 4\pi G \rho_0/3$ 
(cf. Sect.~\ref{analy_spectrum}) and
the particles therefore should oscillate. If no initial velocity is
considered, then the motion of a particle should be, as predicted by PLT
\begin{equation}
\label{oscil-eq}
\ve u(\ve R,t) = \pm \delta \cos\left(\om_0\ t\right)  \, \hat{\ve y},
\end{equation}
with a frequency $\om_0=\sqrt{0.156\cdot 4\pi G\rho_0/3}$.

To observe these oscillations numerically, we have written a special
code to integrate the equations of motion with FG, rather than using
the same code (\textsc{Gadget}) as in the preceeding subsections. The
reason we do this is that it is very difficult numerically to observe
them with such a code. Indeed the simulation of oscillating modes
would be an interesting and challenging test for the precision of
gravitational N-body codes. The primary difficulty is that the
negative eigenvalues are small compared to most of positive ones
(cf. Fig.~\ref{fig1} in Sect.~\ref{analy_spectrum}), so that an even
much smaller amplitude perturbation in any of these modes can grow as
an instability on a time scale much smaller than the period of
oscillation. Such small perturbations are clearly created by numerical
imprecision, but also, by the intrinsic non-linearities in the FG,
i.e., there is a coupling of modes which is neglected in PLT. Thus one
must work not only with great numerical precision but also at
extremely low amplitude to avoid ``contamination'' by other modes
through the non-linearities on the relevant (long) time-scale.

The code that we have used has been built specifically for the
particular initial displacements~\eqref{eq:oscillationdispl} taking
into account the following considerations. From the 
symmetries of the configuration, it follows that the {\it full} 
gravitational force on a particle is only along the $y$ axis, and
modulo a change of sign, all the particles move the same way. 
Moreover, since the distribution of particles can be seen as 
two perfect rectangular sub-lattices with lattice spacing 
$\ell$ in the $y$ and $z$ directions and $2\ell$ in the $x$
direction\,\footnote{One lattice is constituted by the particles 
which are initially displaced by an amount $+\delta$ and the other one by the
particles displaced by $-\delta$.} the force on a given particle is only
due to the particles in the sub-lattice which does not contain this
particle\,\footnote{The force from the particles in the same sub-lattice is
 zero by symmetry.}.  In order to evolve the system, it is therefore sufficient
to integrate the equation of motion of a single particle in one
dimension with a force coming from half of the particles of the system
(the force is calculated by using the Ewald summation formula).  The
method used for the integration is the \emph{embedded
Runge-Kutta-Fehlberg (4, 5) method}, implemented in the GNU scientific
library\,\footnote{http://www.gnu.org/software/gsl/.} and more precise
than the standard leap-frog method used in cosmological N-body
simulations. One can also note that, due to the periodicity of the
system\,\footnote{The systems which we consider are always periodic at
the level of the box but here the distribution inside the box is
itself periodic.}, the number of particles is not important as long as
$N^{1/3}$ is even: $N^{1/3}=2$ is actually enough.

\begin{figure}
\includegraphics[width=0.5\textwidth]{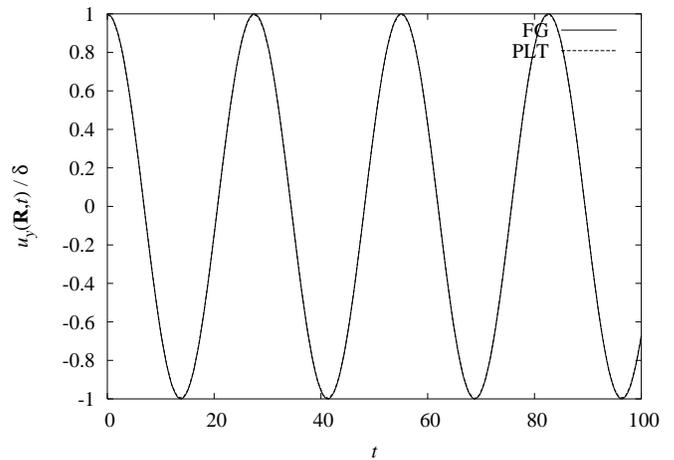}
\caption{Oscillations of a particle with
  initial displacement~\eqref{eq:oscillationdispl} and $\delta=0.004\ell$
  according  to FG (solid line) and PLT (dashed line). The curves are
  actually indistinguishable on the plot. The time is in units of $1/\sqrt{4\pi G\rho_0}$ and the displacement is normalized to the value of
  $\delta$. Details of the FG simulation are given in the text. 
  \label{fig:oscillation1}}
\end{figure}

\begin{figure}
\includegraphics[width=0.5\textwidth]{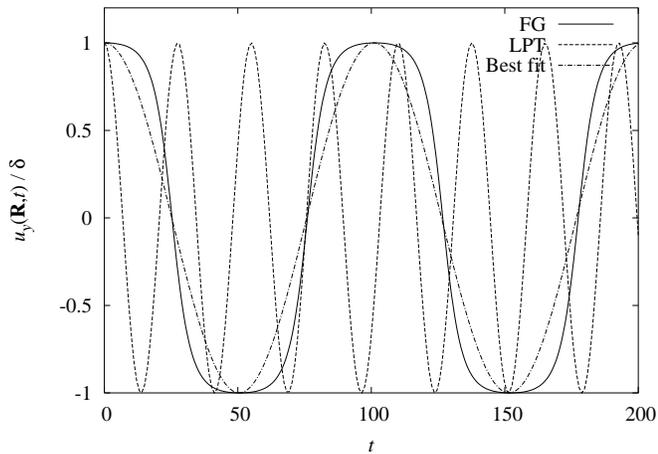}
\caption{Oscillations of a particle with
  initial displacement~\eqref{eq:oscillationdispl} and $\delta=0.248\ell$
  according  to FG (solid line) and PLT (dashed line). The last curve
  (``Best fit'', dashed-dotted line) represents a sine with frequency
  $\sqrt{0.0116 \cdot 4\pi G \rho_0/3}$. It allows one to see that the
  frequency of the oscillations is smaller than the one predicted by
  PLT and that the functional behavior is slightly different from an
  exact sine. The time is in units of $1/\sqrt{4\pi G\rho_0}$ and the displacement is normalized to the value of
  $\delta$. Details of the FG simulation are given in the text. 
  \label{fig:oscillation2}}
\end{figure}

Figure \ref{fig:oscillation1} shows the oscillations along the $y$
axis of a particle according to FG (obtained by using the code we have
just described) and the prediction of PLT. The value of $\delta$ used is
$0.004\, \ell$, for which we find a perfect agreement between FG and
PLT. It is interesting to note that for larger value of $\de$, the
frequencies of the oscillations measured in the FG simulations
decrease and the functional behavior of the oscillations is less and
less close to exactly sinusoidal as can be seen in
Fig.~\ref{fig:oscillation2}. This trend towards a decreasing
frequency [rather than the constant frequency $\om_0$ as 
in Eq.~\eqref{oscil-eq}] as the amplitude increases has a 
simple explanation:  when $\delta = \ell/4$ the full force exactly 
vanishes as the perturbed configuration is in this case again a 
perfect sc lattice. This is also true when $\delta = \ell/2$, but in that case the 
resulting  distribution is just the initial sc lattice. It follows
that if $\delta/\ell \in ]1/4,1/2[$, one
observes the same types of oscillation as for $\delta/\ell \in ]0,1/4[$ due
to the invariance of the system under transformation of the type
$\delta \to j\ell/2 \pm \delta$ with $j$ an integer.
    
\section{Conclusions}

In this paper we have described in detail a new perturbative treatment
which describes the evolution of $N$ self-gravitating particles of
equal mass initially perturbed off a perfect lattice, subject to
periodic boundary conditions, both in a static spacetime and in a
cosmological (expanding) background.  We have reported specifically
the spectrum of eigenvectors and eigenvalues for the modes of the
displacement field on a {\it simple cubic} lattice, which is the case
of relevance in cosmology. While the fluid limit ($N \rightarrow
\infty$) is recovered for long wavelength modes, the full spectrum for
the finite $N$ system contains both modes with negative eigenvalues,
corresponding to oscillations, and modes with exponents greater than
in the fluid limit. Further the eigenvalues depend explicitly not only
on the modulus of the wave-vector $\bk$, but also on its orientation
with respect to the axes of the lattice. The breaking of rotational
invariance in the lattice is thus  imprinted in the evolution of the
system. We have shown, by detailed comparison with numerical
simulations, that the linear order of the scheme has approximately the
same range of validity as the corresponding (linear) order of the
fluid theory, up to when particles come very close to one another
(i.e. up to ``shell-crossing'' in fluid language).  However it traces
the real evolution with greater accuracy than its fluid counterpart.

In the context of cosmological simulations this means that our method
provides a precise tool for quantifying {\it fully}, up to shell
crossing, the effects of discreteness in these simulations: these
effects are nothing other than the difference between the finite $N$
evolution and the fluid limit. In a forthcoming paper
\cite{joyce_06} we will explore more extensively this application, giving
precise quantitative measures of these effects adapted for use in
``correcting'' such N-body simulations.  We conclude this paper by
commenting on a few possible developments of the perturbative theory
we have described here.

Firstly the method can evidently also be extended to higher order,
just as has been done in the analogous treatment of condensed matter
system (see e.g. \cite{carr_61,carr_61b}).  It is straighforward
to generalize Eq.~\eqref{def-dyn} to an expansion to any order. The
$\mu$ component of the force reads:
\bea
\label{expansion_force}
F_{\mu}(\br)&=&\sum_{n=0}^{\infty}\sum_{\bR'\neq \bR}\frac{1}{n!}\mathcal{G}_{\mu,\nu_1\dots\nu_n}^{(n)}(\bR-\bR')\\\nonumber
&\times& [u_{\nu_1}(\bR)-u_{\nu_1}(\bR')]\dots
[u_{\nu_n}(\bR)-u_{\nu_n}(\bR')] .
\eea 
The tensor $\mathcal{G}_{\mu,\nu_1\dots\nu_n}^{(n)}$ is a function only of the
interaction potential $v(\br)$ and is given by:
\be
\label{dynamical_general}
\mathcal{G}_{\mu,\nu_1\dots\nu_n}^{(n)}(\bR)= - \left[ \frac{\partial^{(n+1)}w(\ve r)}{\partial
r_{\mu}\partial r_{\nu_1}\dots\partial r_{\nu_n}} \right]_{\ve r= \ve R}.  
\ee 
The analysis can be followed through at any order in analogy to linear
order. Transforming to reciprocal space the problem simplifies, but
there is now the added complexity of a coupling of modes. A first
calculation of interest would be to map this description, in the fluid
limit, onto the corresponding one at the same order in the fluid
Lagrangian theory. This latter treatment has been explored extensively
in the cosmological literature and compared in detail with numerical
simulations (see e.g. \cite{buchert_93} and references therein). The
study of the corrections to this limit should allow one to get some
insight on the interplay of non-linearity (which is of course also a
feature of the fluid model) and the effects of
discreteness. Non-linearity in gravity involves the transfer of power
from larger to smaller scales, an effect which is often qualitatively
argued to make discreteness effects of less consequence. One might
hope to see such a mechanism at play, if indeed it is there.

The approach presented here may also prove useful in providing insight
about the nature of existing approximations to self-gravitating
systems which go beyond the simple fluid limit, as it provides an
``exact'' evolution of a self-gravitating finite N-body system in a
certain range. For example approximations have been developed to
self-gravitating systems involving pressure terms associated to
velocity dispersion (see
e.g. \cite{buchert_93,buchert_98,tatekawa_02,tatekawa_04b,dominguez_04,buchert_05}). In
principle these terms can be calculated exactly using our perturbative
scheme and the improvement (if this is the case) they allow to the
approximation of the full evolution better understood.

Another direction in which this treatment can be generalised is to the
consideration of other initial configurations. We have analysed here
almost exclusively the case of perturbations from a sc lattice, as
this is the kind of lattice used in cosmological simulations. Our
treatment can easily be generalised to other lattices, and a
comparative study of the discreteness effects should be
straightforward. Without doing any calculation however one simple and
interesting result can be given. It is known that both the bcc and fcc
lattice are stable (or at least meta-stable) configurations for the
Coulomb lattice.  This means that there are no unstable modes. For
gravity this implies that there are \emph{only} unstable modes. There
are therefore no oscillating modes, and thus, by the Kohn sum rule, no
modes with exponent greater than in the fluid limit. Consequently
either of these lattices would appear to be better lattices to use in
N-body simulations, in which one wishes evidently to approach as
closely as possible the fluid evolution. The bcc lattice would appear
to be the more interesting of the two, as it is known to be
\cite{torquato_03} the most densely packed lattice.  It is also a more
isotropic configuration than either the sc or fcc lattice.

Another case of interest to analyse would be that of  ``glassy''
configurations, which are often used as an alternative to the sc
lattice in cosmological simulations
\cite{white_94,jenkins_98,springel_05}. These configurations are
generated by simulating a set of point particles evolving under
negative gravity (i.e. Coulomb forces), with an appropriate damping
term. By doing so one arrives at a configuration in which the forces
are very small, but which is more isotropic than the lattice (in which
the forces are exactly zero). In the approximation that the initial
forces are negligible one could, in principle, carry out the same kind
of analysis as given here. The only difference is that the $3N$
eigenmodes of the displacement field will not be plane waves, which
greatly complicates the analysis compared to the case of the
lattice. Numerically however such a solution should be feasible (for
any specified glassy configuration), and it would be necessary if this
method is to be used to give a precise quantification of discreteness
effects from these IC as we can now give, using the
analysis presented here, for the case of a lattice. We do not expect
the results, however, to be very different (either qualitatively or
quantitively): the effects described here are essentially sampling
effects which depend on the sampling scale (the lattice spacing $\ell$
above) and not on the precise nature of the sampling.  The particular
manifestation of anisotropy which we have observed in the sc lattice
will necessarily be quite different, and likely less pronounced, in
the glassy case, but the average slowing down of growth at smaller
scales would be expected to be very similar in magnitude.

\acknowledgments{

We thank the ``Centro Ricerche e Studi E. Fermi'' (Roma) for the use of
a super-computer for numerical calculations, the EC grant No. 517588
``Statistical Physics for Cosmic Structures" and the MIUR-PRIN05
project on ``Dynamics and Thermodynamics of systems with long range
interactions" for financial supports.  M.\,J. thanks the Instituto dei
Sistemi Complessi for its kind hospitality during October
2006. Finally we thank B. Jancovici, D. Levesque and L. Pietronero for
useful discussions and suggestions.

}

\appendix

\section{Ewald sum of the dynamical matrix $\mD(\bR)$}
\label{ewald-app}

The Ewald sum for the dynamical matrix is given
from \eqref{expr_D} using the Ewald sum for the potential
\eqref{pot_ewald}:
\begin{equation}
\label{dynamical_ewald2}
\mathcal{D}(\bR)=\mathcal{D}^{(r)}(\bR)+\mathcal{D}^{(k)}(\bR)
\end{equation}
with
\begin{eqnarray}
\label{dynamical_ewald_linear_r}
\nonumber
&&\mD_{\mu\nu}^{(r)}(\bR\ne\mathbf0)=-Gm\sum_{\bn}\left[\frac{(\bR-\bn L)_\mu (\bR-\bn L)_\nu}{|\bR-\bn L|^2}\right]\\\nonumber
&\times&\frac{4\alpha^3}{\sqrt{\pi}}\exp(-\alpha^2|\bR-\bn L|^2)\\
&+&Gm\sum_{\bn}\left[\frac{\delta_{\mu\nu} }{|\bR-\bn L|^3}-3\frac{(\bR-\bn L)_\mu (\bR-\bn L)_\nu}{|\bR-\bn L|^5}\right]\\\nonumber
&\times&\left[\mathrm{erfc}(\alpha|\bR-\bn L|)+\frac{2\alpha}{\sqrt{\pi}}\exp(-\alpha^2|\bR-\bn L|^2)|\bR-\bn L|\right]
\end{eqnarray}
and 
\begin{equation}
\label{dynamical_ewald_linear_k}
\mD_{\mu\nu}^{(k)}(\bR)=\frac{4\pi G m}{V_B}\sum_{\bk\ne0}\frac{1}{|\bk|^2}\exp\left(-\frac{|\bk|^2}{4\alpha^2}\right)\cos\left(\bk\cdot\bR\right)k_\mu k_\nu.
\end{equation}
The $\bR=0$ term is 
\be
\label{Ddiag}
\mD(\bR=\mathbf 0)=-\sum_{\bR\ne\mathbf 0} \mD(\bR).
\ee
 Note that, by symmetry, only the first term of the r.h.s. of
\eqref{dynamical_ewald_linear_r} and Eq.~\eqref{dynamical_ewald_linear_k} contribute in the sum of Eq.~\eqref{Ddiag}. In the case of pure gravity the result of the sum \eqref{Ddiag} is given by Eq. \eqref{4pi3}.

\section{Kohn sum rule}
\label{kohn-app}
We derive here the Kohn sum rule \eqref{kohn}. Multiplying
Eq.~(\ref{D_of_phi}) by $(\mathbf{\hat e}_n(\bk))_\mu
(\mathbf{\hat e}_n(\bk))_\nu$ and summing over $n,\ \mu \text{ and }\nu$ we obtain, with Eq.~\eqref{eigen_equation}: 
\bea
\nonumber
\sum_{n=1}^3\om_{n}^2(\bk)&=&-n_0\sum_{n=1}^3 \bigg\{ \tilde w(\bk)(\bk\cdot \mathbf{\hat e}_n(\bk))^2\\\nonumber
&&+\sum_{\bK\ne0}\tilde w(\bk+\bK)\left[(\bk+\bK)\cdot\mathbf{\hat e}_n(\bk)\right]^2\\
&&-\sum_{\bK\ne0}\tilde w(\bK)\left[\bK\cdot\mathbf{\hat e}_n(\bk)\right]^2 \bigg\}.
\eea
Using the orthogonality relation
\be
\sum_{n=1}^3(\mathbf{\hat e}_n(\bk))_{\mu}(\mathbf{\hat e}_n(\bk))_{\nu}=\delta_{\mu\nu},
\ee
we get finally \cite{pines_63}
\bea
\label{sum_rule}
\sum_{i=1}^3\om_{n}^2(\bk)&=&-n_0 k^2\tilde w(\bk)\\
\nonumber
&&-n_0\sum_{\bK\ne0}\left(|\bk+\bK|^2\tilde w(\bk+\bK)-K^2\tilde w(\bK)\right).
\eea
In the case of gravity, using the same approximation as in Eq.~\eqref{FT-grav} we conclude that
\be
\sum_{n=1}^3\om_{n}^2(\bk)=- n_0k^2\tilde w(\bk)=4\pi G\rho_0.
\ee
%

%\bibliography{longpaper}

\begin{thebibliography}{56}
\expandafter\ifx\csname natexlab\endcsname\relax\def\natexlab#1{#1}\fi
\expandafter\ifx\csname bibnamefont\endcsname\relax
  \def\bibnamefont#1{#1}\fi
\expandafter\ifx\csname bibfnamefont\endcsname\relax
  \def\bibfnamefont#1{#1}\fi
\expandafter\ifx\csname citenamefont\endcsname\relax
  \def\citenamefont#1{#1}\fi
\expandafter\ifx\csname url\endcsname\relax
  \def\url#1{\texttt{#1}}\fi
\expandafter\ifx\csname urlprefix\endcsname\relax\def\urlprefix{URL }\fi
\providecommand{\bibinfo}[2]{#2}
\providecommand{\eprint}[2][]{\url{#2}}

\bibitem[{\citenamefont{Peebles}(1980)}]{peebles_80}
\bibinfo{author}{\bibfnamefont{P.~J.~E.} \bibnamefont{Peebles}},
  \emph{\bibinfo{title}{The Large-Scale structure of the Universe}}
  (\bibinfo{publisher}{Princeton University Press},
  \bibinfo{address}{Princeton}, \bibinfo{year}{1980}).

\bibitem[{\citenamefont{Binney and Tremaine}(1987)}]{binney_87}
\bibinfo{author}{\bibfnamefont{J.}~\bibnamefont{Binney}} \bibnamefont{and}
  \bibinfo{author}{\bibfnamefont{S.}~\bibnamefont{Tremaine}},
  \emph{\bibinfo{title}{Galactic Dynamics}} (\bibinfo{publisher}{Princeton
  Series in Astrophysics}, \bibinfo{address}{Princeton}, \bibinfo{year}{1987}).

\bibitem[{\citenamefont{Sahni and Coles}(1995)}]{sahni_95}
\bibinfo{author}{\bibfnamefont{V.}~\bibnamefont{Sahni}} \bibnamefont{and}
  \bibinfo{author}{\bibfnamefont{P.}~\bibnamefont{Coles}},
  \bibinfo{journal}{Phys. Rept.} \textbf{\bibinfo{volume}{262}},
  \bibinfo{pages}{1} (\bibinfo{year}{1995}), \eprint{astro-ph/9505005}.

\bibitem[{\citenamefont{Bernardeau et~al.}(2002)\citenamefont{Bernardeau,
  Colombi, Gaztanaga, and Scoccimarro}}]{bernardeau_02}
\bibinfo{author}{\bibfnamefont{F.}~\bibnamefont{Bernardeau}},
  \bibinfo{author}{\bibfnamefont{S.}~\bibnamefont{Colombi}},
  \bibinfo{author}{\bibfnamefont{E.}~\bibnamefont{Gaztanaga}},
  \bibnamefont{and}
  \bibinfo{author}{\bibfnamefont{R.}~\bibnamefont{Scoccimarro}},
  \bibinfo{journal}{Phys. Rept.} \textbf{\bibinfo{volume}{367}},
  \bibinfo{pages}{1} (\bibinfo{year}{2002}), \eprint{astro-ph/0112551}.

\bibitem[{\citenamefont{Bagla and Padmanabhan}(1997)}]{bagla_97}
\bibinfo{author}{\bibfnamefont{J.~S.} \bibnamefont{Bagla}} \bibnamefont{and}
  \bibinfo{author}{\bibfnamefont{T.}~\bibnamefont{Padmanabhan}},
  \bibinfo{journal}{Pramana} \textbf{\bibinfo{volume}{49}},
  \bibinfo{pages}{161} (\bibinfo{year}{1997}), \eprint{astro-ph/0411730}.

\bibitem[{\citenamefont{{Bertschinger}}(1998)}]{bertschinger_98}
\bibinfo{author}{\bibfnamefont{E.}~\bibnamefont{{Bertschinger}}},
  \bibinfo{journal}{Annu. Rev. Astron. Astrophys.}
  \textbf{\bibinfo{volume}{36}}, \bibinfo{pages}{599} (\bibinfo{year}{1998}).

\bibitem[{\citenamefont{Hockney and Eastwood}(1999)}]{hockney_99}
\bibinfo{author}{\bibfnamefont{R.~W.} \bibnamefont{Hockney}} \bibnamefont{and}
  \bibinfo{author}{\bibfnamefont{J.~W.} \bibnamefont{Eastwood}},
  \emph{\bibinfo{title}{{Computer simulation using particles}}}
  (\bibinfo{publisher}{IOP}, \bibinfo{year}{1999}).

\bibitem[{\citenamefont{Springel et~al.}(2005)}]{springel_05}
\bibinfo{author}{\bibfnamefont{V.}~\bibnamefont{Springel}}
  \bibnamefont{et~al.}, \bibinfo{journal}{Nature}
  \textbf{\bibinfo{volume}{435}}, \bibinfo{pages}{629} (\bibinfo{year}{2005}),
  \eprint{astro-ph/0504097}.

\bibitem[{\citenamefont{Power et~al.}(2003)}]{power_03}
\bibinfo{author}{\bibfnamefont{C.}~\bibnamefont{Power}} \bibnamefont{et~al.},
  \bibinfo{journal}{Mon. Not. Roy. Astron. Soc.}
  \textbf{\bibinfo{volume}{338}}, \bibinfo{pages}{14} (\bibinfo{year}{2003}),
  \eprint{astro-ph/0201544}.

\bibitem[{\citenamefont{Diemand et~al.}(2004)\citenamefont{Diemand, Moore,
  Stadel, and Kazantzidis}}]{diemand_04}
\bibinfo{author}{\bibfnamefont{J.}~\bibnamefont{Diemand}},
  \bibinfo{author}{\bibfnamefont{B.}~\bibnamefont{Moore}},
  \bibinfo{author}{\bibfnamefont{J.}~\bibnamefont{Stadel}}, \bibnamefont{and}
  \bibinfo{author}{\bibfnamefont{S.}~\bibnamefont{Kazantzidis}},
  \bibinfo{journal}{Mon. Not. Roy. Astron. Soc.}
  \textbf{\bibinfo{volume}{348}}, \bibinfo{pages}{977} (\bibinfo{year}{2004}),
  \eprint{astro-ph/0304549}.

\bibitem[{\citenamefont{Baertschiger and Sylos~Labini}(2002)}]{baertschiger_02}
\bibinfo{author}{\bibfnamefont{T.}~\bibnamefont{Baertschiger}}
  \bibnamefont{and}
  \bibinfo{author}{\bibfnamefont{F.}~\bibnamefont{Sylos~Labini}},
  \bibinfo{journal}{Europhys.Lett.} \textbf{\bibinfo{volume}{57}},
  \bibinfo{pages}{322} (\bibinfo{year}{2002}).

\bibitem[{\citenamefont{Dominguez and Knebe}(2002)}]{knebe_02}
\bibinfo{author}{\bibfnamefont{A.}~\bibnamefont{Dominguez}} \bibnamefont{and}
  \bibinfo{author}{\bibfnamefont{A.}~\bibnamefont{Knebe}},
  \bibinfo{journal}{Publ.Astron.Soc.Pac.} \textbf{\bibinfo{volume}{20}},
  \bibinfo{pages}{1} (\bibinfo{year}{2002}).

\bibitem[{\citenamefont{Dominguez and Knebe}(2003)}]{knebe_03}
\bibinfo{author}{\bibfnamefont{A.}~\bibnamefont{Dominguez}} \bibnamefont{and}
  \bibinfo{author}{\bibfnamefont{A.}~\bibnamefont{Knebe}},
  \bibinfo{journal}{Europhys. Lett.} \textbf{\bibinfo{volume}{63}},
  \bibinfo{pages}{631} (\bibinfo{year}{2003}), \eprint{astro-ph/0309381}.

\bibitem[{\citenamefont{Baertschiger and Sylos~Labini}(2003)}]{baertschiger_03}
\bibinfo{author}{\bibfnamefont{T.}~\bibnamefont{Baertschiger}}
  \bibnamefont{and}
  \bibinfo{author}{\bibfnamefont{F.}~\bibnamefont{Sylos~Labini}},
  \bibinfo{journal}{Europhys.Lett.} \textbf{\bibinfo{volume}{63}},
  \bibinfo{pages}{633} (\bibinfo{year}{2003}).

\bibitem[{\citenamefont{Joyce and Marcos}(2004)}]{joyce_04}
\bibinfo{author}{\bibfnamefont{M.}~\bibnamefont{Joyce}} \bibnamefont{and}
  \bibinfo{author}{\bibfnamefont{B.}~\bibnamefont{Marcos}}
  (\bibinfo{year}{2004}), \eprint{astro-ph/0410451}.

\bibitem[{\citenamefont{Joyce et~al.}(2005{\natexlab{a}})\citenamefont{Joyce,
  Levesque, and Marcos}}]{joyce_04b}
\bibinfo{author}{\bibfnamefont{M.}~\bibnamefont{Joyce}},
  \bibinfo{author}{\bibfnamefont{D.}~\bibnamefont{Levesque}}, \bibnamefont{and}
  \bibinfo{author}{\bibfnamefont{B.}~\bibnamefont{Marcos}},
  \bibinfo{journal}{Phys. Rev.} \textbf{\bibinfo{volume}{D72}},
  \bibinfo{pages}{103509} (\bibinfo{year}{2005}{\natexlab{a}}),
  \eprint{astro-ph/0411607}.

\bibitem[{\citenamefont{Melott and Shandarin}(1993)}]{melott_93}
\bibinfo{author}{\bibfnamefont{A.}~\bibnamefont{Melott}} \bibnamefont{and}
  \bibinfo{author}{\bibfnamefont{S.}~\bibnamefont{Shandarin}},
  \bibinfo{journal}{Ap. J.} \textbf{\bibinfo{volume}{410}},
  \bibinfo{pages}{469} (\bibinfo{year}{1993}).

\bibitem[{\citenamefont{Melott et~al.}(1997)\citenamefont{Melott, Splinter, and
  Shandarin}}]{melott_97}
\bibinfo{author}{\bibfnamefont{A.~L.} \bibnamefont{Melott}},
  \bibinfo{author}{\bibfnamefont{R.~J.} \bibnamefont{Splinter}},
  \bibnamefont{and} \bibinfo{author}{\bibfnamefont{S.~F.}
  \bibnamefont{Shandarin}}, \bibinfo{journal}{Ap. J. Lett.}
  \textbf{\bibinfo{volume}{479}}, \bibinfo{pages}{79} (\bibinfo{year}{1997}).

\bibitem[{\citenamefont{Splinter et~al.}(1998)\citenamefont{Splinter, Melott,
  Shandarin, and Suto}}]{splinter_98}
\bibinfo{author}{\bibfnamefont{R.~J.} \bibnamefont{Splinter}},
  \bibinfo{author}{\bibfnamefont{A.~L.} \bibnamefont{Melott}},
  \bibinfo{author}{\bibfnamefont{S.~F.} \bibnamefont{Shandarin}},
  \bibnamefont{and} \bibinfo{author}{\bibfnamefont{Y.}~\bibnamefont{Suto}},
  \bibinfo{journal}{Astrophys. J.} \textbf{\bibinfo{volume}{497}},
  \bibinfo{pages}{38} (\bibinfo{year}{1998}), \eprint{astro-ph/9706099}.

\bibitem[{\citenamefont{Baertschiger et~al.}(2002)\citenamefont{Baertschiger,
  Joyce, and Sylos~Labini}}]{baertschiger_02t}
\bibinfo{author}{\bibfnamefont{T.}~\bibnamefont{Baertschiger}},
  \bibinfo{author}{\bibfnamefont{M.}~\bibnamefont{Joyce}}, \bibnamefont{and}
  \bibinfo{author}{\bibfnamefont{F.}~\bibnamefont{Sylos~Labini}},
  \bibinfo{journal}{Astrophys. J.} \textbf{\bibinfo{volume}{581}},
  \bibinfo{pages}{L63} (\bibinfo{year}{2002}), \eprint{astro-ph/0203087}.

\bibitem[{\citenamefont{Binney and Knebe}(2002)}]{binney_02}
\bibinfo{author}{\bibfnamefont{J.}~\bibnamefont{Binney}} \bibnamefont{and}
  \bibinfo{author}{\bibfnamefont{A.}~\bibnamefont{Knebe}},
  \bibinfo{journal}{Mon. Not. Roy. Astron. Soc.}
  \textbf{\bibinfo{volume}{333}}, \bibinfo{pages}{378} (\bibinfo{year}{2002}),
  \eprint{astro-ph/0105183}.

\bibitem[{\citenamefont{Joyce et~al.}(2005{\natexlab{b}})\citenamefont{Joyce,
  Marcos, Gabrielli, Baertschiger, and Labini}}]{jmgbsl_05}
\bibinfo{author}{\bibfnamefont{M.}~\bibnamefont{Joyce}},
  \bibinfo{author}{\bibfnamefont{B.}~\bibnamefont{Marcos}},
  \bibinfo{author}{\bibfnamefont{A.}~\bibnamefont{Gabrielli}},
  \bibinfo{author}{\bibfnamefont{T.}~\bibnamefont{Baertschiger}},
  \bibnamefont{and} \bibinfo{author}{\bibfnamefont{F.~S.}
  \bibnamefont{Labini}}, \bibinfo{journal}{Phys. Rev. Lett.}
  \textbf{\bibinfo{volume}{95}}, \bibinfo{pages}{011304}
  (\bibinfo{year}{2005}{\natexlab{b}}), \eprint{astro-ph/0504213}.

\bibitem[{\citenamefont{Joyce et~al.}()}]{joyce_06}
\bibinfo{author}{\bibfnamefont{M.}~\bibnamefont{Joyce}} \bibnamefont{et~al.},
  \bibinfo{note}{in preparation}.

\bibitem[{\citenamefont{{Efstathiou} et~al.}(1985)\citenamefont{{Efstathiou},
  {Davis}, {White}, and {Frenk}}}]{efstathiou_85}
\bibinfo{author}{\bibfnamefont{G.}~\bibnamefont{{Efstathiou}}},
  \bibinfo{author}{\bibfnamefont{M.}~\bibnamefont{{Davis}}},
  \bibinfo{author}{\bibfnamefont{S.~D.~M.} \bibnamefont{{White}}},
  \bibnamefont{and} \bibinfo{author}{\bibfnamefont{C.~S.}
  \bibnamefont{{Frenk}}}, \bibinfo{journal}{Astrophys. J. Supp.}
  \textbf{\bibinfo{volume}{57}}, \bibinfo{pages}{241} (\bibinfo{year}{1985}).

\bibitem[{\citenamefont{Navarro et~al.}(1997)\citenamefont{Navarro, Frenk, and
  White}}]{navarro_96}
\bibinfo{author}{\bibfnamefont{J.~F.} \bibnamefont{Navarro}},
  \bibinfo{author}{\bibfnamefont{C.~S.} \bibnamefont{Frenk}}, \bibnamefont{and}
  \bibinfo{author}{\bibfnamefont{S.~D.} \bibnamefont{White}},
  \bibinfo{journal}{Astrophys. J.} \textbf{\bibinfo{volume}{490}},
  \bibinfo{pages}{493} (\bibinfo{year}{1997}).

\bibitem[{\citenamefont{Smith et~al.}(2003)}]{smith_03}
\bibinfo{author}{\bibfnamefont{R.~E.} \bibnamefont{Smith}} \bibnamefont{et~al.}
  (\bibinfo{collaboration}{The Virgo Consortium}), \bibinfo{journal}{Mon. Not.
  Roy. Astron. Soc.} \textbf{\bibinfo{volume}{341}}, \bibinfo{pages}{1311}
  (\bibinfo{year}{2003}), \eprint{astro-ph/0207664}.

\bibitem[{\citenamefont{Clark}(1957)}]{clark_57}
\bibinfo{author}{\bibfnamefont{C.~B.} \bibnamefont{Clark}},
  \bibinfo{journal}{Phys. Rev.} \textbf{\bibinfo{volume}{109}},
  \bibinfo{pages}{1133} (\bibinfo{year}{1957}).

\bibitem[{\citenamefont{Pines}(1963)}]{pines_63}
\bibinfo{author}{\bibfnamefont{D.}~\bibnamefont{Pines}},
  \emph{\bibinfo{title}{Elementary excitations in solids}}
  (\bibinfo{publisher}{Benjamin}, \bibinfo{year}{1963}).

\bibitem[{\citenamefont{Buchert}(1992)}]{buchert_92}
\bibinfo{author}{\bibfnamefont{T.}~\bibnamefont{Buchert}},
  \bibinfo{journal}{Mon. Not. R. astr. Soc.} \textbf{\bibinfo{volume}{254}},
  \bibinfo{pages}{729} (\bibinfo{year}{1992}).

\bibitem[{\citenamefont{Kiessling}(1999)}]{kiessling_99}
\bibinfo{author}{\bibfnamefont{M.~K.~H.} \bibnamefont{Kiessling}}
  (\bibinfo{year}{1999}), \eprint{astro-ph/9910247}.

\bibitem[{\citenamefont{Gabrielli and et~al}()}]{gabrielli_05}
\bibinfo{author}{\bibfnamefont{A.}~\bibnamefont{Gabrielli}} \bibnamefont{and}
  \bibinfo{author}{\bibnamefont{et~al}}, \bibinfo{note}{{\tt astro-ph/}}.

\bibitem[{\citenamefont{Ashcroft and Mermin}(1976)}]{ashcroft_76}
\bibinfo{author}{\bibfnamefont{N.~W.} \bibnamefont{Ashcroft}} \bibnamefont{and}
  \bibinfo{author}{\bibfnamefont{N.~D.} \bibnamefont{Mermin}},
  \emph{\bibinfo{title}{Solid State Physics}} (\bibinfo{publisher}{W.B.
  Saunders}, \bibinfo{address}{Philadelphia}, \bibinfo{year}{1976}).

\bibitem[{\citenamefont{Ziman}(1972)}]{ziman_72}
\bibinfo{author}{\bibfnamefont{J.~M.} \bibnamefont{Ziman}},
  \emph{\bibinfo{title}{Theory of Solids (2nd ed.)}}
  (\bibinfo{publisher}{Cambridge Univ. Press}, \bibinfo{year}{1972}).

\bibitem[{\citenamefont{Ewald}(1921)}]{ewald_21}
\bibinfo{author}{\bibfnamefont{P.}~\bibnamefont{Ewald}}, \bibinfo{journal}{Ann.
  Phys.} \textbf{\bibinfo{volume}{64}}, \bibinfo{pages}{253}
  (\bibinfo{year}{1921}).

\bibitem[{\citenamefont{De~Leeuw et~al.}(1980)\citenamefont{De~Leeuw, Perram,
  and Smith}}]{deleeuw_80}
\bibinfo{author}{\bibfnamefont{S.~W.} \bibnamefont{De~Leeuw}},
  \bibinfo{author}{\bibfnamefont{J.~W.} \bibnamefont{Perram}},
  \bibnamefont{and} \bibinfo{author}{\bibfnamefont{E.~R.} \bibnamefont{Smith}},
  \bibinfo{journal}{Proc. R. Soc. Lond. A} \textbf{\bibinfo{volume}{373}},
  \bibinfo{pages}{27} (\bibinfo{year}{1980}).

\bibitem[{\citenamefont{Nijboer and De~Wette}(1957)}]{nijboer_57}
\bibinfo{author}{\bibfnamefont{B.~R.~A.} \bibnamefont{Nijboer}}
  \bibnamefont{and} \bibinfo{author}{\bibfnamefont{F.~W.}
  \bibnamefont{De~Wette}}, \bibinfo{journal}{Physica}
  \textbf{\bibinfo{volume}{XXIII}}, \bibinfo{pages}{309}
  (\bibinfo{year}{1957}).

\bibitem[{\citenamefont{{Hernquist} et~al.}(1991)\citenamefont{{Hernquist},
  {Bouchet}, and {Suto}}}]{hernquist_91}
\bibinfo{author}{\bibfnamefont{L.}~\bibnamefont{{Hernquist}}},
  \bibinfo{author}{\bibfnamefont{F.~R.} \bibnamefont{{Bouchet}}},
  \bibnamefont{and} \bibinfo{author}{\bibfnamefont{Y.}~\bibnamefont{{Suto}}},
  \bibinfo{journal}{Astrophys. J. Supp.} \textbf{\bibinfo{volume}{75}},
  \bibinfo{pages}{231} (\bibinfo{year}{1991}).

\bibitem[{\citenamefont{Coldwell-Horsfall and Maradunin}(1960)}]{coldwell_60}
\bibinfo{author}{\bibfnamefont{R.~A.} \bibnamefont{Coldwell-Horsfall}}
  \bibnamefont{and} \bibinfo{author}{\bibfnamefont{A.~A.}
  \bibnamefont{Maradunin}}, \bibinfo{journal}{Journal of Mathematical Physics}
  \textbf{\bibinfo{volume}{1}}, \bibinfo{pages}{395} (\bibinfo{year}{1960}).

\bibitem[{\citenamefont{Fuchs}(1935)}]{fuchs_35}
\bibinfo{author}{\bibfnamefont{K.}~\bibnamefont{Fuchs}},
  \bibinfo{journal}{Proc. Roy. Soc.} \textbf{\bibinfo{volume}{A151}},
  \bibinfo{pages}{585} (\bibinfo{year}{1935}).

\bibitem[{\citenamefont{Carr}(1961)}]{carr_61}
\bibinfo{author}{\bibfnamefont{H.}~\bibnamefont{Carr}}, \bibinfo{journal}{Phys.
  Rev.} \textbf{\bibinfo{volume}{{122}}}, \bibinfo{pages}{1437}
  (\bibinfo{year}{1961}).

\bibitem[{\citenamefont{Bertschinger}(1995)}]{bertschinger_95}
\bibinfo{author}{\bibfnamefont{E.}~\bibnamefont{Bertschinger}}
  (\bibinfo{year}{1995}), \eprint{astro-ph/9506070}.

\bibitem[{\citenamefont{Gabrielli et~al.}(2002)\citenamefont{Gabrielli, Joyce,
  and Sylos~Labini}}]{glass}
\bibinfo{author}{\bibfnamefont{A.}~\bibnamefont{Gabrielli}},
  \bibinfo{author}{\bibfnamefont{M.}~\bibnamefont{Joyce}}, \bibnamefont{and}
  \bibinfo{author}{\bibfnamefont{F.}~\bibnamefont{Sylos~Labini}},
  \bibinfo{journal}{Phys. Rev. D} \textbf{\bibinfo{volume}{65}},
  \bibinfo{pages}{083523} (\bibinfo{year}{2002}).

\bibitem[{\citenamefont{Gabrielli et~al.}(2005)\citenamefont{Gabrielli,
  Sylos~Labini, Joyce, and Pietronero}}]{GSJP_05}
\bibinfo{author}{\bibfnamefont{A.}~\bibnamefont{Gabrielli}},
  \bibinfo{author}{\bibfnamefont{F.}~\bibnamefont{Sylos~Labini}},
  \bibinfo{author}{\bibfnamefont{M.}~\bibnamefont{Joyce}}, \bibnamefont{and}
  \bibinfo{author}{\bibfnamefont{L.}~\bibnamefont{Pietronero}},
  \emph{\bibinfo{title}{Statistical Physics for Cosmic Structures}}
  (\bibinfo{publisher}{Springer-Verlag}, \bibinfo{address}{Berlin},
  \bibinfo{year}{2005}).

\bibitem[{\citenamefont{Gabrielli}(2004)}]{gabrielli_04}
\bibinfo{author}{\bibfnamefont{A.}~\bibnamefont{Gabrielli}},
  \bibinfo{journal}{Phys. Rev. E} \textbf{\bibinfo{volume}{70}},
  \bibinfo{pages}{066131} (\bibinfo{year}{2004}).

\bibitem[{\citenamefont{{Springel} et~al.}(2001)\citenamefont{{Springel},
  {Yoshida}, and {White}}}]{gadget}
\bibinfo{author}{\bibfnamefont{V.}~\bibnamefont{{Springel}}},
  \bibinfo{author}{\bibfnamefont{N.}~\bibnamefont{{Yoshida}}},
  \bibnamefont{and} \bibinfo{author}{\bibfnamefont{S.~D.~M.}
  \bibnamefont{{White}}}, \bibinfo{journal}{New Astronomy}
  \textbf{\bibinfo{volume}{6}}, \bibinfo{pages}{79} (\bibinfo{year}{2001}).

\bibitem[{\citenamefont{Gabrielli et~al.}(2003)\citenamefont{Gabrielli, Joyce,
  Marcos, and Viot}}]{causality}
\bibinfo{author}{\bibfnamefont{A.}~\bibnamefont{Gabrielli}},
  \bibinfo{author}{\bibfnamefont{M.}~\bibnamefont{Joyce}},
  \bibinfo{author}{\bibfnamefont{B.}~\bibnamefont{Marcos}}, \bibnamefont{and}
  \bibinfo{author}{\bibfnamefont{P.}~\bibnamefont{Viot}}
  (\bibinfo{year}{2003}), \eprint{astro-ph/0303169}.

\bibitem[{\citenamefont{Carr et~al.}(1961)\citenamefont{Carr, Codwell-Horsfall,
  and Fein}}]{carr_61b}
\bibinfo{author}{\bibfnamefont{W.}~\bibnamefont{Carr}},
  \bibinfo{author}{\bibfnamefont{R.~A.} \bibnamefont{Codwell-Horsfall}},
  \bibnamefont{and} \bibinfo{author}{\bibfnamefont{A.}~\bibnamefont{Fein}},
  \bibinfo{journal}{Phys. Rev.} \textbf{\bibinfo{volume}{124}},
  \bibinfo{pages}{747} (\bibinfo{year}{1961}).

\bibitem[{\citenamefont{Buchert and Weiss}(1993)}]{buchert_93}
\bibinfo{author}{\bibfnamefont{T.}~\bibnamefont{Buchert}} \bibnamefont{and}
  \bibinfo{author}{\bibfnamefont{A.~G.} \bibnamefont{Weiss}},
  \bibinfo{journal}{Proc. IAP Colloq.} \textbf{\bibinfo{volume}{9}},
  \bibinfo{pages}{517} (\bibinfo{year}{1993}), \eprint{astro-ph/9310022}.

\bibitem[{\citenamefont{Buchert and Dom\'{\i}nguez}(1998)}]{buchert_98}
\bibinfo{author}{\bibfnamefont{T.}~\bibnamefont{Buchert}} \bibnamefont{and}
  \bibinfo{author}{\bibfnamefont{A.}~\bibnamefont{Dom\'{\i}nguez}},
  \bibinfo{journal}{Astron. Astrophys.} \textbf{\bibinfo{volume}{335}},
  \bibinfo{pages}{395} (\bibinfo{year}{1998}).

\bibitem[{\citenamefont{Tatekawa et~al.}(2002)\citenamefont{Tatekawa, Suda,
  Maeda, Morita, and Anzai}}]{tatekawa_02}
\bibinfo{author}{\bibfnamefont{T.}~\bibnamefont{Tatekawa}},
  \bibinfo{author}{\bibfnamefont{M.}~\bibnamefont{Suda}},
  \bibinfo{author}{\bibfnamefont{K.-i.} \bibnamefont{Maeda}},
  \bibinfo{author}{\bibfnamefont{M.}~\bibnamefont{Morita}}, \bibnamefont{and}
  \bibinfo{author}{\bibfnamefont{H.}~\bibnamefont{Anzai}},
  \bibinfo{journal}{Phys. Rev.} \textbf{\bibinfo{volume}{D66}},
  \bibinfo{pages}{064014} (\bibinfo{year}{2002}), \eprint{astro-ph/0205017}.

\bibitem[{\citenamefont{Tatekawa}(2004)}]{tatekawa_04b}
\bibinfo{author}{\bibfnamefont{T.}~\bibnamefont{Tatekawa}},
  \bibinfo{journal}{Phys. Rev.} \textbf{\bibinfo{volume}{D70}},
  \bibinfo{pages}{064010} (\bibinfo{year}{2004}), \eprint{astro-ph/0408120}.

\bibitem[{\citenamefont{Dominguez and Melott}(2004)}]{dominguez_04}
\bibinfo{author}{\bibfnamefont{A.}~\bibnamefont{Dominguez}} \bibnamefont{and}
  \bibinfo{author}{\bibfnamefont{A.~L.} \bibnamefont{Melott}},
  \bibinfo{journal}{Astron. Astrophys.} \textbf{\bibinfo{volume}{419}},
  \bibinfo{pages}{425} (\bibinfo{year}{2004}), \eprint{astro-ph/0310693}.

\bibitem[{\citenamefont{Buchert and Dom\'{\i}nguez}(2005)}]{buchert_05}
\bibinfo{author}{\bibfnamefont{T.}~\bibnamefont{Buchert}} \bibnamefont{and}
  \bibinfo{author}{\bibfnamefont{A.}~\bibnamefont{Dom\'{\i}nguez}},
  \bibinfo{journal}{Astronomy \& Astrophysics}  (\bibinfo{year}{2005}).

\bibitem[{\citenamefont{Torquato and Stillinger}(2003)}]{torquato_03}
\bibinfo{author}{\bibfnamefont{S.}~\bibnamefont{Torquato}} \bibnamefont{and}
  \bibinfo{author}{\bibfnamefont{F.~H.} \bibnamefont{Stillinger}},
  \bibinfo{journal}{Phys. Rev. E} \textbf{\bibinfo{volume}{68}},
  \bibinfo{pages}{041113} (\bibinfo{year}{2003}).

\bibitem[{\citenamefont{White}(1994)}]{white_94}
\bibinfo{author}{\bibfnamefont{S.~D.~M.} \bibnamefont{White}}
  (\bibinfo{year}{1994}), \eprint{astro-ph/9410043}.

\bibitem[{\citenamefont{Jenkins et~al.}(1998)}]{jenkins_98}
\bibinfo{author}{\bibfnamefont{A.}~\bibnamefont{Jenkins}} \bibnamefont{et~al.}
  (\bibinfo{collaboration}{Virgo Consortium}), \bibinfo{journal}{Astrophys. J.}
  \textbf{\bibinfo{volume}{499}}, \bibinfo{pages}{20} (\bibinfo{year}{1998}),
  \eprint{astro-ph/9709010}.

\end{thebibliography}

\end{document}